\documentclass[reqno,titlepage]{amsart}

\usepackage{xr}
\usepackage[colorlinks=true,citecolor=black,linkcolor=black,urlcolor=black]{hyperref}

\usepackage{appendix}
\usepackage[longnamesfirst,round]{natbib}
\usepackage{multibib}

\usepackage[scaled]{helvet}
\usepackage{amsmath, amssymb, amsthm} 
\usepackage{dsfont} 
\usepackage{graphicx} 
\usepackage{mathrsfs}
\usepackage{setspace}
\usepackage{upgreek} 
\usepackage{ushort}
\usepackage{bm}

\usepackage[svgnames]{xcolor}
\usepackage{pgf}
\usepackage{tikz}

\usepackage{soul}
\usepackage{mathrsfs}
\usepackage{stmaryrd}
\usepackage{layout}
\usepackage{wasysym}
\usepackage{paralist}
\usepackage{multirow}

\usepackage{todonotes}		% [disable]  % margin notes
\usepackage{footnote}

\usepackage{chngcntr}
\usepackage{apptools}
\usepackage{subcaption}

 \usepackage{bm}

\newcommand{\lowmathcal}[1]{\operatorname{\text{\usefont{U}{BOONDOX-cal}{m}{n}#1}}}

\usepackage[english]{babel}
\usepackage[longnamesfirst,round]{natbib}

\usepackage{bookmark} % [colorlinks=true,citecolor=black,linkcolor=black]

\usepackage[affil-it]{authblk}
\makesavenoteenv{tabular}
\makesavenoteenv{table}

\newtheorem{assump}{Assumption}

\newtheorem{defin}{Definition}
\newtheorem{lemma}{Lemma}

\newtheorem{alemma}{Lemma}
\AtAppendix{\counterwithin{alemma}{section}}

\AtAppendix{\counterwithin{aassump}{section}}

\newtheorem{theorem}{Theorem}

\newtheorem{example}{Example}
\newtheorem{subexample}{Example}[example]

\newtheorem{innercustomgeneric}{\customgenericname}
\providecommand{\customgenericname}{}
\newcommand{\newcustomtheorem}[2]{%
  \newenvironment{#1}[1]
  {%
   \renewcommand\customgenericname{#2}%
   \renewcommand\theinnercustomgeneric{##1}%
   \innercustomgeneric
  }
  {\endinnercustomgeneric}
}

\newcustomtheorem{customexample}{Example}

\DeclareMathAlphabet{\mathpzc}{OT1}{pzc}{m}{it}

\newcommand{\mailingaddressSJ}{Department of Economics, Penn State University, 619 Kern Graduate Building, University Park, PA 16802}
\newcommand{\mailingaddressFZ}{College of Business (Department of Economics), University of Nebraska--Lincoln, 730 N 14th St., HLH 525J, Lincoln, NE 68588}
\newcommand{\emailaddress}[1]{\href{mailto:#1}{#1}
}

\newcommand{\Prob}{{\mathbb{P}}}
\newcommand{\E}{{\mathbb{E}}}
\renewcommand{\Pr}{\mathbb{P}}

\newcommand{\convp}{\stackrel{p}{\rightarrow}}
\newcommand{\convd}{\stackrel{d}{\rightarrow}}

\DeclareMathOperator*{\argmax}{argmax}
\DeclareMathOperator*{\argmin}{argmin}

\newcommand\raiseT[2]{\raisebox{0.25ex}{$#1#2$}}
\newcommand\tr{{\mathpalette\raiseT{\intercal}}}

\newcommand{\one}{\mathds{1}}
\newcommand{\Exp}{\mathbb{E}}

\newcommand{\abstraction}{\noindent\textbf{Abstract.}
We model unobserved confounding through an unknown finite number of latent types. This assumption induces finite-mixture representations of the treated and control outcome distributions. Using the identified mixture components, we characterize the sharp identified set for the number of latent types and derive the sharp identified set for the average treatment effect (ATE) corresponding to each admissible value, thereby providing a natural framework for sensitivity analysis. We further obtain a cutoff beyond which the identified set for the ATE coincides with a version of the Manski bounds, whereas below the cutoff it is strictly smaller. This cutoff grows only linearly with the numbers of mixture components in the treated and control groups, although the maximum admissible number of latent types grows quadratically. We also provide estimation and inference procedures with asymptotic guarantees and illustrate our methodology using LaLonde's data. 
}

\newcommand{\BASELINESTRETCH}{1.5}
\begin{document}

\bibliographystyle{econometrica}

\begin{titlepage}

\renewcommand{\thefootnote}{\fnsymbol{footnote}}

\title{TITLE}

\author[Jun and Zincenko]{Sung Jae Jun and Federico Zincenko}

\begin{center}
    \Large\textsc{Sensitivity Analysis for the Average Treatment Effect under Discrete Unobserved Confounders\footnote{We thank Jacob Dorn, Jin Hahn, Rosa Matzkin, Myungkou Shin, and Danny Tannenbaum for their helpful comments and discussions, Izzat Ahmad Adly for excellent research assistance, and participants at the 2025 UCLA Econometrics Mini Conference, 2025 Midwest Econometrics Group Meeting, and 2026 NASM of the Econometric Society.}}
     \vspace{1ex}
   \end{center}

\vspace{.3cm}

\begin{center}
	\begin{tabular}{ccc}
	 \large\textsc{Sung Jae Jun\footnote{\mailingaddressSJ, \emailaddress{suj14@psu.edu}}} 
	 &
	 & 
	 \large\textsc{Federico Zincenko\footnote{\mailingaddressFZ, \emailaddress{fzincenko2@unl.edu}}} \\  
	 \small\textsc{Penn State Univ.}
	 &
	 &
	 \small\textsc{Univ.\ of Nebraska--Lincoln}		
	\end{tabular}
\end{center}

\begin{center}
\vspace{.3cm}
\today \\
\vspace{.3cm}
\end{center}

\noindent
\abstraction

\bigskip

\noindent\textbf{Keywords:} Average treatment effect, Discrete unobserved confounders, Finite mixture, Partial identification, Potential outcome, Sensitivity analysis.

\bigskip

\noindent\textbf{JEL classification codes:} C10, C21, C51.

\end{titlepage}

\renewcommand{\thefootnote}{\arabic{footnote}\hspace{0.1ex}}
\setcounter{footnote}{0}

\raggedbottom

%%%%%%%%%%%%%%%%%%%%%%%%%%%%%%%%%%%%%%%%%%%%%%%%%%%

\section{Introduction}

Randomization or unconfoundedness is probably the most important assumption for causal inference: see e.g., \citet{rubin1974estimating,rubin1978bayesian,lalonde1986evaluating,hahn1998role,hirano2003efficient,ix25lalonde}. However, it can be too strong or even unrealistic in observational studies, in which case it is a common approach to rely on instrumental variables. But, it can be difficult to find credible instruments, and more importantly, they often require the researcher change the target population of interest, depending on the specific instruments available: see e.g., \citet{imbens1994late}. Therefore, focusing on the case of a binary treatment, we develop a new methodology to learn about the average treatment effect (ATE) without relying on unconfoundedness or instrumental variables. 

Our starting point is to note that the presence of an unobserved confounder is a key source of the problem. One possibility to proceed is to exploit a proxy variable for the unobserved confounder as in \citet{chalak2019identification}. But, proxies are not always available, and therefore we focus on the case where we observe only the outcome and treatment, possibly along with exogenous covariates. 

Following \citet{imbens2003sensitivity}, we start with assuming that there exists an unobserved random variable $G$ such that the treatment assignment and potential outcomes become independent once we condition on $G$: there may be additional observed covariates that need to be controlled for, but we suppress them from our discussion for simplicity.  So, $G$ represents an unobserved confounder. Our key assumption is then that $G$ has finite support, where the number $\bar G$ of the mass points of $G$ is unknown: as $\bar G$ increases, the population exhibits a higher degree of unobserved heterogeneity.  Therefore, our work builds upon the sensitivity analysis of \citet{imbens2003sensitivity}, as well as modern techniques that exploit discretization of unobserved heterogeneity as in e.g., \citet{bonhomme2022discretizing}.

Sensitivity analyses for causal inference in this context have been discussed by several authors: e.g., \citet{rosenbaum2002sensitivity,imbens2003sensitivity,masten2018identification,yadlowsky2022bounds,bonvini2022sensitivity}. The ideas are all similar: i.e., the distribution of the potential outcomes and treatment assignment is allowed to deviate from complete independence, and the identified set for ATE is traced out as a function of the degree of the deviation. The last four references are particularly relevant for us.

\citet{imbens2003sensitivity} and \citet{yadlowsky2022bounds} formulate sensitivity analysis through explicit models of unobserved confounders. By contrast, \citet{masten2018identification} and \citet{bonvini2022sensitivity} remain agnostic about the underlying confounding mechanism. Specifically, \citet{masten2018identification} bound the discrepancy between the marginal probability of treatment and its conditional counterpart given the potential outcomes. Meanwhile, \citet{bonvini2022sensitivity} posit that unconfoundedness may fail for a subset of the population and treat the proportion of such individuals as a sensitivity parameter.

\citet[pp 321--322]{masten2018identification} give suggestions for how to interpret their sensitivity parameter, and by following the view advocated in \citet[pp 325]{rosenbaum2002rejoinder}, they also argue that interpretations of sensitivity parameters are not always necessary. In this view,  economic interpretability or specific reasons for deviations from the baseline model are not essential for sensitivity analysis; what matters is simply that the baseline model may be false for some reason. We take a different stance. We believe that it is important to understand what a particular deviation represents. A researcher may want to assess how plausible a deviation is, or how seriously she should take potential deviations in a given empirical setting. Such reasoning is facilitated when deviations from the baseline model have a clear economic interpretation. Moreover, an interpretable framework enables the researcher to gauge the relative likelihood of different deviations and, consequently, to ``weight'' the resulting causal inferences accordingly. Other authors have also advocated the importance and usefulness of interpretability of sensitivity analysis. For instance, \citet[pp 91]{heckman2000causal} wrote, ``The bounding and sensitivity analysis movement is likely to be more influential if it relies on explicit economic models and uses economically interpretable models to conduct semiparametric bounding and sensitivity analyses.''   See also \citet[pp 65]{cinelli2020making} from the statistics community.

We therefore prefer an explicit approach on the presence of unobserved confounders, as it makes it easier to provide an economic interpretation of deviations from the baseline model. However, both of the two explicit frameworks we mentioned earlier have their own limitations. \citet{imbens2003sensitivity} adopts a fully parametric approach, where the complete joint distribution of the potential outcomes, treatment assignment, and unobserved confounder needs to be specified. This can be restrictive. \citet{yadlowsky2022bounds}'s approach is nonparametric but is less transparent since it restricts the odds-ratio of the treatment assignment given the unobserved confounder. Therefore, our goal is to develop an explicit and easy-to-interpret framework that is less restrictive than \citet{imbens2003sensitivity}.

Similarly to \citet{imbens2003sensitivity}, we begin by assuming Gaussianity for the potential outcome distribution, but our setup is more flexible. Specifically, we assume that the potential outcome $Y(d)$ is Gaussian with mean $\mu_g(d)$ conditional on $G=g$, where $d\in\{0,1\}$ indicates the treatment status, and $G$ is the only source of potential confounding. The variable $G$ is assumed to be discrete, but the number $\bar G$ of its support points is unknown. To ensure that the group label $g$ represents a well-defined type, we assume that $\bigl( \mu_g(0), \mu_g(1) \bigr) \neq  \bigl( \mu_{\tilde g}(0), \mu_{\tilde g}(1) \bigr)$ whenever $g \neq \tilde g$, where the ordering of the group label $g$ is normalized by the lexicographic ordering of the mean vectors of the potential outcomes; this ordering remains well-defined even if $G$ is multi-dimensional, provided that its elements are all discrete. For notational convenience, we label the support of $G$ as $\{1,2,\cdots, \bar G\}$.  Finally, we leave the dependence between $G$ and the treatment assignment unspecified.

In this formulation, heterogeneity in treatment effects arises from two components: mean heterogeneity across latent types $G$ and an idiosyncratic residual component, although only the former is relevant for ATE.  While this structure is shared with \citet{imbens2003sensitivity}, we do not impose functional-form restrictions on the mean heterogeneity. In particular, the location component of $Y(d)$ given $G$, i.e., $\sum_{g=1}^{\bar G}\one(G=g)\mu_g(d)$, is unrestricted apart from its finite support. As a result, the Gaussianity assumption on the residual component imposes relatively weak restrictions on the marginal distribution of $Y(d)$. Indeed, finite Gaussian mixtures are dense in a broad class of distributions, so they can approximate a wide range of distributions with arbitrary accuracy as $\bar G$ becomes sufficiently large.

It is an old idea to decompose heterogeneity in treatment effects into two components. For example, \citet{imbens2003sensitivity} incorporates this through a fully parametric structure for sensivity analysis, while \citet{abadie2024instrumental} use a similar decomposition to improve the asymptotic mean squared error of linear instrumental variable estimators. We adopt this decomposition for causal inference in the absence of instrumental variables. An important consequence of our formulation is that the resulting version of the ``Manski bounds'' on ATE remains finite even when the outcome variable has unbounded support.

We note that \citet{gardner2020identification} also exploits discrete unobserved confounders for causal inference. However, \citet{gardner2020identification} implicitly assumes that the number of latent types in the treatment group always coincides with that in the control group, and that the researcher knows how the types revealed in one group correspond to those revealed in the other; see Footnote \ref{fn:gardener} for details. As a result, \citet{gardner2020identification} point-identifies ATE regardless of how large $\bar G$ is. We view these assumptions as too strong. Instead, we derive sharp identified sets for ATE as a function of a candidate value of $\bar G$ in its identified set. We then characterize how these sets vary with $\bar G$ and compare them with the Manski bounds.

The exact economic interpretation of $G$ depends on the context, but we can generally view $\bar G$, the number of mass points of $G$, as the number of ``unobserved types'' in the population that may affect both the treatment assignment and the potential outcomes: e.g., different skill levels or various industries.  In essence, $\bar G$ is a key parameter that describes the unknown degree of heterogeneity in the population. Our sensitivity analysis then involves expressing the sharp identified set for ATE as a function of $\bar G$. However, we do not need to trace out all integers for a complete sensitivity analysis. Indeed, we show that $\bar G$ is partially identified, and therefore there are only finitely many values of $\bar G$ that are consistent with the data.  Further, we show that there is a special cutoff value in the identified set of $\bar G$ such that whenever $\bar G$ is larger than the cutoff, the sharp identified set for ATE becomes a version of the Manski bounds \citep{manski2003partial,manski2010partial}: we say that it is a version of the Manski bounds because we do not assume that the support of the outcome is bounded, but our bounds have the same structure as the Manksi bounds except that they use the means of the mixture component distributions of the outcome given the treatment status.

The fact that the unobserved confounder $G$ has finitely many mass points leads to finite mixture models of the observed outcome given the treatment status. Finite mixtures have been frequently used in econometrics: e.g., \citet{ichimura1998maximum,arci03fin,kasahara2009nonparametric,hen14par,compiani2016using}. In our framework, however, they serve as a tool for causal inference. To be more specific, let $D$ denote the observed treatment status, and let $Y = D Y(1) + (1-D) Y(0)$ be the observed outcome. Then, the distribution of $Y$ given $D=d$ is a finite Gaussian mixture. Consequently, the data generate two finite mixture models, one for each treatment status, whose parameters can be identified using standard methods.  The Gaussianity assumption is introduced solely to guarantee identification of the mixture parameters, although similar identification results can be obtained under alternative assumptions: see e.g., \citet{yak68id,mp00finmix}. Indeed, we show how the baseline model can be extended to incorporate exogenous covariates, account for sample selection, and accommodate non-Gaussian distributions of $Y(d)$ given $G$. The combination of distributional assumptions and discrete unobserved heterogeneity has also been employed by \citet{bonhomme2022discretizing} in the context of panel data models. Our contribution is to exploit this combination for causal inference and sensitivity analysis.

One of the challenges for causal inference here is that different types in the population do not always correspond to distinct component distributions of $Y$ given $D=d$. Therefore, neither the number $\check G(1)$ of mixing components in the treatment group, nor that in the control group denoted by $\check G(0)$, generally reveals the true value of $\bar G$. Nevertheless, we show that the sharp identified set for $\bar G$ can be expressed as a function of $\check G(1)$ and $\check G(0)$, both of which can be estimated at an arbitrarily fast rate: see e.g., \citet{chen09order}.

A complete sensitivity analysis can then be conducted by tracing out the sharp identified set for ATE for each value of $\bar G$ in its identified set. The number of admissible values of $\bar G$ increases quadratically in general as $\check G(1)$ and $\check G(0)$ increases. However, our identification results show that there exists a cutoff value $\bar G_C$ such that $\bar G_C$ increases linearly as $\check G(1)$ and $\check G(0)$ increase, and that the sharp identified set for ATE corresponds to (a version of) the Manski bounds whenever $\bar G$ is larger than $\bar G_C$. Therefore, the total number of admissible values of $\bar G$ that need to be checked for a complete sensitivity analysis is finite and increases only linearly as $\check G(0)$ and $\check G(1)$ increase. 

For the purpose of estimation, we take a plug-in approach. Specifically, we first estimate the mixture orders of the observed outcome for the treatment and control groups, for which we suggest using the method of \citet{chen09order}. We then estimate the mixture parameters by MLE, replacing the unknown mixture orders with their first-step estimates. Finally, for each admissible value of $\bar G$, we plug these estimators into the expression for the identified set for ATE. In the supplement, we establish the asymptotic properties of the proposed estimators, including $\sqrt{n}$-consistency, and develops valid inference procedures.

The proposed method provides applied researchers with a tool for program evaluation that does not depend on either the unconfoundedness assumption nor the availability of good instruments: see, e.g., the surveys by \citet{iw09survey,ix25lalonde}. For instance, even when the treatment group comes from experimental data and the control group from observational data, our approach continues to provide a valid method for learning about ATE, although we refer interested readers to \citet{yang2025cross} for a more systematic approach to integrating experimental and observational data for causal inference. In addition, our approach is not tied to particular research designs such as difference-in-differences or regression discontinuity designs, nor does it require the design-specific identification assumptions underlying those approaches, such as parallel trends or exogenous running variables.

Finally, we illustrate the practical value of our approach by applying it to the widely studied dataset of \citet{lalonde1986evaluating}. The experimental control group in this dataset contains many observations with zero earnings. Excluding these observations raises concerns about selection bias. To address this issue, we apply both our baseline model and its modified version for selection. The results are not overly different. The benchmark Heckit regression results, which can be interpreted as the case of $\bar G=1$, show that the coefficient of the training indicator is approximately 0.07. However, when we allow the presence of an unobserved confounder, the admissible set of $\bar G$ is $\{2,3,4\}$. Among these values, $\bar G\geq 3$ produces the Manski bounds, whereas $\bar G= 2$ yields sharp identified $\{ 0.28, 0.56 \}$ for the average effect of job training.

The rest of the paper is organized as follows. Section \ref{sec:section2} describes the setup, and introduces the identifiable mixture parameters and group clustering issues. Section \ref{sec:identification} presents our identification results, characterizing how the identified set of ATE changes as a function of $\bar G$. Section \ref{sec:esti} describes the estimation procedure. Section \ref{sec:exten} develops extensions to allow for covariates, selection, and non-Gaussian distributions. Section \ref{sec:empirical example} provides an empirical illustration and, finally, Section \ref{sec:conclude} concludes. All proofs are collected in the Appendix. In addition, we provide a Supplement that includes the proofs of the auxiliary lemmas used in the Appendix and the asymptotic properties of the estimator together with the inference procedures.

Before we proceed, we make brief comments on our notation.

\subsection*{Notation and Terminology}

An array with its elements separated by commas such as $v =  (v_1,\cdots, v_m)$ is always considered a column vector, as well as any element of $\mathbb{R}^m$. Given two vectors $v \in \mathbb{R}^{m_1}$ and $w \in \mathbb{R}^{m_2}$, the array $( v , w)$ is considered a (column) vector in $\mathbb{R}^{m_1 + m_2}$. For any subset of $\mathbb{R}^m$, the lexicographical order is our default order relation, and we use the symbol $\prec$ to denote it. For any subset of matrices, its elements are lexicographically ordered by using the row-wise vectorization. Given an ordered set $S = \{ s_1 \prec   \cdots  \prec s_m \}$, we write $( v_s )_{s \in S} = ( v_{s_1}, \cdots  , v_{s_m} )$. The super-script $^\tr$ means transpose. In addition, we say that a tuple $(S_1, S_2, \cdots, S_k)$ of subsets of $S$ is an ordered partition of a set $S$ if it satisfies: (i) $S_i \neq \emptyset$ for all $i=1,\dots,k$; (ii) $S_i \cap S_j = \emptyset$ for all $i \neq j$; and (iii) $\bigcup_{i=1}^k S_i = S$.

We employ the usual notation $N ( a , b)$ to represent a normal distribution with mean $a \in \mathbb{R}$ and variance $b >0$, while $N ( a , 0 )$ denotes a degenerate distribution that assigns probability one to the value $a$. With a slight abuse of notation, for $A \in \mathbb{R}^m$ and a \emph{positive semi-definite} matrix $B \in \mathbb{R}^{m \times m}$, we write $Y \sim N ( A , B  )$ to mean that $v^\tr Y \sim  N ( v^\tr A , v^\tr B  v )$ for all $v \in \mathbb{R}^m$. The symbols $\convp$ and $\convd$ denote convergence in probability and distribution, respectively, while w.p.a.1 abbreviates ``with probability approaching one.''

\section{Endogenous Treatments, Unobserved Confounders, and Gaussian Mixtures} \label{sec:section2}

In this section, we show how causal effects can be identified when treatment assignment may be correlated with potential outcomes, but only through discrete confounding factors.  Specifically, we assume that there are only finitely many unobserved innate types, where the number of types is unknown. Further, we assume that the potential outcomes of interest follow Gaussian mixtures, where the number of mixing components, which is unknown as well, is constrained by that of the innate types. Gaussianity is not essential for our discussion, but it is convenient because identifiability of Gaussian mixtures is well understood \cite[]{yak68id}. An extension to a selection model is discussed in Section \ref{sec:sampleselect}, while extensions to other non-Gaussian mixture models are presented in Section \ref{sec:extensions}. The treatment assignment can be correlated with the unobserved types, and therefore the types are unobserved confounding factors. The causal parameter of interest is the average treatment effect (ATE).

\subsection{The Setup}		\label{sec:setup}

Let $D \in \{ 0 , 1\}$ be a binary treatment, and let $Y(d)$ for $d\in \{0,1\}$ be potential outcomes.  The observed outcome is $Y = DY(1) + (1-D) Y(0)$. For the purpose of identification analysis, we assume that the joint distribution of $(Y,D)$ is known. We assume that there are finitely many innate types, which are represented by $G$.  The econometrician does not observe $G$: all she knows about it is that it is discrete with finite support: i.e., $G$ takes a value from $\{1,2,\cdots, \bar G\}$, where $\bar G \geq 2$ is unknown.  There may be observable covariates $X$, but we suppress them throughout the discussion for expositional simplicity. Therefore, all statements should be understood as conditional on $X=x$ when such covariates are present. We also discuss an alternative approach to incorporating covariates in Section \ref{sec:sampleselect}, which is more restrictive but more pragmatic for implementation.  

We now assume that the potential outcomes can be modeled by Gaussian mixtures.
\begin{assump} \label{ass:setup}
For $d\in \{0,1\}$, we have 
\[
 Y(d) = \sum_{g=1}^{\bar G} \one(G=g)\mu_g(d) + \epsilon(d),
\]
where $\bar G\geq 2$ is unknown, $\epsilon(d) \sim N\bigl( 0, \sigma^2(d) \bigr)$ is independent of $(D,G)$ for $d\in \{0,1\}$, and $\bm{\mu}_g : = \bigl( \mu_g(0), \mu_g(1) \bigr)$'s are distinct and lexicographically ordered across $g = 1, \cdots , \bar G$. Further, we have $0 <  \Prob ( G = g, D = d ) < 1$ for all $(g ,d) \in \{1,2,\cdots, \bar G\}\times \{0,1\}$. 
\end{assump}  

Assumption \ref{ass:setup} implies that $Y(d)$ is independent of $D$ once we condition on $G$. Therefore, $G$ plays the role of the unobserved confounder. A similar idea has been employed in \citet{imbens2003sensitivity} for a sensitivity analysis.  However, Assumption \ref{ass:setup} provides a more flexible environment than the fully parametric approach of \citet{imbens2003sensitivity}. All that is assumed for the distribution of $G$ is that $\bar G$ is finite. Assumption \ref{ass:setup} is silent about dependence between $D$ and $G$: the only restriction on their joint distribution is that all types and treatment statuses are realized with positive probability. Also, the conditional mean of $Y(d)$ given $G$ is nearly unrestricted: $\sum_{g=1}^{\bar G}\one(G=g)\mu_g(d)$ has a multinomial distribution with an unknown number of mass points, where the multinomial distribution can approximate any distribution arbitrarily well if $\bar G$ is sufficiently large. Finally, it is worth noting that $\bar G$ often admits an economic interpretation, although the precise interpretation depends on the context. This feature facilitates sensitivity analyses in the spirit of \citet{imbens2003sensitivity} and may help motivate additional restrictions that the researcher may want to impose on $\bar G$, or more generally on the distribution of $G$.

We emphasize a few important features of Assumption \ref{ass:setup}. Assumption \ref{ass:setup} does not require that all $\mu_g(d)$'s be distinct for a given $d\in\{0,1\}$, albeit that $\bm{\mu}_1,\cdots, \bm{\mu}_{\bar G}$ must be distinct as vectors. In other words, the mean vectors are all distinct across different types, but that does not necessarily imply that all types are revealed by the marginals.  In fact, Assumption \ref{ass:setup} allows $Y(1)$ and $Y(0)$ to have different numbers of mixing locations, where the number of distinct values in $\mu_1(d),\cdots, \mu_{\bar G}(d)$ for a given $d\in \{0,1\}$ can be strictly smaller than $\bar G$. Assumption \ref{ass:setup} imposes lexicographic ordering on the mean vectors such that 
\begin{equation} \label{eq:norep}
\bm{\mu}_1 \prec  \bm{\mu}_2 \prec \cdots \prec \bm{\mu}_{\bar G}.   
\end{equation}
This is purely about how we label the innate types $1,2,\cdots, \bar G$, and there is no loss of generality. Finally, Assumption \ref{ass:setup} focuses on Gaussian location mixtures for the distribution of $Y(d)$, but this imposes only few restrictions. Specifically, as we commented earlier, the mixture mean $\sum_{g=1}^{\bar G}\one(G=g)\mu_g(d)$ can approximate any distribution arbitrarily well as long as $\bar G$ is sufficiently large. Then, the distribution of $Y(d)$ is obtained by adding an independent Gaussian noise to it, where the Gaussian noise is homoskedastic across different types. The homoskedastic error across types is to rule out the possibility that two distinct types share the same mean for both $d=0$ and $d=1$.  

We present a simple example of the data-generating process (DGP) described in Assumption \ref{ass:setup}.

\begin{example} \label{exa:simple}
Suppose that there are $\bar G  = 3$ types that are characterized by the following means: $\bm{\mu}_1  = ( 0 , 1)$, $\bm{\mu}_2  = ( 0 , 2)$, and $\bm{\mu}_3  = ( 1 , 2)$.  The DGP can be fully determined by setting, e.g., $\sigma^2(0) = 3, \sigma^2(1) = 4$, $\Prob ( D = 0 ) = 1/2$, and $\Prob ( G = g \mid  D = d ) = 1/3$ for all $(g,d)$, so that $D$ and $G$ are independent.
\end{example}

\subsection{Clustering, Data Distributions, and Identifiable Parameters}		\label{sec:cluster}

For a given $d\in\{0,1\}$, not all of $\mu_1(d),\cdots, \mu_{\bar G}(d)$ need to be distinct. Let $\check{G}(d)$ be the number of distinct elements among $\mu_1(d),\cdots, \mu_{\bar G}(d)$: note that Assumption \ref{ass:setup} allows either $\check{G}( 0) = 1$ or $\check{G}( 1) = 1$, but not both at the same time. We will denote the distinct elements by $\check{\mu}_1(d) < \cdots < \check{\mu}_{\check G (d)}(d)$.  The relationship between $\mu_g(d)$'s and $\check{\mu}_g(d)$'s determines an \emph{ordered} partition on $\{1,2,\cdots, \bar G\}$ of the form
\begin{equation}\label{eq:overlap}
\mathcal{P} (d ) : =  \left( \mathcal{G}_{1} (d) , \cdots ,  \mathcal{G}_{\check{G}(d)} (d) \right)  \ \ \text{with} \ \  \mathcal{G}_{j} (d)  : = \left\{ g \in  \{1 , \cdots , \bar{G} \}   :   \  \mu_g (d) = \check{\mu}_{j} (d)  \right\} ,
\end{equation}
noting that $ \bigcup_{j =1}^{\check{G}(d)} \mathcal{G}_j (d) = \{ 1 , \cdots , \bar{G} \} $. Essentially, each $\mathcal{G}_j(d)$ is a cluster of the types that correspond to the distinct value $\check{\mu}_j(d)$ and the resulting order of the tuple $ ( \mathcal{G}_{1} (d) , \cdots ,  \mathcal{G}_{\check{G}(d)} (d) )$ is given by such distinct values. We remark that, in contrast to an unordered partition, the order in which the subsets $\mathcal{G}_j (d) $ are listed matters.

An important implication of Assumption \ref{ass:setup} is that, for both $d=0$ and $d=1$, there exist ordered partitions satisfying Equation \eqref{eq:overlap}. In particular, every type $g$ must be represented in both the treatment and control groups. If the treatment and control groups were drawn from different populations, then comparing them for causal inference would be simply nonsensical.

The ordered partitions of the innate types satisfy several properties as a consequence of the normalization imposed by the lexicographic ordering of the distinct mean vectors $\bm{\mu}_1,\cdots, \bm{\mu}_{\bar G}$. First, we always have $\max\ \mathcal{G}_j(0) < \min\ \mathcal{G}_k(0)$ whenever $j<k$. In words, this means that $\mathcal{P} (0)$ partitions $\{1 , \cdots , \bar{G}\}$ into $\check{G}( 0 )$ consecutive blocks. However, this is not necessarily the case for the treatment group: i.e., for $\mathcal{P} ( 1 ) $, $j<k$ does not necessarily imply that $\max \mathcal{G}_{j} (1) < \min \mathcal{G}_{k} (1)$, nor even $\max \mathcal{G}_{j} (1)  \leq \min \mathcal{G}_{k} (1)$. Finally, we remark that $\mathcal{G}_j(0)\cap \mathcal{G}_k(1)$ must be either empty or a singleton for any $j, k$.

We continue to consider Example \ref{exa:simple} to illustrate the ideas. 
\begin{subexample}[Cont.] \label{exa:1a}
Recall that $\bar G = 3$ with $\bm{\mu}_1 = (0,1), \bm{\mu}_2 = (0,2)$, and $\bm{\mu}_3 = (1,2)$. The number of distinct means for each group is $\check{G}(0) = \check{G}(1) = 2$, while the distinct means are $\check{\mu}_{1} (0) = 0, \check{\mu}_{2} (0) = 1, \check{\mu}_{1} (1) = 1$, and $\check{\mu}_{2} (1) = 2$. The partitions describing the mean clustering are as follows.
\begin{itemize}
\item Control group: $\{ 1 , 2 , 3\} =  \mathcal{G}_{1} (0)  \cup \mathcal{G}_{2} (0)$ with $\mathcal{G}_{1} (0)  = \{1 ,2 \}$, and $\mathcal{G}_{2} (0)  =  \{  3 \} $.
\item Treatment group: $\{ 1 , 2 , 3\} =  \mathcal{G}_{1} (1)  \cup \mathcal{G}_{2} (1)$ with $\mathcal{G}_{1} (1)  = \{1 \}$ and $\mathcal{G}_{2} (1)  =  \{  2 , 3 \} $.
\end{itemize}
Here, we do have $\max \mathcal{G}_{1} (1)  \leq \min \mathcal{G}_{2} (1)$, but this may not hold in general. 
\end{subexample}

Define $\pi_{j} (d , d^\prime)  :=  \Prob\{  G \in  \mathcal{G}_{j} (d) \mid  D  = d^\prime \}$ for $( d , d^\prime ) \in \{ 0 , 1 \}^2$. The distribution of the observed outcome, conditional on $D$, will follow a Gaussian mixture model such that 
\begin{eqnarray} 
    Y\mid D=0  &  \sim &  \sum_{j =1}^{\check G(0)}  \pi_{j} (0 , 0 )   \times N\bigl( \check{\mu}_{j} (0), \sigma^2(0) \bigr)  ,  
    \label{eq:mix0} \\
    Y\mid D=1  &  \sim & \sum_{j=1}^{\check G(1)} \pi_{j} (1 , 1) \times N\bigl( \check{\mu}_{j}(1), \sigma^2(1) \bigr) .
    \label{eq:mix1} 
\end{eqnarray}
By Proposition 2 in \cite{yak68id}, the parameters that are directly identified from the Gaussian mixtures are
\begin{align}
\check G(0),  \ \bigl\{ \bigl( \pi_{j}(0,0), \check\mu_{j}(0) \bigr): j = 1,2,\cdots, \check G(0) \bigr\}, \  \sigma^2(0), 
\label{eq:parameters0}
\\
\check G(1), \  \bigl\{ \bigl( \pi_{j}(1,1), \check\mu_{j}(1) \bigr): j = 1,2,\cdots, \check G(1) \bigr\} , \  \sigma^2(1).
\label{eq:parameters1}
\end{align}
Moreover, $\Prob (D = 1)$ is trivially identified. Therefore, we will assume that the parameters in \eqref{eq:parameters0} and \eqref{eq:parameters1} as well as the treatment probability $\Prob(D=1)$ are known when we discuss identification of the average treatment effect. In other words, the researcher knows that there are ordered partitions $\mathcal{P} (d) = ( \mathcal{G}_{1} (d) , \cdots ,  \mathcal{G}_{\check{G}(d)} (d)  )$ of the population, $d=0,1$, where each $\mathcal{G}_j(d)$ corresponds to the mixture component mean $\check{\mu}_j(d)$, but the exact compositions of $\mathcal{G}_j(d)$'s are unidentified, and thus the researcher does not know which of the mixture components a specific type $g$ belongs to.\footnote{\label{fn:gardener}\citet{gardner2020identification} considers a similar setup and approach with a focus on panel data. However, his analysis implicitly relies on strong assumptions. In particular, it assumes that the types revealed by mixtures in the treatment and control groups are exactly matched. For illustration, suppose that the mixture model of the treatment group reveals two distinct types, say $\{1,2\}$. \citet{gardner2020identification} then implicitly assumes that the mixture model of the control group also reveals exactly two distinct types, say $\{a,b\}$, and that the econometrician knows that $\{1\}$ and $\{a\}$ represent the same type in the population, and likewise for $\{2\}$ and $\{b\}$: i.e., it is implicitly assumed that $\check G(0) = \check G(1)$, and more importantly, the types in the population that are revealed by $\check\mu_j(1)$ and $\check\mu_j(0)$ always coincide, from which it is guaranteed that $\mu_j(d) = \check\mu_j(d)$ for both $d\in \{0,1\}$. Consequently, \citet{gardner2020identification} obtains point identification of the average treatment effect regardless of the number of mass points of the unobserved confounder. This ``exact matching'' assumption appears to be too restrictive and arbitrary in realistic settings.}

Using Example \ref{exa:simple}, the situation is illustrated below. 
\begin{subexample}[Cont.]   %{\ref{exa:simple} (cont.)}
    Recall the clustering partitions described in Example \ref{exa:1a}, and we have  
    \begin{equation*} 
    \begin{aligned} 
        \pi_{1} (0 , 0)    &=   \Prob \{  G \in  \{ 1 , 2 \} \mid  D  = 0 \}  = 2/3,
        &&
        \pi_{1} (1 , 0)    =   \Prob \{  G \in  \{ 1 \} \mid  D  = 0 \} = 1/3,
        \\
        \pi_{1} (0 , 1)    &=   \Prob \{  G \in  \{ 1 , 2 \} \mid  D  = 1 \}  = 2/3,
        &&
        \pi_{1} (1 , 1)    =   \Prob \{  G \in \{ 1\} \mid D =1  \} = 1/3,
    \end{aligned}
    \end{equation*}
    where we trivially have $\pi_{2} (d , d^\prime) = 1 -  \pi_{1} (d , d^\prime) $ for all $(d , d^\prime) \in \{ 0 , 1\}^2$.  Hence,  
    \begin{align} 
        Y\mid D=0  &\sim \frac{2}{3}   \times N ( 0,   3 )  + \frac{1}{3}   \times N ( 1,   3 ),
        \label{eq:mixture_eg0}
        \\
        Y\mid D=1  &\sim  \frac{1}{3}   \times N ( 1,   4 )  + \frac{2}{3}   \times N ( 2,   4 ) 
        \label{eq:mixture_eg1}
    \end{align}
    are the identifiable mixtures.
    \end{subexample}

Unlike the parameters in \eqref{eq:parameters0} and \eqref{eq:parameters1}, the clustering partitions $\mathcal{P} (0 ) $ and $\mathcal{P} (1 )  $ are unidentified.  Without additional assumptions, we do not know, e.g., whether any $\mathcal{G}_{j} (0)$ and $\mathcal{G}_{k} (1)$ share any elements (types) in common. Formally, this means that the composition of each set $ \mathcal{G}_{j} (d) $ is unknown to the researcher. Furthermore, $ \pi_{j} (0 , 1 )$ and $ \pi_{j} (1 , 0 )$ are unidentified as well. However, the total number of innate types $\bar G$ is partially identified, as we show in Lemma \ref{def:matpos0} below.

To facilitate the subsequent identification analysis, we show in the next subsection that there is a succinct way of representing the \emph{admissible} partitions of innate types, i.e., the ordered partitions that are consistent with Assumption \ref{ass:setup}.  Specifically, we will show that, under Assumption \ref{ass:setup}, each potential candidate for the latent pair of ordered partitions $\bigl( \mathcal{P} (0), \mathcal{P} (1) \bigr)$ can be uniquely represented by a $\check{G}(0)\times \check{G}(1)$ matrix.

\subsection{Admissible partitions of types and their matrix representation}\label{sec:admin}

For $(m_1 , m_2)  \in \mathbb{N}^2$, let $\{ 0 , 1\}^{m_1\times m_2}$ denote the set of $m_1\times m_2$ binary matrices, where each entry is either zero or one.

\begin{defin} \label{def:matpos0}
  Let $\mathscr{I}$ be the set of matrices $\bm{\iota} = (   \iota_{j , k}  ) \in \{ 0 , 1\}^{\check{G}(0) \times \check{G}(1) }$ that satisfy the following requirements:
  \begin{enumerate}
  
    \item $\forall j \in \{ 1, \cdots ,  \check{G}(0) \}$, $\sum_{k=1}^{\check{G}(1)} \iota_{j , k} \geq 1$.
    
    \item $\forall k  \in \{  1, \cdots ,  \check{G}(1) \}$, $\sum_{j=1}^{\check{G}(0)} \iota_{j , k} \geq 1$. 
    
    \end{enumerate}
\end{defin}

Condition (1) mandates that each row of $\bm{\iota} \in \mathscr{I}$ contains at least one non-zero entry, while (2) requires the same for each column. The set $\mathscr{I}$ is clearly nonempty, and it is a singleton if and only if either $\check{G}(0) = 1$ or $\check{G}(1) = 1$, in which case the only element of $\mathscr{I}$ is a matrix with ones in all its entries.

The idea behind Definition \ref{def:matpos0} is as follows. Let $( \lowmathcal{p} ( 0 ) ,   \lowmathcal{p} ( 1 ) )$ be an admissible candidate for the (true) latent pair of underlying partitions $( \mathcal{P}(0) ,   \mathcal{P} ( 1 )  )$, where $ \lowmathcal{p} ( 0 )  = (  \lowmathcal{g}_{1} ( 0 ) , \cdots ,  \lowmathcal{g}_{ \check{G}(0) } ( 0 ) )$ and $ \lowmathcal{p} ( 1)  = ( \lowmathcal{g}_{1} ( 1 ) , \cdots ,  \lowmathcal{g}_{\check{G}(1)} ( 1 ) )$.
Note first that the intersections $\lowmathcal{g}_j(0)\cap \lowmathcal{g}_k(1)$ for $j\in\{1,\cdots, \check{G}(0)\}$ and $k\in \{ 1,\cdots, \check{G}(1)\}$ are all that matters to completely characterize the candidate $( \lowmathcal{p} ( 0 ) ,   \lowmathcal{p} ( 1 ) )$ because $\cup_{j=1}^{\check{G}(0)}  \lowmathcal{g}_j(0)\cap \lowmathcal{g}_k(1) = \lowmathcal{g}_k(1)$ and $\cup_{k=1}^{\check{G}(1)}  \lowmathcal{g}_j(0)\cap \lowmathcal{g}_k(1)  = \lowmathcal{g}_j(0)$. However, under Assumption \ref{ass:setup}, $\lowmathcal{g}_j(0)\cap \lowmathcal{g}_k(1)$ is always either empty or a singleton, and Conditions (1) and (2) in Definition \ref{def:matpos0} trace the non-emptiness patterns in those intersections. Specifically, the matrix $\bm{\iota} \in \{ 0 , 1\}^{\check{G}(0) \times \check{G}(1) }$ associated with the candidate $ ( \lowmathcal{p} ( 0 ) ,   \lowmathcal{p} ( 1 ) )$ is constructed such that $\iota_{j,k}$ equals one if and only if $\lowmathcal{g}_j(0)\cap \lowmathcal{g}_k(1) $ is nonempty.

The following lemma formalizes the preceding discussion.

\begin{lemma}
\label{lem:partsgen}
There exists a bijection between $\mathscr{I}$ and the set of all admissible pairs of ordered partitions, i.e., those pairs that are allowed under Assumption \ref{ass:setup}.
\end{lemma}

Lemma \ref{lem:partsgen} follows as a direct consequence of the lexicographic ordering and the distinct mean vector condition specified in Assumption \ref{ass:setup}. A specific closed-form expression for the bijection is provided in Section \ref{sec:mainproofs} in the Appendix. Due to the existence of such a bijection, hereafter, we will use $\bm{\iota} \in \mathscr{I}$ to refer to a candidate for the (true) latent pair of ordered partitions $( \mathcal{P}(0) ,   \mathcal{P} ( 1 )  )$.

By using Lemma \ref{lem:partsgen}, we can show that $\bar G$, the number of innate types, is partially identified. 

\begin{lemma}
\label{lem:Gbar}
Under Assumption \ref{ass:setup}, the sharp identified set for $\bar G$ is given by $\{ m\in \mathbb{N}:\ \bar G_L \leq m \leq \bar G_U\}$, where $\bar G_L:= \max\{ \check{G}( 0) , \check{G}( 1) \}$ and $\bar G_U := \check{G}( 0) \, \check{G}( 1)$.
\end{lemma}

Lemma \ref{lem:Gbar} shows that $\bar G$ is point identified if and only if $\min\{ \check{G}(0), \check{G}(1) \}$ is equal to $1$. In general, $\bar G$ belongs to a set of finitely many integers. The sharp identified set for $\bar G$ is obtained by noting that the number of types associated with the partition represented by $\bm{\iota} = ( \iota_{j,k} ) \in \mathscr{I}$ is given by the sum of its entries, i.e.\ $\sum_j \sum_k \iota_{j,k}$. Hence, the sharp identified set for $\bar G$ becomes $\{ \sum_j \sum_k \iota_{j,k}  :  \bm{\iota}= ( \iota_{j,k} )  \in \mathscr{I} \}$. Intuitively, the lower bound $\bar G_L$ is trivial, and it corresponds to the case of no hidden type in either the treatment or the control group: the total number of innate types cannot be smaller than the number of distinct types in either the treatment or control group. The upper bound is a consequence of the normalization of the lexicographic ordering, and it corresponds to the case of maximal clustering.  Specifically, given $\check{G}(0)$ and $\check{G}(1)$, we know that $\check{\mu}_1(0)<\dots<\check{\mu}_{\check{G}(0)}(0)$ and $\check{\mu}_1(1)<\dots<\check{\mu}_{\check{G}(1)}(1)$. Therefore, the largest number of different ways of lexicographically pairing them is obtained when all the means of the treatment group are paired with each of the means in the control group.

Before proceeding to the next section, we revisit Example \ref{exa:simple} to illustrate the ideas discussed above. In particular, we emphasize that it is necessary to consider \emph{ordered} partitions to derive the bijection established in Lemma \ref{lem:partsgen}.

\begin{subexample}[Cont.] \label{eg:partitions}
From the mixture distributions in \eqref{eq:mixture_eg0} and \eqref{eq:mixture_eg1}, $\check{G}(0) = \check{G}(1) = 2$ are identified, but the clusters $\mathcal{G}_1(0)$, $\mathcal{G}_2(0)$, $\mathcal{G}_1(1)$, and $\mathcal{G}_2(1)$ remain unrevealed. However, from Lemma \ref{lem:Gbar}, we know that $2 \leq \bar G\leq 4$, and hence there are three values $\bar G$ can take, yielding seven possibilities of different partitions: two for the case $\bar G = 2$, four when $\bar G = 3$, and one when $\bar G =  4$. We describe all of them in the following tables and provide the corresponding matrices $\bm{\iota}\in \mathscr{I}$. Recall that $\bm{\mu}_g = \bigl( \mu_g(0), \mu_g(1) \bigr)$ for type $g\in \{1,2,\cdots, \bar G\}$. Below we write $\mu(d) = \bigl( \mu_1(d), \cdots, \mu_{\bar G}(d) \bigr)$ for $d\in\{0,1\}$. 

\begin{enumerate} 
\item \textbf{Case $\bar G = 2$.} There are two possibilities. 
\begin{enumerate} 
  \item\label{enum:redundant} $\mu(0) = (0,1)$ and $\mu(1) = (1,2)$, leading to the ordered partitions given by 
\[
\textrm{ 
\begin{tabular}{c|cc|c}
$\mathcal{G}_j(0)  \cap \mathcal{G}_k(1) $ &   $k = 1$    &   $k = 2 $     &   $\mathcal{G}_j(0)$\\ 
\hline
$ j = 1 $                                 &   $\{ 1 \}$       & $ \emptyset$  &  $\{1\}$    \\
$ j = 2 $                                 &   $ \emptyset$    &  $\{ 2  \}$   &  $\{2\}$    \\
\hline 
$\mathcal{G}_k(1) $    &   $\{1\}$         & $\{2\}$ &  
\end{tabular}
}
\Leftrightarrow \
\bm{\iota}_1 
= 
\left(
\begin{array}{cc}
  1  &  0 \\
  0  & 1  
\end{array}
\right).
\]

\item $\mu(0) = (0,1)$ and $\mu(1) = (2,1)$, leading to the ordered partitions given by 
\[
  \textrm{
  \begin{tabular}{c|cc|c}
  $\mathcal{G}_j(0)  \cap \mathcal{G}_k(1) $ &   $k = 1$    &   $k = 2 $     &  $\mathcal{G}_j(0) $ \\ 
  \hline
  $ j = 1 $                                 &   $\emptyset$ & $ \{1\}$  &  $\{1\}$    \\
  $ j = 2 $                                 &   $\{2\}$     &  $\emptyset$   &  $\{2\}$    \\
  \hline 
$\mathcal{G}_k(1) $    &   $\{2\}$ & $\{1\}$ &  
  \end{tabular} 
  }
  \Leftrightarrow \
  \bm{\iota}_2
  = 
  \left(
    \begin{array}{cc} 
      0  & 1 \\
      1  & 0
    \end{array}
  \right)
\]
\end{enumerate}

In both cases (a) and (b), we have $\mathcal{P} ( 0) = \bigl(\{ 1 \} , \{ 2 \}  \bigr)$. The distinction arises for $\mathcal{P} (1)$. In case (a), we have $\mathcal{P} ( 1)  = \bigl( \{1 \} , \{ 2 \} \bigr) $, whereas in case (b), $\mathcal{P} ( 1)  = \bigl(  \{ 2 \} , \{1 \} \bigr) $. Because the partitions are ordered, these two objects are distinct. If the ordering were ignored, however, cases (a) and (b) would correspond to the same partition of $\{ 1 , 2 \}$, namely, $\bigl\{ \{1 \} , \{ 2 \} \bigr\}  = \bigl\{ \{ 2 \}   , \{1 \} \bigr\}$.

\item \textbf{Case $\bar G = 3$.} There are four possibilities. 
\begin{enumerate} 
\item $\mu(0) = (0,0,1)$ and $\mu(1) = (1,2,1)$, leading to the ordered partitions given by
\[
  \textrm{
    \begin{tabular}{c|cc|c}
    $\mathcal{G}_j(0)  \cap \mathcal{G}_k(1) $ &   $k = 1$    &   $k = 2 $     &  $\mathcal{G}_j(0) $ \\ 
    \hline
    $ j = 1 $                                 &  $\{1\}$      &  $\{2\}$        &  $\{1,2\}$    \\
    $ j = 2 $                                 &   $\{3\}$     &  $\emptyset$   &  $\{3\}$    \\
    \hline 
  $\mathcal{G}_k(1) $    &   $\{1,3\}$ & $\{2\}$ &  
    \end{tabular} 
    }
    \Leftrightarrow \
    \bm{\iota}_3
    = 
    \left(
      \begin{array}{cc} 
        1  & 1   \\
        1  & 0
      \end{array}
    \right)
\]
\item\label{enum:true partitions} $\mu(0) = (0,0,1)$ and $\mu(1) = (1,2,2)$, leading to the \textbf{true} ordered partitions given by 
\[
  \textrm{
    \begin{tabular}{c|cc|c}
    $\mathcal{G}_j(0)  \cap \mathcal{G}_k(1) $ &   $k = 1$    &   $k = 2 $     &  $\mathcal{G}_j(0)$ \\ 
    \hline
    $ j = 1 $                                 &  $\{1\}$      &  $\{2\}$        &  $\{1,2\}$    \\
    $ j = 2 $                                 &   $\emptyset$     &  $\{3\}$   &  $\{3\}$    \\
    \hline 
  $\mathcal{G}_k(1)$    &   $\{1\}$ & $\{2,3\}$ &  
    \end{tabular} 
    }
    \Leftrightarrow \
    \bm{\iota}_4
    = 
    \left(
      \begin{array}{cc} 
        1  & 1   \\
        0  & 1
      \end{array}
    \right)
\]
\item $\mu(0) = (0,1,1)$ and $\mu(1) = (1,1,2)$, leading to the ordered partitions given by
\[
  \textrm{
    \begin{tabular}{c|cc|c}
    $\mathcal{G}_j(0)  \cap \mathcal{G}_k(1) $ &   $k = 1$    &   $k = 2 $     &  $\mathcal{G}_j(0)$ \\ 
    \hline
    $ j = 1 $                                 &  $\{1\}$      &  $\emptyset$        &  $\{1\}$    \\
    $ j = 2 $                                 &   $\{2\}$     &  $\{3\}$   &  $\{2,3\}$    \\
    \hline 
  $\mathcal{G}_k(1)$    &   $\{1,2\}$ & $\{3\}$ &  
    \end{tabular} 
    }
    \Leftrightarrow \
    \bm{\iota}_5
    = 
    \left(
      \begin{array}{cc} 
        1  & 0   \\
        1  & 1
      \end{array}
    \right)
\]
\item $\mu(0) = (0,1,1)$ and $\mu(1) = (2,1,2)$, leading to the ordered partitions given by
\[
  \textrm{
    \begin{tabular}{c|cc|c}
    $\mathcal{G}_j(0)  \cap \mathcal{G}_k(1) $ &   $k = 1$    &   $k = 2 $     &  $\mathcal{G}_j(0)$ \\ 
    \hline
    $ j = 1 $                                 &  $\emptyset$      &  $\{1\}$        &  $\{1\}$    \\
    $ j = 2 $                                 &   $\{2\}$     &  $\{3\}$   &  $\{2,3\}$    \\
    \hline 
  $\mathcal{G}_k(1)$    &   $\{2\}$ & $\{1,3\}$ &  
    \end{tabular} 
    }
    \Leftrightarrow \
    \bm{\iota}_6
    = 
    \left(
      \begin{array}{cc} 
        0  & 1   \\
        1  & 1
      \end{array}
    \right)
\]
\end{enumerate}

\item \textbf{Case $\bar G = 4$.} There is only one possibility. 
\begin{enumerate} 
\item $\mu(0) = (0,0,1,1)$ and $\mu(1) = (1,2,1,2)$, corresponding to the ordered partitions given by
\[
  \textrm{
    \begin{tabular}{c|cc|c}
    $\mathcal{G}_j(0)  \cap \mathcal{G}_k(1) $ &   $k = 1$    &   $k = 2 $     &  $\mathcal{G}_j(0)$ \\ 
    \hline
    $ j = 1 $                                 &  $\{1\}$      &  $\{2\}$        &  $\{1,2\}$    \\
    $ j = 2 $                                 &   $\{3\}$     &  $\{4\}$   &  $\{3,4\}$    \\
    \hline 
  $\mathcal{G}_k(1)$    &   $\{1,3\}$ & $\{2,4\}$ &  
    \end{tabular} 
    }
    \Leftrightarrow \
    \bm{\iota}_7
    = 
    \left(
      \begin{array}{cc} 
        1  & 1   \\
        1  & 1
      \end{array}
    \right)
\]
\end{enumerate} 

\end{enumerate} 

Part (\ref{enum:true partitions}) represents the true underlying partition in this example, and the others show admissible candidate partitions of the innate types with the same values of the identifiable parameters as in the truth. Below is an example that Assumption \ref{ass:setup} rules out though. Suppose that $\bar G= 3$ with $\mu(0) = (0,1,1)$ and $\mu(1) = (1,2,2)$. In this case, $\bm{\mu}_2 = \bm{\mu}_3 = (1,2)$, and therefore, the requirement of distinct mean vectors of Assumption \ref{ass:setup} is violated. This case can be described by the following partition: 
\[
\textrm{
    \begin{tabular}{c|cc|c}
    $\mathcal{G}_j(0)  \cap \mathcal{G}_k(1) $ &   $k = 1$    &   $k = 2 $     &  $\mathcal{G}_j(0)$ \\ 
    \hline
    $ j = 1 $                                 &  $\{1\}$      &  $\emptyset$        &  $\{1\}$    \\
    $ j = 2 $                                 &   $\emptyset$     &  $\{2,3\}$   &  $\{2,3\}$    \\
    \hline 
  $\mathcal{G}_k(1)$    &   $\{1\}$ & $\{2,3\}$ &  
    \end{tabular} 
}.
\]
However, this situation is not distinguishable from the one described in Part (\ref{enum:redundant}), and therefore the zero-one patterns in the entries of $\bm{\iota}_1$ are all that matters here. 
\end{subexample}

\section{Partial Identification of ATE}\label{sec:identification}

\subsection{A Simple Approach}

Our parameter of interest is ATE. Since 
\begin{equation*}
\E \left\{ Y (d)  \right\}  =   \sum_{j =1}^{\check{G}(d)}  \Prob \{  G \in  \mathcal{G}_{j} (d)  \}  \check{\mu}_{j} (d)  \quad \text{for}  \  d \in \{ 0 , 1\},
 \end{equation*}
where $\Prob \{  G \in  \mathcal{G}_{j} (d)  \}  =  \pi_{j} (d , 0 )  \Prob ( D = 0) + \pi_{j} (d , 1 ) \Prob ( D = 1)$ by the law of total probability, we can express ATE as follows: 
\begin{multline}\label{eq:ategr} 		
\tau  
: =      
\E \left\{ Y (1)  - Y (0)  \right\}  
  =      
\sum_{k =1}^{\check{G}(1)}    \{ \pi_k (1 , 0 )  \Prob ( D = 0) + \pi_k (1 , 1 ) \Prob ( D = 1)  \}\check{\mu}_k (1) 	
\\
-  \sum_{j =1}^{\check{G}(0)}  \{  \pi_j (0 , 0 )  \Prob ( D = 0) + \pi_j (0 , 1 ) \Prob ( D = 1)     \}  \check{\mu}_j (0) ,
 \end{multline}
 where $\{\pi_k (1 , 0 ): k=1,\cdots,\check{G}(1)\}$ and $\{ \pi_j (  0 , 1 ): j=1,\cdots, \check{G}(0)\}$ are unidentified conditional probabilities. Assumption \ref{ass:setup} does not directly impose any restrictions on these unidentified parameters except that they are non-degenerate probabilities such that $\sum_{j=1}^{\check{G}(1)}\pi_j(1,0) = \sum_{j=1}^{\check{G}(0)} \pi_j(0,1) = 1$. Therefore, it is not too difficult to obtain the sharp identifiable set of $\tau$ by searching over the right simplexes.  Specifically, let $\Pi_j  = \{ \textbf{p} \in \mathbb{R}_{++}^j :  \sum_{l=1}^j  p_l= 1 \}$ for $j \in \mathbb{N}$, and define 

\begin{multline*}
\mathcal{T}  
:= 
 \Biggl\{    
   \sum_{k=1}^{\check{G}(1)}  \bigl\{ \Prob ( D = 0)  q_k  +  \Prob ( D = 1)   \pi_k (1 , 1 ) \bigr\}\check{\mu}_k (1) 
\\ 
-
\sum_{j =1}^{\check{G}(0)}  \bigl\{   \Prob ( D = 0)  \pi_{j} (0 , 0 ) + \Prob ( D = 1)   p_{j}   \bigr\}  \check{\mu}_{j} (0) :  
 \  ( \mathbf{p}   ,  \mathbf{q}  ) \in  \Pi_{\check{G}(0)}   \times  \Pi_{\check{G}(1)}  \Biggr\} . 
\end{multline*}

\begin{theorem}
\label{thm:simple}
Suppose that Assumption \ref{ass:setup} holds. Then, $\mathcal{T}$ is the sharp identifiable set for $\tau$, which can be equivalently expressed as 
\begin{multline}
\label{eq:charaT}
  \mathcal{T}
    = 
  \bigg\{  \Prob ( D = 1)  \sum_{j=1}^{\check{G}(1)} \pi_{j} (1 , 1 )  \check{\mu}_{j} (1)   
  -   \Prob ( D = 0)\sum_{j =1}^{\check{G}(0)} \pi_{j} (0 , 0 )   \check{\mu}_{j} (0)  + t  : 
  \\
  t   \in 
  \Bigl( \Prob ( D = 0)  \check{\mu}_1 (1)   - \Prob ( D = 1) \check{\mu}_{\check G (0)} (0),\   
    \Prob ( D = 0)  \check{\mu}_{\check G (1)} (1)   - \Prob ( D = 1) \check{\mu}_1 (0)   \Bigr)    \bigg\}.
\end{multline}
\end{theorem}

Theorem \ref{thm:simple} is straightforward and easy to understand: estimating $\mathcal{T}$ does not involve any optimization beyond estimating the identified mixture parameters.  It is worth noting that $\mathcal{T}$ is reminiscent of the Manski bounds \citep[e.g.][]{manski2003partial,manski2010partial}: indeed,   
\begin{equation}
\label{eq:manski}
    \Exp( Y \mid D=1) = \sum_{j=1}^{\check{G}(1)} \pi_{j} (1,1) \check{\mu}_{j} (1)
    \quad\text{and}\quad 
    \Exp(Y\mid D=0) = \sum_{j=1}^{\check{G}(0)} \pi_{j} (0,0) \check{\mu}_{j} (0),
\end{equation}
whereas, for $d=0,1$, we have 
\begin{equation*}
\left\{
\begin{aligned}
\Exp\{ Y(d) \mid D=1-d \} &= \check{\mu}_{1}(d)  && \text{if} \  \check{G}(d)=1  \\
\Exp\{ Y(d) \mid D=1-d \} &\in \bigl( \check{\mu}_{1}(d),\ \check{\mu}_{\check{G}(d)}(d) \bigr)  && \text{if}\  \check{G}(d) >1 , 
\end{aligned}
\right.
\end{equation*}
for which we recall that either $\check{G}(0)=1$ or $\check{G}(1)=1$ is allowed. The fact that we have open intervals in the case of $\check{G}(d) > 1$ is because all of $\check{\mu}_j(d)$'s for $d=0,1$ and $j=1,2,\cdots, \check{G}(d)$ have non-zero weights under Assumption \ref{ass:setup}. Since the support of $Y(d)$ for $d=0,1$ is unrestricted under Assumption \ref{ass:setup}, the na\"{i}ve Manski bounds would be unbounded. However, Assumption \ref{ass:setup} does impose finite mixture structures on $Y(d)$, which are exploited in Theorem \ref{thm:simple}. Specifically, the distributions of $Y$ given $D=d$ for $d=0,1$ are also finite mixtures, and Theorem \ref{thm:simple} formally establishes sharp bounds based on the mixing components that are identified from the distribution of $Y$ given $D=d$.

For the sake of illustration, we have computed $\tau$ and $\mathcal{T}$ in the setup of Example \ref{exa:simple} below. 
\begin{subexample}[Cont.]
\label{exa:taupid}
  The ATE and its identified set are $\tau = 4/3$ and $\mathcal{T} = (2/3 , 5/3)$, respectively. Therefore, the sharp lower bound on $\tau$ as a conservative measure is $2/3$.
\end{subexample}

Theorem \ref{thm:simple} takes a completely agnostic stance about $\bar G$, which can be extreme. For instance, the researcher may want to conduct a sensitivity analysis for various values of $\bar G$, or she may be willing to impose further restrictions on $\bar G$. Theorem \ref{thm:simple} does not provide a useful path for these purposes.  $\bar G$ is a key parameter that summarizes how much heterogeneity there is in the population, and the researcher knows that it is an integer between $\bar G_L$ and $\bar G_U$ by Lemma \ref{lem:Gbar}. As $\bar G$ moves away from $\bar G_L$, the situation becomes more challenging in that the number of types that are not revealed in either the treatment or the control group becomes larger.
The researcher may be willing to rule out extreme possibilities of ``too many clusterings,'' and to focus on the case where the total number of innate types in the population is revealed by either the treatment or the control group.  Alternatively, the researcher may consider using a weighted-averaging scheme for different possibilities for the value of $\bar G$.  Theorem \ref{thm:simple} is not convenient to open up those possibilities. Therefore, in the following subsection, we will characterize the sharp identified set for ATE as a function of $\bar G$.

\subsection{Using the Value of $\bar{G}$}	\label{sec:barG}

In order to describe the sharp identified set for $\tau$ for a given value of $\bar G$, we first characterize the identified set for a given $\bm{\iota}\in\mathscr{I}$.  Here, $\bm{\iota}\in\mathscr{I}$ represents a pair of specific partitions on $\{1,2,\cdots, \bar G\}$ that are allowed under Assumption \ref{ass:setup}, whereas $\bar G$ is an economic parameter given by nature.  We start with introducing the following set of matrices.

\begin{defin} \label{def:pmats}
For $\bm{\iota} = ( \iota_{j,m}  )\in \mathscr{I}$, we define $\mathcal{M}_{\bm{\iota}}(0)$ and $\mathcal{M}_{\bm{\iota}}(1)$ as follows. 
\begin{enumerate}
\item  Let $\mathcal{M}_{\bm{\iota}} (0)$ be the subset of matrices $\mathbf{p} = (  p_{j , k} ) \in  \mathbb{R}_+^{\check{G}(0) \times \check{G}(1)}$ such that
\[
\begin{aligned}
&(a) \quad \forall \  j =1 , \cdots ,\check{G}(0) \ \text{and}  \  k =1 , \cdots ,\check{G}(1),  & &  p_{j , k} > 0  \text{  if and only if  }  \iota_{j ,k } = 1 ,  \\
&(b) \quad \forall \  j =1 , \cdots ,\check{G}(0) , &&  \sum_{k=1}^{\check{G}(1)} p_{j,k} =  \pi_{j} ( 0 , 0) .
\end{aligned}
\]

\item Let $\mathcal{M}_{\bm{\iota}} (1)$ be the subset of matrices $\mathbf{q} = (  q_{j , k} ) \in  \mathbb{R}_+^{\check{G}(0) \times \check{G}(1)}$ such that 
\[ 
\begin{aligned}
&(a) \quad \forall \  j =1 , \cdots ,\check{G}(0) \ \text{and}  \  k =1 , \cdots ,\check{G}(1), && q_{j , k} > 0   \text{  if and only if  }  \iota_{j ,k } = 1 , \\
&(b) \quad  \forall \  k =1 , \cdots ,\check{G}(1)  , && \sum_{j=1}^{\check{G}(0)} q_{j,k} =  \pi_{k} ( 1 , 1)  .
\end{aligned}
\] 
\end{enumerate}
\end{defin}

The sets $\mathcal{M}_{\bm{\iota}} (0)$ and $\mathcal{M}_{\bm{\iota}} (1)$ are collections of nonnegative matrices, where the nonnegative entries are in accordance with those in $\bm{\iota}\in\mathscr{I}$, and their columns and rows are restricted to sum to $\pi_j(0,0)$ and $\pi_k(1,1)$, respectively. The purpose of these sets is to provide a systematic way to track all possible values that the conditional probability $\Prob\{ G \in \lowmathcal{g}_{\bm{\iota}, j} (0) \cap \lowmathcal{g}_{\bm{\iota}, k} (1)  \mid    D = d \}$ can take for $d\in \{0,1\}$ and $(j , k ) \in \{  1, \cdots  , \check{G}(0)   \} \times \{  1 , \cdots  ,   \check{G}(1) \}$, where $\lowmathcal{p}_{\bm{\iota}}(d) := \bigl( \lowmathcal{g}_{\bm{\iota}, 1}(d),\cdots, \lowmathcal{g}_{\bm{\iota}, \check{G}(d)} (d) \bigr )$ denotes the ordered partition associated with $\bm{\iota}$ in group $d$. 

To be specific, for a given matrix $\text{$\mathbf{p}$} = ( p_{j,k}   ) \in \mathcal{M}_{\bm{\iota}} (0)$, the entry $p_{j,k}$ represents a possible value of $\Prob \{  G \in \lowmathcal{g}_{ \bm\iota , j} (0) \cap \lowmathcal{g}_{\bm\iota , k} (1)  \mid    D = 0 \}$ if $( \lowmathcal{p}_{\bm{\iota}} (0) ,   \lowmathcal{p}_{\bm{\iota}} (1) )$ were the underlying partition. The restrictions on the column-wise and row-wise sums are because $\lowmathcal{p}_{\bm{\iota}}(d)$ is a partition on $\{1,\cdots,\bar G\}$: therefore, taking the union of $\lowmathcal{g}_{ \bm\iota , j} (0) \cap \lowmathcal{g}_{\bm\iota , k} (1)$ across $j$ (or $k$) leads to $\lowmathcal{g}_{\bm\iota , k} (1)$ (or $\lowmathcal{g}_{\bm\iota , j} (0)$). For example, Condition (b) of Definition \ref{def:pmats}.(1) ensures that the values of $\Prob\{  G \in \lowmathcal{g}_{\bm{\iota}, j} (0) \cap \lowmathcal{g}_{\bm{\iota}, k} (1)  \mid    D = 0 \}$, $k =1 , \dots ,  \check{G}(1)$ are logically consistent with that of the identified conditional probability $\pi_{j} ( 0 , 0) $ as
\begin{equation*}
\Prob \{  G \in \lowmathcal{g}_{\bm{\iota},j} (0)   \mid    D = 0 \}  
=  
\sum_{k=1}^{\check{G}(1)}  
 \Prob \{  G \in \lowmathcal{g}_{\bm{\iota}, j} (0) \cap \lowmathcal{g}_{\bm{\iota}, k} (1)  \mid    D = 0\} 
= 
\sum_{k=1}^{\check{G}(1)} p_{j,k}  =  \pi_{j} ( 0 , 0).
\end{equation*}
We also note that the zero entries of the matrix $\mathbf{p}$ are associated with the case where $\lowmathcal{g}_{\bm{\iota},j} (0) \cap \lowmathcal{g}_{\bm{\iota}, k} (1) $ is empty.  In addition, we trivially have that $\sum_j \sum_k p_{j,k} = 1$ due to Condition (b).  A symmetric argument can be applied to the components of $\mathbf{q}\in \mathcal{M}_{\bm{\iota}}(1)$, which are associated with possible values of $\Prob\{ G \in \lowmathcal{g}_{\bm{\iota},j}(0) \cap \lowmathcal{g}_{\bm{\iota}, k} (1)  \mid    D = 1 \}$.

For $\bm{\iota}\in\mathscr{I}$, define 
\begin{multline*}
  \mathcal{T}_{\bm{\iota}}
   := 
  \Biggl\{  \sum_{k =1}^{\check{G}(1)} 
       \biggl\{  \Prob ( D = 0) \sum_{j=1}^{\check{G}(0)} p_{j, k}  
       + \Prob ( D = 1) \pi_{k} (1 , 1 )   \biggr\} \check{\mu}_{k} (1) 
  \\ 
  -
  \sum_{j =1}^{\check{G}(0)}  \biggl\{  \Prob ( D = 0) \pi_{j} (0 , 0 )   +   \Prob ( D = 1)  \sum_{k=1}^{\check{G}(1)} q_{j, k}       \biggr\}  \check{\mu}_{j} (0) : \
   ( \text{$\mathbf{p}$}   , \text{$\mathbf{q}$}  ) \in \mathcal{M}_{\bm{\iota}} (0) \times \mathcal{M}_{\bm{\iota}} (1)  \Biggr\},
\end{multline*}
and note that $\mathcal{T}_{\bm{\iota}}$, which is a subset of $\mathcal{T}$, is always an interval because $\mathcal{M}_{\bm{\iota}} (d)$'s are convex. Let $\boldsymbol{\iota}_{\mathcal{P}} \in \mathscr{I}$ represent the unknown latent pair of ordered partitions $( \mathcal{P} (0) , \mathcal{P} (1) )$, i.e., the true pair. 

\begin{lemma} \label{lem:characT}
Suppose that Assumption \ref{ass:setup} holds. Then, $\mathcal{T}_{\iota_\mathcal{P}}$ would be the sharp identified set for $\tau$ if $\boldsymbol{\iota}_{\mathcal{P}}$ were known.
\end{lemma}

An immediate consequence of Theorem \ref{thm:simple} and Lemma \ref{lem:characT} is that $\cup_{\bm{\iota} \in \mathscr{I}} \mathcal{T}_{\bm{\iota}} =   \mathcal{T}$. In fact, we can divide the set $\mathcal{T}$ by using potential values of the economic parameter $\bar G$ instead of $\bm{\iota}$. We will discuss this issue in detail later in this section. 

Before we proceed, we revisit Example \ref{exa:simple} to illustrate the ideas.

\begin{subexample}[Cont.] 
\label{eg:Tiota}
Consider the cases discussed in Example \ref{eg:partitions}.
\begin{enumerate} 
  \item \textbf{Case $\bar G = 2$.} There are two $\bm{\iota}$ matrices to consider: $\bm{\iota}_1, \bm{\iota}_2$. 
  \begin{enumerate} 
  \item For $\bm{\iota}_1$: 
  \[ 
  M_{\bm{\iota}_1}(0) = \left\{ 
    \left(
    \begin{array}{cc}
      2/3  &  0 \\
      0  & 1/3  
    \end{array}
    \right)  \right\}, 
    M_{\bm{\iota}_1}(1) = \left\{ 
      \left(
      \begin{array}{cc}
        1/3  &  0 \\
        0  & 2/3  
      \end{array}
      \right)  \right\}.  
  \]
  Therefore, 
  \[
  \mathcal{T}_{\bm{\iota}_1} = \{ 1 \}.
  \]
  \item For $\bm{\iota}_2$: 
  \[
  M_{\bm{\iota}_2}(0) = \left\{ 
    \left(
    \begin{array}{cc}
      0  &  2/3 \\
      1/3  & 0  
    \end{array}
    \right)  \right\}, 
    M_{\bm{\iota}_2}(1) = \left\{ 
      \left(
      \begin{array}{cc}
        0  &  2/3 \\
        1/3  & 0  
      \end{array}
      \right)  \right\}.
  \]
  Therefore, 
  \[
    \mathcal{T}_{\bm{\iota}_2} = \{ 4/3  \}.
  \]
  \end{enumerate}

\item \textbf{Case $\bar G = 3$.} There are four $\bm{\iota}$ matrices to consider: $\bm{\iota}_3, \bm{\iota}_4, \bm{\iota}_5, \bm{\iota}_6$. 
      \begin{enumerate} 
      \item For $\bm{\iota}_3$: 
      \begin{align*} 
        M_{\bm{\iota}_3}(0) 
        &= 
        \left\{ 
          \left(
            \begin{array}{cc}
              p  &  2/3 - p \\
              1/3  & 0  
            \end{array}
          \right): 0 < p  < 2/3 \right\},
        \\ 
        M_{\bm{\iota}_3}(1) 
        &= 
        \left\{ 
            \left(
              \begin{array}{cc}
                q  &  2/3 \\
                1/3-q  & 0  
              \end{array}
            \right): 0< q< 1/3  \right\}.   
      \end{align*} 
      Therefore, 
      \[
      \mathcal{T}_{\bm{\iota}_3} = \{ 4/3 + (q-p)/2:\ 0<p<2/3, 0< q<1/3 \} = (1 ,  3/2).
      \]
      \item For $\bm{\iota}_4$:
        \begin{align*} 
          M_{\bm{\iota}_4}(0) 
          &= 
          \left\{ 
            \left(
              \begin{array}{cc}
                p  &  2/3 - p \\
                0  & 1/3  
              \end{array}
            \right): 0 < p  < 2/3 \right\},
          \\ 
          M_{\bm{\iota}_4}(1) 
          &= 
          \left\{ 
              \left(
                \begin{array}{cc}
                  1/3  &  q \\
                  0  &   2/3 - q
                \end{array}
              \right): 0< q< 2/3  \right\}.   
        \end{align*} 
      Therefore, 
      \[
      \mathcal{T}_{\bm{\iota}_4} = \{ 4/3 + (q-p)/2  :\ 0< p<2/3, 0<q<2/3  \} = ( 1  , 5/3  ).
      \]

      \item For $\bm{\iota}_5$: 
      \begin{align*} 
        M_{\bm{\iota}_5}(0) 
        &= 
        \left\{ 
          \left(
            \begin{array}{cc}
              2/3  &  0 \\
              p  & 1/3-p  
            \end{array}
          \right): 0 < p  < 1/3 \right\},
        \\ 
        M_{\bm{\iota}_5}(1) 
        &= 
        \left\{ 
            \left(
              \begin{array}{cc}
                q  &  0 \\
                1/3 -q  &   2/3 
              \end{array}
            \right): 0< q< 1/3  \right\}.   
      \end{align*} 
      Therefore, 
      \[
      \mathcal{T}_{\bm{\iota}_5} = \{ 5/6 + (q-p)/2:\ 0<p<1/3, 0<q<1/3 \} = ( 2/3 , 1). 
      \]
      \item For $\bm{\iota}_6$: 
      \begin{align*} 
        M_{\bm{\iota}_6}(0) 
        &= 
        \left\{ 
          \left(
            \begin{array}{cc}
              0  &  2/3 \\
              p  & 1/3-p  
            \end{array}
          \right): 0 < p  < 1/3 \right\},
        \\ 
        M_{\bm{\iota}_6}(1) 
        &= 
        \left\{ 
            \left(
              \begin{array}{cc}
                0  &  q \\
                1/3  &   2/3 - q
              \end{array}
            \right): 0< q< 2/3  \right\}.   
      \end{align*}
      Therefore, 
      \[
      \mathcal{T}_{\bm{\iota}_6} = \{ 7/6 + (q-p)/2:\ 0< p< 1/3, 0<q<2/3  \} =  (1 ,  3/2).
      \]
      \end{enumerate}
\item \textbf{Case $\bar G = 4$.} There is one $\bm{\iota}$ matrix to consider: $\bm{\iota}_7$.
      \begin{enumerate} 
      \item For $\bm{\iota}_7$: 
      \begin{align*} 
        M_{\bm{\iota}_7}(0) 
        &= 
        \left\{ 
          \left(
            \begin{array}{cc}
              p_1  &  2/3-p_1 \\
              p_2  & 1/3-p_2  
            \end{array}
          \right): 0 < p_1  < 2/3, 0< p_2<1/3 \right\},
        \\ 
        M_{\bm{\iota}_7}(1) 
        &= 
        \left\{ 
            \left(
              \begin{array}{cc}
                q_1  &  q_2 \\
                1/3-q_1  &   2/3 - q_2
              \end{array}
            \right): 0< q_1 < 1/3, 0< q_2<2/3 \right\}.   
      \end{align*}
      Therefore, 
      \begin{align*} 
      \mathcal{T}_{\bm{\iota}_7} 
      &= \{ 7/6 + (q_1+q_2 - p_1 - p_2)/2: 0<p_1<2/3, 0<p_2<1/3, 0<q_1<1/3, 0<q_2<2/3 \} \\
      &= (2/3,  5/3 ).
      \end{align*}
      \end{enumerate} 
\end{enumerate}

\end{subexample}

All the $\mathcal{T}_{\bm{\iota}}$'s described in Example \ref{eg:Tiota} are either a singleton or an open interval. It can be shown that this is always the case in general: see Lemma \ref{alem:shapeT} in the Appendix. Specifically, $\mathcal{T}_{\bm{\iota}}$ is a singleton if and only if $\sum_{j,k} \iota_{j,k} = \check{G} (0) =  \check{G} (1)$. We also note that Lemma \ref{alem:shapeT} in the Appendix provides closed-form expressions for the infimum and supremum of $\mathcal{T}_{\bm{\iota}}$. Therefore, there is no need to solve an optimization problem to compute $\mathcal{T}_{\bm{\iota}}$.

We now consider grouping $\bm{\iota}\in \mathscr{I}$ by using the potential values of $\bar G$, i.e., the number of innate types. As we argued earlier, $\bar G$ is an economic parameter summarizing the amount of unobserved heterogeneity in the population, whereas $\bm{\iota}\in\mathscr{I}$ is a consequence of Assumption \ref{ass:setup} given $\bar G$.   

For an integer $m$ satisfying $ \bar G_L \leq  m  \leq  \bar G_U$, we define the sets 
\begin{equation*}
 \mathscr{I}( m )
    := 
    \left\{ 
        \bm{\iota} \in \mathscr{I}:\ 
        \sum_{j,k}  \iota_{j,k} = m
    \right\} 
    \quad   \text{and} \quad 
   \mathcal{T} (  m )   := \cup_{\bm{\iota} \in  \mathscr{I}( m )}  \mathcal{T}_{\bm{\iota}}. 
\end{equation*}
The next theorem establishes that $ \mathcal{T} (  \bar{G} ) $ would be the sharp identified set for $\tau$ if $\bar{G}$ were known, and it also characterizes the shape of $ \mathcal{T} (  m ) $ for different values of $m$. Define $\bar G_C := \check{G}(0) + \check{G}(1) - 1$.

\begin{theorem} \label{thm:main0}
Suppose that Assumption \ref{ass:setup} is satisfied. Then, the following statements hold.

\begin{enumerate}
\item $ \mathcal{T} ( \bar{G}  ) $ would be the sharp identified set for $\tau$ if $\bar{G} $ were known.
\item For any $m \in \mathbb{N}$ satisfying $\bar G_L  \leq m \leq  \bar G_U $, we have
  \begin{enumerate}
  \item $\mathcal{T} (m) =    \mathcal{T} $ if $ m \geq 2\bar G_C - 2$; 
  \item $\mathcal{T} \backslash \mathcal{T} (m)$ is finite if $\bar G_C \leq  m$ and $m  <  2\bar G_C - 2$; %and therefore $\inf \mathcal{T}(m) = \inf \mathcal{T}$ and $\sup \mathcal{T}(m) = \sup \mathcal{T}$;\FZtodo{added to thm for the sake of clarity}
  \item $\inf \mathcal{T} (  m+1)   <   \inf \mathcal{T} (  m )  \leq  \sup \mathcal{T} (  m )  <   \sup \mathcal{T} (  m+1 ) $ if $ m  <  \bar G_C $.
  \end{enumerate}
\end{enumerate}
\end{theorem}

Theorem \ref{thm:main0} immediately implies that
\begin{equation*}
\bigcup_{m=\bar G_L}^{\bar G_U} \mathcal{T}  ( m ) =  \mathcal{T}.
\end{equation*}
Also, the cut-off value $\bar G_C$ is noteworthy. For all $m\geq \bar G_C$, $\mathcal{T}(m)$ is essentially $\mathcal{T}$: they are equal, or the difference is at most finite when  $\bar G_C \leq  m  <  2\bar G_C - 2$.  The case where $\mathcal{T}\backslash \mathcal{T}(m)$ is a nonempty finite set should be considered a technical nuisance that arises only because $\mathcal{T}(m)$ may be a union of disjoint open intervals, whereas $\mathcal{T}$ is always an open interval. Even in this case, it is guaranteed that $\mathcal{T}$ and $\mathcal{T}(m)$ share the same closure so that $\inf\mathcal{T} = \inf\mathcal{T}(m)$ and $\sup\mathcal{T} = \sup\mathcal{T}(m)$ for all $m\geq \bar G_C$. For all $m < \bar G_C$, $\mathcal{T}(m)$ is a strictly smaller set than $\mathcal{T}$, and $\mathcal{T}(m)$ strictly shrinks as $m$ becomes smaller. Therefore, if we restrict $\bar G$ such that $\bar G<\bar G_C$, then the sharp identified set for $\tau$ will be strictly smaller than $\mathcal{T}$.  

The cutoff value $\bar G_C$ naturally arises as the largest value allowed for $\bar G$ in some situation. One such case is when there is rank invariance in the two groups: i.e., for any $g, \tilde g\in \{1,2,\cdots, \bar G\}$, we have 
\begin{equation}\label{eq:ri}
\mu_g(0) \leq \mu_{\tilde g}(0) \quad \textrm{if and only if} \quad  \mu_g(1) \leq \mu_{\tilde g}(1).
\end{equation}
In other words, the treatment does not work \emph{against} any particular type: if the high-talent type does better than the low-talent type on average without the treatment, then the high type does better on average with the treatment as well, and vice versa. We note that the rank invariance does not rule out the possibility that the treatment may be more beneficial on average to one type than to another: it is the relative rankings that do not change. It can be shown that if the rank invariance condition holds, then the sharp identified set for $\bar G$ becomes $\bigl\{m\in\mathbb{N}:  \bar G_L \leq m \leq \bar G_C \bigr\}$.\footnote{To see this, note that under rank invariance, for both $d = 0 ,1$, $\mathcal{P} ( d ) $ partitions $\{ 1 , \cdots , \bar{G} \}$ into $\check{G}(d)$ consecutive blocks such that  $\max\ \mathcal{G}_j(d) < \min\ \mathcal{G}_k(d)$ whenever $j<k$. This immediately rules out $\bar{G} >\bar{G}_C$ because $\mathcal{G}_j(0)\cap\mathcal{G}_k(1)$ is either empty or a singleton for all $j,k$ by Assumption \ref{ass:setup}. However, $\bar{G} = \bar{G}_C$ can be achieved, e.g., by choosing $\iota$ so that all entries in the first row and the last column are equal to one, and all remaining entries are zero.} Therefore, $\bar G = \bar G_C$ corresponds to the worst-case scenario under the rank invariance.

The causal effect $\tau$ is not point-identified in general because of potential confounding from unobserved heterogeneity. However, Theorem \ref{thm:main0} suggests that there are several conservative measures of interest. For example, the researcher may present $\ell_m := \inf \mathcal{T}(m)$ for all integers $m$ between $\bar G_L$ and $\bar G_C$, which will provide a complete picture on the lower bounds on the causal effect $\tau$. We may do the same for the upper bounds if aggressive measures are desired.  If more succinct summaries are wanted, then we may consider
\begin{align} 
\ell_{\bar{G}_C}  & = \min_{m} \ell_m = \inf \mathcal{T},\\ 
\tau_{ALB} &:= \sum_m w_m \ell_m,
\end{align} 
where $w_m$'s are weights chosen by the researcher. Observe that $\ell_{\bar{G}_C} $ represents the Minimum Lower Bounds (MLB) on $\tau$, whereas $\tau_{ALB}$ shows the Average Lower Bounds (ALB).  For example,
\[
\tilde{\tau}_{ALB} 
:= \frac{1}{\bar G_C - \bar G_L + 1}\sum_{m=\bar G_L}^{\bar G_C} \ell_m 
\]
is the Maximum Entropy ALB for $\tau$ in that it uses the uniform weights on the sharp identified set of $\bar G$ that reflects the rank invariance condition in \eqref{eq:ri}. For the use of the maximum entropy principle under partial identification in different contexts, see e.g., \citet{jun2024information}.  We can similarly define the Maximum Upper bounds (MUB) and the Average Upper Bounds (AUB).

The MLB and ALB have their own advantages and disadvantages. For instance, the MLB on $\tau$ is easier to interpret, and it does not require the researcher choose a prior on $\bar G$. But it may be overly conservative. The ALB, which is the average of the lower bounds on $\tau$ for different values of $\bar G$, is always less conservative than MLB since it reflects the researcher's attitude about uncertainty on $\bar G$.  We do not advocate a particular stance here.

Before proceeding, we provide the value of each $\mathcal{T}(m)$ in Example \ref{exa:simple} as an illustration.
\begin{subexample}[Cont.]
Recall from Examples \ref{eg:partitions} and \ref{exa:taupid} that the sharp identified sets for $\bar{G}$ and $\tau$ are $\{2  , 3 , 4\}$ and $ \mathcal{T} = (2/3, 5/3)$, respectively. Then, for a given $\bar{G} \in \{ 2 , 3, 4 \}$, the sharp identified sets for $\tau$ are $\mathcal{T}(2) = \{ 1 , 4/3 \}$, $\mathcal{T} ( 3 ) = ( 2/3, 1 ) \cup (1 , 5/3) = \mathcal{T}\setminus \{1\}$, and $\mathcal{T}( 4 ) = (2/3, 5/3) =\mathcal{T} $.
Moreover, since $\bar G_L = 2$ and $\bar G_C = 3$, the MLB is $\ell_{\bar{G}_C} = 2/3$ and the Maximum Entropy ALB is $ \tilde{\tau}_{ALB} =5/6$. Note that here $\tilde{\tau}_{ALB} < \tau = 4/3$, but this may not always be the case.
\end{subexample}

To conclude this section, we highlight that additional restrictions on the partitions can be systematically incorporated into our framework to construct the sharp identified set for $\tau$. For instance, if we invoke rank invariance, the sharp identified set becomes 
\[
 \mathcal{T}_{\mathrm{ri}}  : = \bigcup_{\bm{\iota} \in \mathscr{I}_{\mathrm{ri}}} \mathcal{T}_{\bm{\iota}}  ,
\]
where $ \mathscr{I}_{\mathrm{ri}} \subset  \mathscr{I}$ denotes the subset of admissible partitions under Assumption \ref{ass:setup} and rank invariance. Specifically, it can be shown that $\mathscr{I}_{\mathrm{ri}}  =  \{ \bm{\iota} \in  \mathscr{I}: \  \iota_{j^\prime m^\prime} = 0  \ \text{if} \  \iota_{j m} = 1  \ \text{for some} \ j < j^\prime , m < m^\prime \}$ from inspecting the proof of Lemma \ref{lem:partsgen}. In addition, for a given $m \in \{ \bar G_L, \cdots , \bar G_C  \}$, we can construct the sharp identified set for $\tau$ under the rank invariance as follows:
\[
 \mathcal{T}_{\mathrm{ri}}   (m): = \bigcup_{\bm{\iota} \in \mathscr{I}_{\mathrm{ri}} ( m ) } \mathcal{T}_{\bm{\iota}}   \quad \text{with} \ \  \mathscr{I}_{\mathrm{ri}} ( m ) =  \mathscr{I} ( m )  \cap \mathscr{I}_{\mathrm{ri}}.
\]

We remark that if $\bar G=\check G(0)=\check G(1)$, then $\mathscr{I}(\bar G)\cap \mathscr{I}_{\mathrm{ri}}$ contains only the identity matrix. Consequently, $\mathcal{T}_{\mathrm{ri}}(\bar G)$ is a singleton. This observation helps explain the point-identification result of \citet{gardner2020identification}. Reformulating the discussion in Footnote \ref{fn:gardener}, his framework implicitly imposes both rank invariance and the restriction $\bar G=\check G(0)=\check G(1)$, which together imply point identification of $\tau$.

While extending Theorem \ref{thm:main0} to the sets $\mathcal{T}_{\mathrm{ri}}$ and $\mathcal{T}_{\mathrm{ri}}(m)$ is beyond the scope of this paper, the following example provides an illustrative special case.

\begin{subexample}[Cont.] 
If we impose rank invariance, the sharp identified set for $\bar{G}$ is $\{ 2 , 3\}$, while $\mathscr{I}_{\mathrm{ri}} = \{\bm{\iota}_1, \bm{\iota}_4, \bm{\iota}_5\}$.
Therefore, the sharp identified sets for $\tau$ become $\mathcal{T}_{\mathrm{ri}}(2) = \mathcal{T}_{\bm{\iota}_1} = \{ 1 \}$, $\mathcal{T}_{\mathrm{ri}}(3) = \mathcal{T}_{\bm{\iota}_4} \cup  \mathcal{T}_{\bm{\iota}_5} = (2/3, 1 ) \cup ( 1 , 5/3) = \mathcal{T}\setminus \{1\}$. Consequently, $\mathcal{T}_{\mathrm{ri}} = \mathcal{T}$. That is, rank invariance yields point identification when $\bar G = 2$, but it does not generally improve the Manksi bounds.
\end{subexample}

\section{The Estimator} \label{sec:esti}

In this section, we propose an estimator of $\ell:=( \ell_{\bar G_L}, \cdots, \ell_{\bar G_C} )$, while we provide its asymptotic properties in the Supplement together with the inference procedures. We do not discuss the cases where $m>\bar G_C$, but there is no loss of generality here because Theorem \ref{thm:main0} implies $\ell_{\bar G_C} = \ell_m$ whenever $\bar G_C \leq m  \leq \bar G_U$. Also, we focus on the lower bounds $\ell_m$, but the methodology and results can be symmetrically applied to the upper bounds $\sup \mathcal{T}(m)$ as well.

We consider a random sample $\{ (Y_1, D_1), \dots, (Y_n, D_n) \}$ drawn from the distribution of $(Y, D)$ under Assumption \ref{ass:setup}. Estimating $\ell$ requires preliminary estimators of the identified parameters, for which we present the following two-step procedure.

In the first step, for each $d=0,1$, we estimate the mixture order $ \check G(d)$ by using any method that produces an estimator $\hat G(d) \in \mathbb{N}$ such that $\hat G(d) = \check{G}(d)$ w.p.a.1, i.e.,
\begin{equation}
\label{eq:conG}
\Pr\{ \hat G(d) = \check{G}(d)\} \rightarrow 1 .
\end{equation}
For instance, one can use \cite{chen09order}'s estimator, which involves the use of penalized maximum likelihood with a cluster-based penalty function that follows \citet{fan01var}. In the second step, for each $d=0,1$, we estimate the vector of the mixture parameters
\[
  \gamma(d) := \bigl( \pi_1(d,d),\cdots, \pi_{\check{G}(d)-1}(d,d), \check{\mu}_1(d),\cdots, \check{\mu}_{\check{G}(d)}, \sigma^2(d)\bigr)
\]  
by maximum likelihood, using $\hat{G}(d)$ in lieu of $\check{G}(d)$, and we denote the resulting estimator by
\[
\hat \gamma(d) := \bigl( \hat\pi_1 ( d , d ),\cdots, \hat\pi_{\hat{G}(d)-1}(d,d) , \hat{\mu}_1(d)  , \cdots , \hat{\mu}_{\hat G(d)} (d) , \hat\sigma^2 (d) \bigr) ,
\]  
emphasizing that $\hat{\mu}_{j} (d)$ is an estimator of $\check{\mu}_{j} (d)$, not ${\mu}_{j} (d)$.\footnote{When $\check{G}(d)=1$, $\gamma(d)$ reduces to $ ( {\mu}_1(d)  , \sigma^2(d) )$, and its estimator $\hat\gamma(d)$ coincides with the usual maximum likelihood estimator under normality.}
To obtain $\hat \gamma(d)$ and ensure its existence w.p.a.1, we remark that the maximization is carried out over a compact subset $\Gamma ( \hat{G} (d) ) \subset \mathbb{R}^{2 \hat{G} (d)} $ specified in the Supplement. We refer to \citet[Chapter 2]{mp00finmix} for further discussion on implementation aspects.

In order to construct the estimator of $\ell$ and to derive its asymptotic properties in the next subsection, we introduce the following notation. Let $\Lambda := \bigl(  \Pr(D=0) , \gamma(0)  , \gamma(1)   \bigr)$ denote the vector of parameters characterizing the joint distribution of $(Y,D)$ and write $\mathscr{I}_{LC} : = \cup_{m=\bar{G}_L}^{\bar{G}_C} \mathscr{I} (m)$. Then, define the vector-valued function $\mathcal{L}(\Lambda)  :=   \bigl( \mathcal{L}_{\bm\iota} (\Lambda)  \bigr)_{\bm\iota \in \mathscr{I}_{LC}}$ by 
\begin{equation}     \label{eq:defL}
\mathcal{L}_{\bm\iota}(\Lambda  )
:= 
\Prob ( D = 1)  \sum_{k=1}^{\check{G}(1)} \pi_k ( 1 , 1) \big\{  \check{\mu}_{k} (1)  -  \check{\mu}_{\bar{j}_k (\bm{\iota})} (0)  \big\} 
-   
\Prob ( D = 0)    \sum_{j=1}^{\check{G}(0)} \pi_j ( 0 , 0)  \big\{ \check{\mu}_{j} (0)   -  \check{\mu}_{\ushort{k}_j(\bm{\iota})} (1)  \big\}  ,
\end{equation} 
where $\bar{j}_k (\bm{\iota})  :=  \max \bigl\{ j  \in \{ 1, \cdots , \check{G} (0)  \} :  \iota_{jk} = 1  \bigr\}$ and $ \ushort{k}_j (\bm{\iota}) :=  \min \bigl\{ k  \in \{ 1, \cdots , \check{G} (1)  \} :  \iota_{jk} = 1  \bigr\}$. Moreover, for a given real vector $v = (v_{\bm\iota}   )_{\bm\iota \in \mathscr{I}_{LC}}$ indexed by $\mathscr{I}_{LC}$, define the function $\mathcal{M}$ by
\begin{equation}
\mathcal{M} (v) : = \left(  \mathcal{M}_{\bar{G}_L}  (v) , \cdots  ,  \mathcal{M}_{\bar{G}_C}   (v)  \right)  \in \mathbb{R}^{\bar{G}_C - \bar{G}_L +1} \  \  \text{with}  \    \mathcal{M}_m ( v) := \min \left\{ v_{\bm{\iota}} :\    \bm{\iota} \in \mathscr{I} (m) \right\} .
\label{eq:defM}
\end{equation}
From Lemma \ref{alem:shapeT} in the Appendix, we can now write
\begin{equation}
\label{eq:ell}
\ell = \mathcal{M} \circ \mathcal{L} \left( \Lambda  \right).
\end{equation}

Now, we consider a plug-in approach to estimate $\ell$ by using \eqref{eq:ell} and the preliminary estimators, i.e., $(\hat G(0), \hat G(1))$ and $(\hat\gamma (0), \hat\gamma (1))$. Let $\hat\Lambda := \bigl(  \hat{\Pr}(D=0) , \hat\gamma(0), \hat\gamma(1)   \bigr)$, where $\hat{\Prob} ( D = d)  : = n^{-1} \sum_{i=1}^n \one ( D_i = d)$. We then define the estimator of $\ell$ by
\begin{equation} \label{eq:estiell}
\hat{\ell} := \hat{\mathcal{M}} \circ \hat{\mathcal{L}} \, \bigl( \hat{\Lambda}  \bigr)   \quad \text{with}  \ \ \hat{\mathcal{L}}( \hat{\Lambda})   : =   \bigl( \hat{\mathcal{L}}_{\bm\iota} (\hat{\Lambda})  \bigr)_{\bm\iota \in \hat{\mathscr{I}}_{LC}} ,
\end{equation}
where $\hat{\mathscr{I}}_{LC}$, $\hat{\mathcal{L}}_{\bm\iota}$, and $\hat{\mathcal{M}}$ are defined analogously to $\mathscr{I}_{LC}$, $\mathcal{L}_{\bm\iota}$, and $\mathcal{M}$, respectively, with $(\hat G(0)  ,\hat G(1))$ replacing $(\check G(0)  ,\check G(1))$.
In this regard, note that $\hat{\mathscr{I}}_{LC} = \mathscr{I}_{LC}$, $\hat{\mathcal{L}} = \mathcal{L}$ and $\hat{\mathcal{M}} = \mathcal{M}$ hold w.p.a.1 due to Eq.\ \eqref{eq:conG}. We remark that computing $\hat{\mathcal{L}}(\hat{\Lambda})$ can be demanding when $\hat{G}(0)$ and $\hat{G}(1)$ are large, since the cardinality of $\hat{\mathscr{I}}_{LC}$ increases rapidly with these quantities. In such cases, one may instead employ approximation methods such as resampling elements from $\hat{\mathscr{I}}_{LC}$. The resulting approximation error can be made statistically negligible by increasing the number of replications.

The function $\mathcal{M}_m$ generally requires that we consider all entries $\mathcal{I}(m)$ except the one corresponding to the MLB $\ell_{\bar{G}_C}$. To see this point, let $\ushort{\bm\iota}$ denote the $\check G(0)  \times \check G(1) $ matrix that has all ones in the first column and in the last row, and zeroes everywhere else, noting that $\ushort{\bm\iota} \in \mathscr{I} ( \bar{G}_{C} )$.
Theorems \ref{thm:simple} and \ref{thm:main0}, as well as Lemma \ref{alem:shapeT} in the Appendix, imply
\begin{multline}\label{eq:lowerl}
{\ell}_{\bar{G}_{C}} 
= 
\mathcal{M}_{\bar{G}_{C}} \circ   \mathcal{L} (\Lambda    )  
= 
\mathcal{L}_{\ushort{\bm\iota}} (\Lambda    ) 
\\
=
\Prob ( D = 1) 
\sum_{k=1}^{\check{G}(1)} \pi_{k} (1 , 1 ) \bigl\{  \check{\mu}_{k} (1) - \check{\mu}_{\check{G} (0)} ( 0 ) \bigr\}
-  
\Prob (D =0) 
\sum_{j=1}^{\check{G}(0)} \pi_{j} (0 , 0 )  \bigl\{ \check{\mu}_{j} (0) - \check{\mu}_1 (1) \bigr\}.
\end{multline} 
As a result, letting $\hat{G}_{C} =  \hat G(0) + \hat G(1) - 1$, $\hat{\ell}_{\hat{G}_{C}}$ can be computed simply by replacing the parameters in \eqref{eq:lowerl} with their corresponding estimators.  Of course, this is equivalent to applying the plug-in principle to the formula of the Manski bounds in \eqref{eq:manski}.

We remark that ${\ell}_{\bar{G}_{C}}$ can also be estimated via an overfitted MLE, i.e., by specifying a mixture order larger than $\check{G}(d)$ rather than estimating it in a first step. To see this, observe that 
\begin{equation*}
{\ell}_{\bar{G}_{C}}   =  \Prob (D =1)   \left\{ \E (Y \mid D =1)   -  \max_j \check{\mu}_{j} (0) \right\} -  \Prob (D =0) \left\{ \E (Y \mid D =0) -  \min_j \check{\mu}_{j} (1)  \right\}  .
\end{equation*}
An alternative plug-in estimator can therefore be constructed by replacing $\Prob (D =d)$ and $\Exp (Y\mid D =d)$ with their sample analogs, and $\max_{j} \check\mu_j (0)$ and $\min_j \check\mu_j (1)$ with the corresponding estimates obtained from the overfitted MLE.

Because the overfitted MLE yields consistent estimates of $\max_j \check\mu_j (0)$ and $\min_j \check\mu_j (1)$ -- a consequence of consistency in Wasserstein distance -- this alternative estimator is also consistent, though converging at a slower rate of $n^{-1/4}$ up to a polylogarithmic factor \cite[][Theorem 4.3]{ho16strong}.  Thus, we do not consider this approach in the asymptotic analysis provided in the Supplement. In practice, however, if the primary objective is to obtain a consistent estimate of the Manski bound ${\ell}_{\bar{G}_{C}}$, this alternative avoids the need to estimate the mixture orders.

\section{Extensions}	\label{sec:exten}

\subsection{Covariates and Sample Selection}		\label{sec:sampleselect}

We have been implicit about covariates $X$, and therefore, our discussion so far can be understood as conditional on $X=x$: i.e., the total number $\bar G$ of types is allowed to be heterogeneous across different subpopulations, and our results should be understood as bounding the conditional ATE (CATE). However, when it comes to estimation and inference, this approach of completely localizing at (or around) $X=x$ may not be practically attractive, especially when the dimension of $X$ is high. Further, censoring or selection is an issue that frequently arises in applications, and it generally makes na\"{i}ve Gaussian models unrealistic. Therefore, we now discuss extensions to address covariates and selection; we will focus on the simplest case of Type I censoring, but extensions to other types of selection are straightforward.

Let $X$ be a random vector of covariates, which does not include a constant, and consider the shifted potential outcome
\begin{equation*}
\tilde{Y}(d)  : = X'\beta(d) + \underbrace{\sum_{g=1}^{\bar G}\one(G=g) Y_g(d)}_{=: Y(d)}
\quad \text{with} \quad 
Y_g(d) : =   \mu_g (d) + \epsilon (d) \ \  \text{and} \  \  d\in \{0,1\},
\end{equation*}
where $\beta(d)$ is a vector of slope coefficients. In this setting, $Y(d) :=  \sum_{g=1}^{\bar G} \one(G=g) Y_g (d)$ represents the potential residual after partialling out the effects of the covariates, where $Y_g (d)$ can be interpreted as a group-specific ``residual.''  

We consider the following assumption about a possible selection mechanism and the residual distributions. Let $\tilde{Y} : = D \tilde{Y}(1) + (1-D) \tilde{Y}(0) $ denote the shifted outcome, and let $\lambda$ be the inverse Mills ratio.

\begin{assump}	\label{ass:sampleselect}
The next conditions hold.
\begin{enumerate}

\item The variance-covariance matrix of $X$ is finite and positive definite.

\item $(S \tilde{Y} , S D, X, S )$ is observed, where $S : =  \one ( \eta_0 +  X' \eta_1 + \nu > 0  )$ is a selection indicator, where $\eta_0$ and $\eta_1 \neq 0$ are parameters, and $\nu$ is an error term that is random. 

\item For $d = 0 , 1$, $\left( X ,\ \lambda( \eta_0 +  X' \eta_1  ) \right)$ has finite and positive definite variance-covariance matrix conditional on $(S , D)=( 1 , d)$. 

\item For $d = 0 , 1$, $( \epsilon (d) , \nu ) $ is independent of $(D ,G , X)$, and it follows a bivariate normal distribution of the form
\begin{equation*}
\left(
\begin{array}{c}
   \epsilon (d)  \\
 \nu  
\end{array}
\right)
\sim N  \left(
\left(
\begin{array}{c}
  0  \\
  0
\end{array}
\right) ,
\left(
\begin{array}{cc}
   \sigma^2 (d) & \sigma_{\epsilon\nu}(d)   \\
 \sigma_{\epsilon \nu}(d) & 1 
\end{array}
\right)
\right)  .
\end{equation*}

\end{enumerate}

\end{assump}

By condition (1), $X$ cannot include a constant. Conditions (1) and (2) guarantee identification of $(\eta_0 , \eta_1)$ via probit, while (3) ensures identification of $\beta(d)$ and $\sigma_{\epsilon\nu}(d)$ via the least squares of $Y$ on $X$ and $\lambda(\eta_0 + X'\eta_1)$ conditional on $D=d$. So, hereafter, $\bigl( \eta_0, \eta_1, \beta(d), \sigma_{\epsilon\nu}(d) \bigr)$ will be treated as known parameters for identification purposes.

In this setup, CATE can still be represented by $\tau$ as defined in equation \eqref{eq:ategr}: i.e., 
\[
\text{CATE}(x)  
= x'\{ \beta(1) - \beta(0)\} + \E\{ Y(1) - Y(0) \}
= x'\{ \beta(1) - \beta(0)\} + \tau. 
\]
Hence, constructing identified sets for CATE reduces to constructing identified sets for $\tau$. In the special case of $\sigma_{\epsilon\nu}(d) = 0$, selection is exogenous, and it is innocuous to focus on ``uncensored observations.'' Then, we can obtain identified sets for $\tau$ by applying our previous results to the \emph{homogenized} outcome $Y = \tilde Y - X'\beta(d)$ given $D=d$. More generally, however, selection necessitates further modification, which we elaborate on below.

The potential residuals $(Y(0), Y(1))$ satisfy Assumption \ref{ass:setup}, and therefore the conditional distribution of $Y = D Y(1) + (1-D)Y(0) = \tilde Y - X'\beta(d)$ given $D=d$ (but unconditional on $S=1$) remains a Gaussian mixture as specified in equations \eqref{eq:mix0}-\eqref{eq:mix1}.  Therefore, Theorem \ref{thm:main0} would still apply if the number $\check G(d)$ of mixing components, and 
\[
  \gamma(d) = \bigl( \pi_1(d,d),\cdots, \pi_{\check{G}(d)-1}(d,d), \check{\mu}_1(d),\cdots, \check{\mu}_{\check{G}(d)}, \sigma^2(d)\bigr) 
\]
were known. Therefore, all we need to establish here is that these parameters are still identified when we condition on $S=1$.  In the following lemma, we show that conditioning on $S=1$ yields a non-Gaussian mixture likelihood, from which we can identify and estimate $\check G(d)$ and $\gamma(d)$.  Let $\Phi$ denote the standard normal distribution function, and let $\phi := \Phi'$ be its density.

\begin{lemma}	\label{lem:sampleselect}
Suppose that Assumption \ref{ass:sampleselect} holds. Then, the conditional probability density function of $\tilde Y$ given $(D,X ,S) = (d, x ,1)$ is 
\begin{equation*}
f_{\tilde Y | D,X ,S} \left( y  \mid  d,x ,1 \right)  =  \sum_{j=1}^{\check{G}(d)}  \pi_j ( d , d)   f_{j} \left( y  \mid  d,x  \right) ,
\end{equation*}
where $f_{j} ( y  \mid  d,x ) $ is equal to
\begin{equation*}
\frac{1}{  \Phi \left( \eta_0 +  x' \eta_1 \right) } \Phi \left( \frac{ \eta_0 +  x' \eta_1  + \frac{\sigma_{\epsilon \nu}(d) }{\sigma^2(d)}\left\{y -  x' \beta(d) - \check{\mu}_j (d)\right\}}{\sqrt{1  - \frac{\sigma_{\epsilon \nu}(d) }{\sigma^2(d)}} } \right)  \frac{1}{ \sigma(d) } \phi \left(  \frac{y - x' \beta(d) - \check{\mu}_j (d)}{\sigma(d)} \right)  .
\end{equation*}
As a result, $\check G(d)$ and $ \gamma(d)$ are identified from $f_{\tilde Y | D,X ,S} ( \cdot  |   d , x ,1) $.
\end{lemma}

From this lemma, estimating the identified sets for $\tau$ is just a matter of estimating the mixture parameters $\check G(d)$ and $ \gamma(d)$; see Section \ref{sec:barG}. With this aim, considering a random sample
\begin{equation*}
\{ ( S_1 \tilde{Y}_1, S_1 D_1 , X_1 , S_1), \cdots, ( S_n \tilde{Y}_n, S_n D_n , X_n , S_n) \} \quad \text{from}  \  \ (S \tilde{Y} , S D , X  , S) ,
\end{equation*}
we suggest two possibilities.

The first one consists in a three-step procedure. In the first two steps, we estimate $\eta_0, \eta_1, \beta(d)$, and $\sigma_{\epsilon\nu}(d)$ by a combination of probit and least squares, often referred to as Heckit. In the third, we first estimate $\check G(d)$, for which we suggest using \citet{chen09order}'s penalization method. Then, we estimate $\gamma(d)$ by maximum likelihood using the selected sample and treating $\check G(d)$ as the true mixture order. Specifically, this procedure can be described as follows.

\begin{enumerate}

\item Run a Probit regression of $S_i$ on an intercept and $X_i$ to obtain $\hat\eta_0$ and $\hat\eta_1$, as well as $\hat \lambda_i := \lambda( \hat\eta_0 + X_i\hat\eta_1)$ for $i = 1 , \cdots , n$.

\item Using the selected sample ($S_i =1$), run an OLS regression of $\tilde{Y}_i$ on an intercept, $D_i$, $X_i$, $\hat \lambda_i$, $D_i X_i$, and $D_i\hat \lambda_i$. Obtain $\hat\beta(d)$ and $\hat\sigma_{\epsilon \nu}(d)$, for $d = 0 ,1$, for which we note that the coefficient of $X_i$ is $\hat \beta(0)$, while that of $X_i D_i$ is $\hat\beta(1) - \hat \beta(0)$; similarly, the coefficient of $\hat\lambda_i$ is $\hat \sigma_{\epsilon \nu}(0)$, while that of $D_i\hat\lambda_i$ is $\hat \sigma_{\epsilon \nu}(1) - \hat \sigma_{\epsilon\nu}(0)$.

\item Using the selected sample ($S_i =1$), for each $d=0,1$, first compute the estimator $\hat{G} (d)$ of $\check{G}(d)$, e.g., by using \cite{chen09order}'s penalization method, and then estimate $\gamma(d)$ by maximum likelihood as follows: define
\begin{eqnarray*}
& &  \hat f_i\bigl( m, s \bigr) \ : = \\
& &   \  \ \frac{1}{\Phi(\hat\eta_0 + X_i'\hat\eta_1)}
\Phi\left( \frac{\hat\eta_0 + X_i\hat\eta_1 + \frac{\hat\sigma_{\epsilon \nu}(d)}{s}\{\tilde Y_i - X_i'\hat\beta(d) - m  \} }{\sqrt{1-\frac{\hat\sigma_{\epsilon \nu}(d)}{s}}}  \right)
\phi\left(\frac{\tilde Y_i - X_i'\hat\beta(d) - m}{s} \right) \frac{1}{s} 
\end{eqnarray*}
and maximize 
\begin{equation}
\label{eq:llselec}
\sum_{i: S_i = 1, D_i = d}  \log\left\{ \sum_{j=1}^{\hat{G} (d)}  p_j ( d )   \hat f_i \bigl(   m_j (d), s (d)  \bigr) \right\}
\end{equation}
with respect to $p_j(d) >0$, $m_j (d) \in \mathbb{R}$ , $s (d) > 0$, where $p_{\hat{G} (d)} ( d ) = 1 - \sum_{j<\hat{G} (d)}  p_j ( d ) $.\footnote{To implement the penalization method of Chan and Khalili in this setting, we suggest using \eqref{eq:llselec} as the log-likelihood in their equation (4), with the mixture order set sufficiently large so as to exceed $\check{G}(d)$.}
\end{enumerate}

The second approach consists in estimating all model parameters jointly via (conditional) maximum likelihood, except for the mixture orders $(\check{G}(0), \check{G}(1))$. Specifically, we first estimate these mixture orders (e.g., by \cite{chen09order}'s  method) and then, treating such estimates as the true mixture orders, we jointly estimate all other parameters via (conditional) maximum likelihood. In this regard, note that the joint log-density of $(S \tilde{Y} , S D , X  , S)$ evaluated at $(s \tilde{y} ,s d,x,s)$ is given by
\begin{equation*}
(1-s)\log\{ 1- \Phi(\eta_0 + x \, \eta_1) \} 
 + 
s \bigl\{ d \log f_{\tilde Y\mid D,X,S}(\tilde{y} \mid 1, x, 1) + (1-d)\log f_{\tilde Y\mid D,X,S}(\tilde y \mid 0, x, 1) \bigr\} ,
\end{equation*}
while the closed-form expression for $f_{\tilde Y\mid D,X,S}$ can be obtained from Lemma \ref{lem:sampleselect}.

The three-step procedure is attractive for its computational simplicity, while the alternative joint estimation approach is asymptotically efficient and attains the efficiency bound. Details on asymptotic theory and inference are provided in the Supplement.

\subsection{Generalization to Non-Gaussian Finite Mixtures}	\label{sec:extensions}

Since we use finite mixtures to deal with unobserved discrete types, we start by introducing an identifiable mixture class. We suppress covariates again for simplicity. Consider a parametric family of univariate distributions $\mathscr{F} := \{ F(\cdot ; \theta):\ \theta \in \Theta \}$, where $\Theta \subseteq \mathbb{R}^K$, for some $K\in \mathbb{N}$, and $F$ satisfies $\int |y| \ dF(y;\theta) <\infty$ for all $\theta \in \Theta$.
Let $\mathscr{M}$ be the class of finite mixtures based on $\mathscr{F}$ such that 
\begin{equation}\label{eq:M}
\mathscr{M} = \Bigl\{ \sum_{j=1}^m p_j F( \cdot ;\theta_j):\  (p_1,\cdots, p_m)\in \Pi_m,\  (\theta_1,\cdots, \theta_m)\in \Omega_m,\  m \in\mathbb{N} \Bigr\},
\end{equation}
where $\Pi_m = \{ \textbf{p} \in \mathbb{R}_{++}^m :  \sum_{l=1}^m  p_l= 1 \}$ and $\Omega_m\subseteq \Theta^m$.

The class $\mathscr{M}$ depends on $\mathscr{F}$ as well as $\{\Pi_m:\ m\in\mathbb{N}\}$ and $\{\Omega_m: m\in\mathbb{N}\}$, but we suppress this dependence for the sake of notational simplicity. The parameter space for the class $\mathscr{M}$ is $\cup_{m=1}^\infty \Pi_m\times \Omega_m$.  This formulation does not require that the true number of mixing components be known: e.g., if $m_0$ is the true number of component distributions, then the true parameters $(p_1,\cdots, p_{m_0})$ and $(\theta_1,\cdots, \theta_{m_0})$ are in $\Pi_{m_0}$ and $\Omega_{m_0}$, respectively. Also, it is worth noting that $\{\Omega_m:\ m\in\mathbb{N}\}$ need not be the same as $\{\Theta^m:\ m\in\mathbb{N}\}$. For example, certain elements of $\theta_m$ may remain constant across $m$, as in the Gaussian location mixture models discussed earlier, in which the variance does not vary across components.

\subsubsection{Identification} Following \citet[Section 1.5]{mb88mix}, we say that $\mathscr{M}$ is an identifiable class when the following condition is satisfied: if 
\begin{equation*}
\forall y\in\mathbb{R},\quad 
\sum_{j=1}^m   p_j F (  y , \theta_j)   =   \sum_{j=1}^{\tilde{m}}   \tilde{p}_j F (  y , \tilde{\theta}_{j}),
\end{equation*}
where $\theta_j  \neq \theta_{j'}$ and $\tilde\theta_j  \neq \tilde\theta_{j'}$ for all $j\neq j'$, then it follows that $\tilde m  = m$ and $\bigl( ( \tilde{p}_1 , \tilde{\theta}_1  ) , \cdots ,  ( \tilde{p}_{\tilde{m}} , \tilde{\theta}_{\tilde{m}}  )  \bigr)$ is a permutation of $\bigl( ( p_1 , \theta_1  ) , \cdots ,  ( p_m , \theta_m  )  \bigr)$. There are many well-known examples of identifiable mixture classes. For instance, the class of Poisson mixtures is identifiable; see \cite{kar01rob} and \citet[Example~2]{chen09order}. Likewise, exponential mixture models are also identifiable. Additional examples and discussions can be found in \citet[Ch.\ 3]{eve13fin} and \citet[Sections 5.8.4 and 5.9]{mp00finmix}.

Now, we discuss extending Assumption \ref{ass:setup} beyond Gaussian mixtures. Let $Y_g(d)$ represent the potential outcome for the treatment status $d\in \{0,1\}$ for an individual who belongs to type $g\in \{1,2,\cdots, \bar G\}$, where $\bar G$ is unknown to the researcher. We assume that the marginal distribution of $Y_g(d)$ is represented by $F\bigl(\cdot; \theta_g(d) \bigr)\in \mathscr{F}$, and denote the mean vector of $\bigl( Y_g(0), Y_g(1)\bigr)$ by $\bm{\mu}_g  : = \bigl( \mu_g(0), \mu_g(1) \bigr) : = \bigl( \Exp\{ Y_g(0)\}, \Exp\{ Y_g(1)\} \bigr)$, where $\mu_g(d)$ depends on $\theta_g(d)$ because of $\mu_g(d) = \int y\ dF\bigl(y;\theta_g(d) \bigr)$.  We now state a generalization of Assumption \ref{ass:setup}.

\begin{assump}		\label{ass:exten}
There exists a possibly unknown $\bar G \in \mathbb{N}$ such that for $d\in \{0,1\}$,
  \begin{equation} \label{eq:ygen}
   {Y}(d) = \sum_{g=1}^{\bar G} \one(G=g) {Y}_g (d),
  \end{equation}
where the following conditions hold. 
\begin{enumerate}
\item\label{cond1} For all $(g,d) \in \{1,2,\cdots, \bar G\}\times\{0,1\}$, we have $0 < \Pr(G=g, D=d)<1$.

\item \label{cond1-1} We have $\bm{\mu}_1 \prec  \bm{\mu}_2 \prec \cdots \prec \bm{\mu}_{\bar G}$, where $\prec$ represents lexicographic ordering. 

\item\label{cond2} For each $g \in \{1,2,\cdots, \bar G\}$, ${Y}_{g} (d)$ is independent of $(D,G)$, and the marginal distribution of $Y_g(d)$ is given by $F\bigl(\ \cdot\ ; \theta_g(d)\bigr) \in \mathcal{F}$ with $\bigl( \theta_1(d),\cdots, \theta_{\bar G}(d) \bigr) \in \Omega_{\bar G} \subseteq \Theta^{\bar G}$ for $d\in \{0,1\}$.    

\item\label{cond3} The class $\mathscr{M}$ of finite mixtures (defined in \eqref{eq:M}) is identifiable.

\item\label{cond4} If $\theta_{g} (d)  \neq \theta_{g^\prime} (d)  $, then $ \mu_g( d) \neq  \mu_{g^\prime} ( d) $.
\end{enumerate}
\end{assump}

Assumption \ref{ass:setup} can be easily nested into Assumption \ref{ass:exten} by setting
\begin{equation*}
Y_g (d) = \mu_g (d)   + \epsilon(d) \quad \text{for} \   g = 1  , \cdots , \bar G,
\end{equation*}
with $\epsilon(d) \sim N(0, \sigma^2(d) )$ being independent of $(D,G)$ so that $\theta_g (d) = ( \mu_g (d),  \sigma^2(d) )$.  
Condition (\ref{cond1}) is simple regularity: there are only finitely many types that are relevant. Condition (\ref{cond1-1}) is an innocuous normalization, provided that the mean vectors are all distinct across different types: i.e., $\bigl( \mu_g(0), \mu_g(1) \bigr) \neq \bigl( \mu_{g'}(0), \mu_{g'}(1) \bigr)$ whenever $g\neq g'$. Condition (\ref{cond2}) allows $Y(d)$ to depend on $G$, but $Y(d)$ becomes independent of $D$ once $G$ is controlled for. Therefore, $G$ is the only source of confounding.  It is worth noting that condition (\ref{cond2}) does not impose any restrictions on the dependence between $G$ and $D$, or between $Y_g(d)$ and $Y_{g'}(d')$ as long as $g\neq g'$. Condition (\ref{cond3}) requires that the class $\mathscr{M}$ be identifiable, but, it does not require that all $\theta_g(d)$'s be distinct for different values of $g$ given a fixed value of $d=0,1$. For example, if $\theta_1(0) = \theta_2(0)$, then the number of mixing components that is identified from the control group will be strictly smaller than $\bar G$.  In fact, neither the treatment group nor the control group may reveal the number $\bar G$ of types in the population. Condition (\ref{cond4}) ensures that different component distributions have distinct means, meaning that we only focus on \emph{type heterogeneity in the mean}.

As we commented above, for any given $d\in \{0,1\}$, not all of $\theta_g(d)$'s need to be distinct for different values of $g$. However, conditions (\ref{cond1-1}) and (\ref{cond4}) impose some restrictions. For instance, if $\theta_g(0) = \theta_{g'}(0)$, then we must have $\theta_g(1) \neq \theta_{g'}(1)$. Otherwise, we would have $\bigl( \mu_g(0), \mu_g(1) \bigr) = \bigl( \mu_{g'}(0), \mu_{g'}(1)\bigr)$, in which case there is no strict lexicographic ordering between the means of types $g$ and $g'$, and therefore, in view of condition (\ref{cond4}), $g$ and $g'$ should simply be treated as the same type. 

Let $\check{\theta}_1(d),\cdots, \check{\theta}_{\check{G}(d)}(d)$ be distinct parameter values among $\theta_1(d),\cdots, \theta_{\bar G}(d)$. Let $\check{\mu}_j(d)$ be the mean corresponding to $\check{\theta}_j(d)$ so that $\check{\mu}_1(d) < \cdots < \check{\mu}_{\check{G}(d)}(d)$, and define  
\[
\mathcal{G}_j(d) 
:=
\bigl\{ g\in \{1,\cdots, \bar G\}:\ \check{\theta}_j(d) = \theta_g(d) \bigr\} 
=
\bigl\{ g\in \{1,\cdots, \bar G\}:\ \check{\mu}_j(d) = \mu_g(d) \bigr\},  
\] 
where the second equality is by condition (\ref{cond4}) in Assumption \ref{ass:exten}. We define  
\[
\pi_j(d,d') := \Pr\{ G \in \mathcal{G}_j(d)\mid D = d'\}
\]
for $(d,d')\in \{0,1\}^2$ and $j = 1,\cdots, \check{G}(d)$, and we obtain the following lemma.  Let $Y = DY(1) + (1-D) Y(0)$ be the observed outcome as before. 

\begin{lemma}		\label{lem:gen}
  Suppose that Assumption \ref{ass:exten} holds. Then, for all $(y,d) \in \mathbb{R}\times\{0.1\}$,   
\begin{equation*}
    \Pr(Y\leq y\mid D=d) =  \sum_{j=1}^{\check{G}(d)}  \pi_j ( d , d)  F\bigl( y; \check\theta_{j} (d) \bigr),
\end{equation*}
from which the mixture parameters $\check{G}(d)$ and $\bigl\{ (  \pi_j ( d , d)  , \check\theta_{j} (d) ) : j = 1 , \cdots , \check{G}(d)  \bigr\}$ are identified. Therefore, all the results using Assumption \ref{ass:setup} such as Lemmas \ref{lem:partsgen}, \ref{lem:Gbar}, and \ref{lem:characT}, and Theorems \ref{thm:simple} and \ref{thm:main0} continue to hold when Assumption \ref{ass:setup} is replaced with Assumption \ref{ass:exten}. 
\end{lemma}

\subsubsection{Estimation}	\label{sec:estiinfexten}
As in Section \ref{sec:esti}, we suggest first estimating $\check G(d)$ by any procedure that yields a consistent estimator in the sense of equation (\ref{eq:conG}). For instance, one may apply \cite{chen09order}'s estimator, or \citet{makha21}'s penalized-likelihood method when the mixing distribution is multidimensional. 

Given the estimates of the mixture orders, the remaining mixture parameters $(\pi_j(d,d), \check{\theta}_j(d))$ can then be estimated by MLE, replacing the true mixture orders with their estimated counterparts, as in Section \ref{sec:esti}. We refer to the Supplement for a detailed discussion on estimation and inference for general mixture models.

\section{An Empirical Illustration}\label{sec:empirical example}

We now illustrate our methodology by using the data from \cite{lalonde1986evaluating} and \cite{imbens2003sensitivity}. We take two approaches: one is to ignore zero earnings to use the Gaussian mixture method of Section \ref{sec:identification}, and the other is to explicitly address the censoring issue by using the model we presented in Section \ref{sec:sampleselect}. For this exercise, we focus on  the experimental treatment and control groups of the data, both of which were drawn from the same population with relatively poor labor market prospects such as ex-drug addicts, ex-criminal offenders, and high-school dropouts. Using the PSID control as a comparison group does not seem to be reasonable because we do not believe that the condition in \eqref{eq:overlap} holds even after controlling for covariates such as age, age-squared, and years of education.  

The treatment $D$ indicates whether the agent attended a training program or not. The (pre-homogenized) outcome $\tilde Y(d)$ is the logarithm of the earnings in 1978, and the convariates we use for homogenization consist of age, age-squared, and the years of education; these are the traditional variables we use in a Mincer equation. We control for the same covariates for the selection equation based on the Probit. 

The (experimental) control group contains a total of 260 observations, which is reduced to 168 after removing observations with zero earnings. The treatment group has a total of 185 observations, and the number becomes 140 after removing observations with zero earnings. Therefore, we have a total of 308 observations to estimate the treatment effect of the job training on the log-earnings, where we control for age, age-squared, and years of education.

Given the limited sample size, we impose the restrictions that $\beta(0) = \beta(1)$ and $\sigma_{\epsilon\nu}(0) = \sigma_{\epsilon\nu}(1)$: i.e., the coefficients of the interactions of the treatment with the covariates and the inverse Mills ratio are all equal to zero. As a benchmark, we first estimate the treatment effect by a simple linear model with and without selection. Estimation results by simple ordinary least squares (OLS) and by two-step Heckit are presented in the following table.

\begin{table}[ht]
\vspace{0.5em}
  \begin{tabular}{  c | c c ccc  | c }
  Selection &   intercept    &   training     &   age &   age-squared   &   education &  $\sigma_{\epsilon\nu}$ \\ 
  \hline
  no     &7.911 & 0.067 & 0.011 & 0.000 & 0.036  &  -  \\
  yes    & 7.808 & 0.068 & -0.039 & 0.001 & 0.047 & 1.364    \\
  \hline
  \end{tabular}  
  \caption{Benchmark Results by OLS and Two-Step Heckit \label{tb:ols}}
  \vspace{0.5em}
\end{table}

All the coefficients (except for the intercept) are individually insignificant at a 10\% significance level; the standard errors are not reported then. However, the point estimates still appear to be reasonable for a Mincer equation. Therefore, we rely on them for the purpose of homogenization. According to these estimates, receiving the job training increases earnings by about 7\%.

We now estimate the mixture models for both the treatment and control groups following the three-step procedure from the previous section. For this purpose, we first apply \cite{chen09order}'s method with and without selection to obtain $\hat G(0) = \hat G(1) = 2$.  We use the minimum suggested value for the tuning parameter (i.e., the smallest penalization for increasing the value of $\hat{G}(d)$), which is suggested in \cite{chen09order}'s Monte Carlo section. Despite that, our estimates $\hat G(d)$ are small. Before we proceed, we present the estimates of the identified parameters from the mixture models. 

\begin{table}[ht]
\vspace{0.5em}
\begin{tabular}{  c | cc | cc | cc | cc }
   & \multicolumn{4}{c|}{Control ($d=0$)} &  \multicolumn{4}{c}{Treatment ($d=1$)} \\
 \hline
 Selection & \multicolumn{2}{c|}{no} & 
\multicolumn{2}{c|}{yes} & 
\multicolumn{2}{c|}{no} & 
\multicolumn{2}{c}{yes} \\
\hline
$ \check{\mu}_{1}\ \pi_{1}(d,d) $  & 5.27 & 0.06 & 5.22 & 0.07 & 6.45 & 0.22 & 6.63 & 0.27 \\
$ \check{\mu}_{2}\ \pi_{2} (d,d) $ & 8.09 & 0.94 & 8.11 & 0.93 & 8.41 & 0.78 & 8.52 & 0.73 \\
\hline
$\sigma  (d) $ & \multicolumn{2}{c|}{0.72} & 
\multicolumn{2}{c|}{1.29} & 
\multicolumn{2}{c|}{0.62} & 
\multicolumn{2}{c}{1.21} 
\\
\hline
\end{tabular}
\caption{Estimates of the Mixture Paramters: $\hat G(0) = \hat G(1) = 2$. \label{tb:mixture}}
\vspace{0.5em}
\end{table}

The estimates reported in Table \ref{tb:mixture} can be used to show how the sharp bounds for the training effect will change as the number of unobserved types changes. Since $\hat G(0) = \hat G(1) = 2$, we have $\bar G\in \{2,3,4\}$. However, it suffices to check $\bar G \in \{2,3\}$, because the Manski bounds will be the best we can obtain whenever $\bar G\geq \check{G}(0) + \check{G}(1) - 1 = 3$. The sharp identified sets for the training effect, under $\bar G = 2$ and $\bar{G} \in \{3,4\}$, are presented in the following table.

\begin{table}[ht]
\vspace{0.5em}
  \begin{tabular}{  l | c  | c  }
  \hline 
         &  No Selection     &   Selection     \\
  \hline  
  $\bar G \in \{ 2\}$         &  $\{ 0.219,\  0.439 \}$   &   $ \{ 0.280,\ 0.564 \} $  \\
  $\bar G\in \{3, 4\}$ &  $ ( -0.846,\  1.500) $   &   $(-0.750,\   1.594)$ \\
  \hline 
  \end{tabular}  
  \caption{Sharp Identified Sets for the Job Training Effect \label{tb:ours}}

\vspace{0.5em}

\begin{minipage}{0.8\linewidth}
\footnotesize
\textit{Note:} If rank invariance is imposed, for $\bar{G}=2$, the point estimates of the training effect become $0.439$ under the no-selection model and $0.564$ under selection. For $\bar{G}=3$, the corresponding set estimates are $(-0.846, 1.500) \backslash \{ 0.439 \}$ and $(-0.750,  1.594)  \backslash \{ 0.564\}$.
\end{minipage}

\vspace{0.5em}

\end{table}

Tables \ref{tb:ols} and \ref{tb:ours} show a complete picture of the effect of the job training and its robustness. The coefficients of ``training'' are the estimates of the effect of the job training with and without addressing selection when we assume that there is no unobserved heterogeneity, which can be interpreted as $\bar G = 1$. In contrast, fitting the mixture models suggests that $\bar G$ may be as large as $4$. If $\bar G \in \{3,4\}$, then we already hit the Manski bounds, in which case we do not learn much about the training effect. If $\bar G = 2$, then the regression estimates in Table \ref{tb:ols} may be substantially underestimating the training effect. For instance, under rank invariance, the estimated effect is $0.439$ in the no-selection model and $0.564$ when selection is accounted for.

\section{Concluding Remarks}	\label{sec:conclude}

We have shown how to learn about the average treatment effect without assuming unconfoundedness by assuming that unobserved confounders are discrete with the number of mass points unknown. As concluding remarks, we make two comments. 

First, it is important for our methodology that both the treatment and control groups are influenced by every latent type in the population. If the treatment and control groups are sampled from the same population, then this requirement is generally plausible. However, it may become problematic if the researcher combines data from different sources. For example, in \citet{lalonde1986evaluating}, the experimental treatment group and the PSID control group are drawn from markedly different populations, making it difficult to justify the assumption that they share the same set of latent types.

Second, we have focused on the average treatment effect. Extending our methodology to other causal parameters, such as quantile treatment effects, appears to be a nontrivial question. This difficulty is partly due to the absence of a law of iterated quantiles analogous to the law of iterated expectations. More recently, \citet{masten2025relaxing} extended the c-dependence framework of \citet{masten2018identification} to sensitivity analysis for a broader class of causal parameters, including quantile treatment effects. Nevertheless, our approach remains unique in that it formulates sensitivity analysis directly through restrictions on latent confounders. Consequently, any substantive prior knowledge or assumptions about the confounders can be incorporated in a transparent and systematic manner.

\clearpage

\appendix

\renewcommand{\thesubsection}{A.\arabic{subsection}}\setcounter{subsection}{0}
\renewcommand{\theequation}{A.\arabic{equation}} \setcounter{equation}{0}

\renewcommand{\thealemma}{A.\arabic{alemma}} \setcounter{alemma}{0}

\section*{Appendix. Proofs}	% \label{sec:app}  		% of the Results in the Main Text

In this section, we provide the proofs of lemmas and theorems stated in the main text. For this purpose, in Subsection \ref{sec:alemmas}, we first state a series of auxiliary lemmas and then, in Subsection \ref{sec:mainproofs}, we present the proofs of the results in the main text. Proofs of the auxiliary lemmas are provided in the Supplement.

\subsection{Auxiliary Lemmas}		\label{sec:alemmas}
The subsection provides a series of auxiliary lemmas. We emphasize that these results do not rely on those presented in the main text. Before proceeding, for a given $\boldsymbol{\iota} \in \mathscr{I}$, define
\begin{equation} \label{eq:indexes}
\begin{aligned}
\ushort{j}_k (\bm{\iota}) 
&:=  
\min\ \bigl\{ j  \in \{ 1, \cdots , \check{G} (0)  \} :  \iota_{jk} = 1  \bigr\}, \\
\bar{j}_k (\bm{\iota}) 
&:=    
\max\ \bigl\{ j  \in \{ 1, \cdots , \check{G} (0)  \} :  \iota_{jk} = 1  \bigr\}, \\
\ushort{k}_j (\bm{\iota}) 
&:=   
\min\ \bigl\{ k  \in \{ 1, \cdots , \check{G} (1)  \} :  \iota_{jk} = 1  \bigr\}, \\
\bar{k}_j (\bm{\iota}) 
&:=  
\max\ \bigl\{ k  \in \{ 1, \cdots , \check{G} (1)  \} :  \iota_{jk} = 1  \bigr\}. 
\end{aligned}
\end{equation}
Observe that, for a given $\bm{\iota}$, $\ushort{j}_k (\bm{\iota}) $ and $ \bar{j}_k (\bm{\iota}) $ identify the first and last positions of the 1’s in column $k$, respectively, while $\ushort{k}_j (\bm{\iota}) $ and $\bar{k}_j (\bm{\iota})$ play the analogous role across row $j$.

Now we are ready to state the first two auxiliary lemmas, which describe $\mathcal{T}_{\bm{\iota}} $ and its boundaries.

\begin{alemma}
\label{alem:shapeT}
Suppose that Assumption \ref{ass:setup} holds and let $\bm{\iota} \in \mathscr{I}$.

\begin{enumerate}

\item $ \mathcal{T}_{\bm{\iota}}$ is an interval whose infimum and supremum are given by
\begin{equation*}
\inf \mathcal{T}_{\bm{\iota}} = \lowmathcal{t} + 
\Prob ( D = 0) \sum_{j=1}^{\check{G}(0)} \pi_j ( 0 , 0)  \check{\mu}_{\ushort{k}_j(\bm{\iota})} (1)
 -   \Prob ( D = 1) \sum_{k=1}^{\check{G}(1)} \pi_k ( 1 , 1)  \check{\mu}_{\bar{j}_k (\bm{\iota})} (0) 
\end{equation*}
and
\begin{equation*}
 \sup \mathcal{T}_{\bm{\iota}}  =  \lowmathcal{t} + 
 \Prob ( D = 0)  \sum_{j=1}^{\check{G}(0)} \pi_j ( 0 , 0)  \check{\mu}_{\bar{k}_j(\bm{\iota})} (1)
 -   \Prob ( D = 1)  \sum_{k=1}^{\check{G}(1)} \pi_k ( 1 , 1)  \check{\mu}_{\ushort{j}_k (\bm{\iota})} (0)    ,
\end{equation*}
respectively, where $ \lowmathcal{t} : = \Prob ( D = 1)  \sum_{k=1}^{\check{G}(1)} \pi_{k} (1 , 1 )  \check{\mu}_{k} (1)   
    -   \Prob ( D = 0)\sum_{j =1}^{\check{G}(0)} \pi_{j} (0 , 0 )   \check{\mu}_{j} (0)$.

\item $ \mathcal{T}_{\bm{\iota}} $ is a singleton if and only if $\sum_{j,k} \iota_{j,k} = \check{G} (0) =  \check{G} (1)$; otherwise, $ \mathcal{T}_{\bm{\iota}}$ is open.

\item The infimum (supremum) of $\mathcal{T}_{\bm{\iota}} $ coincides with that of $ \mathcal{T}$ if and only if all elements in the first (last) column and all elements in the last (first) row of $\bm{\iota}$ are equal to 1. That is,

\begin{enumerate} 

\item $\inf \mathcal{T}_{\bm{\iota}} = \inf  \mathcal{T}$ if and only if $\iota_{j,1} =  \iota_{\check{G}(0) , k} = 1$ for all $(j,k)$.

\item $\sup \mathcal{T}_{\bm{\iota}} = \sup  \mathcal{T}$ if and only if $\iota_{1,k} =  \iota_{j , \check{G}(1) } = 1$ for all $(j,k)$.

\end{enumerate}

\end{enumerate}

\end{alemma}

\begin{alemma}
\label{alem:infsup}
Suppose that Assumption \ref{ass:setup} holds and $\min \{ \check{G} (0 )  , \check{G} (1 )  \}  \geq 2$.

\begin{enumerate}

\item Let $\ushort{\bm{\iota}}^{(m)} \in \mathscr{I}$, $m=0,\cdots,\check{G} (0 ) -1 $, be matrices with entries defined as follows:
\begin{equation*}
\ushort{\iota}_{jk}^{(m)} = \left\{
\begin{array}{ll}
1   &  \text{if} \ \left\{
\begin{array}{l}
k = 1 \ \text{and} \ j \leq \check{G} (0 )-m   , \ or  \\
1 <  k   \leq \check{G} (1) -1    \  \text{and} \  j = \check{G} (0 )      ,  \ or \\
k  =  \check{G} (1)   \  \text{and} \  j  \geq  \check{G} (0 )-m   ;
\end{array}
\right. \\
0 & \text{otherwise} .  
\end{array}
\right.
\end{equation*}
Then, $\inf \mathcal{T}_{\ushort{\bm{\iota}}^{(0)} } = \inf  \mathcal{T}$ and $\inf \mathcal{T}_{\ushort{\bm{\iota}}^{(m)} }  \leq \sup \mathcal{T}_{\ushort{\bm{\iota}}^{(m-1)} } \leq   \sup \mathcal{T}_{\ushort{\bm{\iota}}^{(m)} }$ for $m = 1, \cdots , \check{G} (0 ) -1$.

\item  Let $\bar{\bm{\iota}}^{(m)} \in \mathscr{I}$, $m=0,\cdots,\check{G} (1 ) -1 $, be matrices with entries defined as follows:
\begin{equation*}
\bar{\iota}_{jk}^{(m)} = \left\{
\begin{array}{ll}
1   &  \text{if} \ \left\{
\begin{array}{l}
j = 1 \ \text{and} \ k \leq \check{G} (1 )-m   , \ or  \\
1 <  j \leq \check{G} (0) -1    \  \text{and} \  k = \check{G} (1 )      ,  \ or \\
j  =  \check{G} (0)   \  \text{and} \  k  \geq  \check{G} (1 )-m   ;
\end{array}
\right. \\
0 & \text{otherwise}.  
\end{array}
\right.
\end{equation*}
Then, $\sup \mathcal{T}_{\bar{\bm{\iota}}^{(0)} } = \sup  \mathcal{T}$ and $\inf \mathcal{T}_{\bar{\bm{\iota}}^{(m)} }   \leq \inf \mathcal{T}_{\bar{\bm{\iota}}^{(m-1)} }  \leq \sup \mathcal{T}_{\bar{\bm{\iota}}^{(m)} }  $ for $m = 1, \cdots , \check{G} (1 ) -1$.

\item Moreover, $\inf \mathcal{T}_{\bar{\bm{\iota}}^{(\check{G} (1 ) -1)}} \leq \sup \mathcal{T}_{\ushort{\bm{\iota}}^{(\check{G} (0 ) -1)}}$.

\item $\mathcal{T}$ can be covered by the union of the closures of the sets $\mathcal{T}_{\ushort{\bm{\iota}}^{(m^\prime)}}$ for $m^\prime = 0, \cdots , \check{G}(0)-1$, together with the closures of $\mathcal{T}_{\bar{\bm{\iota}}^{(m^{\prime\prime})}}$ for $m^{\prime\prime} = 0, \cdots , \check{G}(1)-1$; specifically,
\begin{equation*}
\mathcal{T} \subseteq \left( \bigcup_{m^{\prime}=0}^{\check{G} (0)-1}  \mathcal{T}_{\ushort{\bm{\iota}}^{(m^{\prime})} }  \cup \left\{ \inf \mathcal{T}_{\ushort{\bm{\iota}}^{(m^{\prime})} } ,   \sup  \mathcal{T}_{\ushort{\bm{\iota}}^{(m^{\prime})} } \right\} \right)
 \cup 
  \left( \bigcup_{m^{\prime\prime}=0}^{\check{G} (1)-1}  \mathcal{T}_{\bar{\bm{\iota}}^{(m^{\prime\prime})} }  \cup \left\{ \inf \mathcal{T}_{\bar{\bm{\iota}}^{(m^{\prime\prime})} } ,   \sup  \mathcal{T}_{\bar{\bm{\iota}}^{(m^{\prime\prime})} } \right\}  \right)   .
\end{equation*}

\end{enumerate}

\end{alemma}

\subsection{Proofs of the Results in the Main Text}		\label{sec:mainproofs}

This section contains the proofs of the lemmas and theorems stated in the main text.

\subsubsection*{Proof of Lemma \ref{lem:partsgen}}
Observe that a pair of ordered partitions $ \lowmathcal{p}  :=  ( \lowmathcal{p} (0),  \lowmathcal{p} (1) ) $ is admissible under Assumption \ref{ass:setup} if and only if it satisfies the following conditions:
\begin{itemize}

\item[(ad1)] $\lowmathcal{p} (0)$ and $\lowmathcal{p} (1)$ are ordered partitions of the same subset of integers $\{1, 2,\cdots , m  \}$ with $\max\{ \check{G}(0) , \check{G}(1) \} \leq m \leq \check{G}(0) \, \check{G}(1)$.

\item[(ad2)] $\lowmathcal{p} (0)$ is of the form $\lowmathcal{p} (0) =  ( \lowmathcal{g}_{1} ( 0 ) , \cdots ,   \lowmathcal{g}_{ \check{G}(0) } ( 0 ) )$ with $\max \lowmathcal{g}_{j} ( 0 ) < \min  \lowmathcal{g}_{j^\prime} ( 0 )$ for any $j < j^\prime$.

\item[(ad3)] $\lowmathcal{p} (1)$ is of the form $\lowmathcal{p} (1) =  ( \lowmathcal{g}_{1} ( 1 ) , \cdots ,   \lowmathcal{g}_{ \check{G}(1) } ( 1 ) )$ and satisfies: for any $g < g^\prime $ satisfying $g , g^\prime \in  \lowmathcal{g}_j (0)$ for some $j = 1 ,\cdots , \check{G} (0)$, it follows that $g \in \lowmathcal{g}_{k} ( 1 )$ and $g^\prime \in \lowmathcal{g}_{k^\prime} ( 1 )$ with $k < k^\prime$.

\end{itemize}
Condition (ad1) follows from the fact that the control and treatment groups share the same types, while conditions (ad2) and (ad3) arise from the strict lexicographic ordering of the mean vectors $( \mu_g(0), \mu_g(1) )$, $g = 1 , \cdots , \bar{G}$.

Now let $\mathscr{P}$ denote the collection of pairs of ordered partitions  $ \lowmathcal{p}  = ( \lowmathcal{p} (0),  \lowmathcal{p} (1) )$ that satisfy conditions (ad1)-(ad3), and consider the function $\psi : \mathscr{P} \rightarrow \mathscr{I}$ defined such that the $(j , k)$-entry of $\psi (  \lowmathcal{p}  )$ is given by  
\begin{equation*}
\psi   \left(   \lowmathcal{p} \right)_{jk}  = 
 \left\{
\begin{array}{ll}
0 & \text{if} \   \lowmathcal{g}_{j} ( 0 )  \cap \lowmathcal{g}_{k} ( 1 )  =  \emptyset , \\
1 & \text{if} \   \lowmathcal{g}_{j} ( 0 )  \cap \lowmathcal{g}_{k} ( 1 )  \ \text{is a singleton} .
\end{array}
\right.
\end{equation*}
Observe that $\psi$ is well-defined due to Conditions (ad1) and (ad3). In particular, (ad3) implies that $\lowmathcal{g}_{j} ( 0 )  \cap \lowmathcal{g}_{k} ( 1 )$ is either empty or a singleton, while (ad1) guarantees that every row and every column of $\psi   (   \lowmathcal{p}  )$ contains at least one nonzero entry; hence, $\psi   (    \lowmathcal{p}  ) \in \mathscr{I}$. Moreover, note that $ \sum_{j,k}  \psi   (    \lowmathcal{p}  )_{jk}$ coincides with the integer $m$ specified in (ad1).

To complete the proof, it suffices to show that $\psi$ is one-to-one and onto. Starting with the former, pick any $  \lowmathcal{p}  ,  \lowmathcal{p}^\prime  \in \mathscr{P}$ such that $\psi (  \lowmathcal{p} ) =  \psi ( \lowmathcal{p}^\prime ) $ and observe that
\begin{eqnarray*}
\psi (  \lowmathcal{p} ) =  \psi ( \lowmathcal{p}^\prime ) 
&\Longrightarrow  &\forall \ (j ,k)  \in \{ 1 , \cdots ,  \check{G}(0) \} \times  \{ 1 , \dots ,  \check{G}(1) \}  , \      \lowmathcal{g}_{j} ( 0 )  \cap \lowmathcal{g}_{k} ( 1 )    =    \lowmathcal{g}_{j}^\prime ( 0 )  \cap \lowmathcal{g}_{k}^\prime ( 1 )  \\
&\Longrightarrow  &\forall \ j   \in \{ 1 , \cdots ,  \check{G}(0) \}  ,  \  \cup_{k}  \lowmathcal{g}_{j} ( 0 )  \cap \lowmathcal{g}_{k} ( 1 )  = \cup_{k}  \lowmathcal{g}_{j}^\prime ( 0 )  \cap \lowmathcal{g}_{k}^\prime ( 1 ) \\
&\underset{ \text{(ad1)} }{\Longrightarrow}  &\forall \ j   \in \{ 1 , \cdots ,  \check{G}(0) \}  ,  \   \lowmathcal{g}_{j} ( 0 ) =   \lowmathcal{g}_{j}^\prime ( 0 ) ,
\end{eqnarray*}
where $ \lowmathcal{p}^\prime = (   \lowmathcal{p}^\prime (0) ,  \lowmathcal{p}^\prime(1)  )$ and $ \lowmathcal{p}^\prime ( d ) = (  \lowmathcal{g}_{1}^\prime ( d ) , \cdots ,   \lowmathcal{g}_{ \check{G}(d) }^\prime ( d )  ) $ for $d=0,1$. Proceeding in a similar manner, we can also show that $   \lowmathcal{g}_{k} ( 1 ) =   \lowmathcal{g}_{k}^\prime ( 1 )  $ for every $k = 1, \cdots , \check{G}(1)$, from which we conclude that $ \lowmathcal{p}  = \lowmathcal{p}^\prime $ and therefore that $\psi$ is one-to-one.

To prove that $\psi$ is onto, pick any $\bm\iota \in \mathscr{I}$ and consider the pair of ordered partitions $\lowmathcal{q}  : = (\lowmathcal{q} (0) , \lowmathcal{q} (1))$ defined as follows:
\begin{itemize}

\item Construction of $\lowmathcal{q} (0) : = ( \lowmathcal{h}_1 (0) , \cdots ,  \lowmathcal{h}_{ \check{G}(0) } (0)  )$. For each $j = 1,\dots,\check{G}(0)$, define
\begin{equation*}
n_j := \sum_{k=1}^{\check{G}(1)} \mathbf{1}(\iota_{j,k}=1),
 \  \ 
N_0 := 0,
 \  \ 
N_j := \sum_{l=1}^j n_l , 
\end{equation*}
and set $\lowmathcal{h}_{j}( 0 ) := \{ N_{j-1}+1,\dots,N_j \}$, noting that $\# \lowmathcal{h}_{j}( 0 )$ coincides with $n_j$, i.e., the number of ones in the $j$th row of $\bm\iota$.

\item Construction of $\lowmathcal{q} (1) : = (  \lowmathcal{h}_1 (1) , \cdots ,  \lowmathcal{h}_{ \check{G}(1) } (1)  )$. For each $j = 1,\dots,\check{G}(0)$, write the elements of $ \lowmathcal{h}_{j}(  0 )$ in increasing order as $ \{ h_{j,1} < \cdots < h_{j,n_j} \} : = \lowmathcal{h}_{j}(  0 )$.
Then, for each $k = 1,\dots,\check{G}(1)$, define
\begin{equation*}
 \lowmathcal{h}_{k}( 1 )
:=
\cup_j
\left\{ h_{j,\, r_{j,k}} \right\}  ,
\end{equation*}
where the (disjoint) union $\cup_j$ is taken over $j = 1,\cdots,\check{G}(0)$ satisfying $\iota_{j,k}=1$,
while $r_{j,k} := \sum_{s=1}^k \mathbf{1}(\iota_{j,s}=1) $ counts the number of ones in row $j$ up to column $k$, for which we remark that $r_{j,k} \geq 1$ when $\iota_{j,k}=1$.

\end{itemize}

Now, we need to show that $\lowmathcal{q}  \in \mathscr{P} $ and $\psi (  \lowmathcal{q} ) = \bm\iota$.  Starting with the former, note that condition (ad1) is satisfied because each $ \lowmathcal{q} ( d )$, $d = 0 , 1$, is an ordered partition of $\{ 1 , 2 , \cdots ,  \sum_{j,k} \iota_{j,k} \}$. Moreover, (ad2) holds because $\lowmathcal{q} (0) $ is defined by partitioning $\{ 1 , 2 , \cdots ,  \sum_{j,k} \iota_{j,k} \}$ into consecutive blocks. To see that (ad3) is also satisfied, pick any distinct $h_{j , m_1} , h_{j , m_2} \in \lowmathcal{h}_{j}(  0 )$ with $1 \leq m_1 < m_2 \leq n_j$, which implies $h_{j , m_1} < h_{j , m_2}$, and note that there exists a unique couple of integers $k_1 < k_2 $ such that $r_{j,k_1}  = m_1$ and $r_{j,k_2}  = m_2$. As a result, $h_{j , m_1}\in \lowmathcal{h}_{k_1}(  1 )$ and $ h_{j , m_2} \in \lowmathcal{h}_{k_2}(  1 )$ with $k_1 < k_2$. Finally, to show that $\psi (  \lowmathcal{q} )  = \bm\iota$, pick any $(j,k)$ and note that $\lowmathcal{h}_{j}( 0 ) \cap \lowmathcal{h}_{k}( 1 ) =  \{  h_{j,\, r_{j,k}} \} $ if and only if $\iota_{j,k}=1$ by construction of $\lowmathcal{h}_{k}( 1 )$, and therefore $\psi (  \lowmathcal{q} )_{j,k} = \iota_{j,k}$.
\qed

\subsubsection*{Proof of Lemma \ref{lem:Gbar}}
Just note that the number of types associated with the ordered partition represented by $\bm{\iota} = ( \iota_{j,k} ) \in \mathscr{I}$ is given by the sum of its entries, $\sum_j \sum_k \iota_{j,k}$. See also the proof of Lemma \ref{lem:partsgen} for further details.
\qed

\subsubsection*{Proof of Theorem \ref{thm:simple}}
On the one hand, to prove that $\tau \in \mathcal{T}$, choose $\mathbf{q} = (  \pi_1 ( 1 , 0) , \cdots , \pi_{\check{G}(1)} ( 1 , 0) ) \in \Pi_{\check{G}(1)}$ and $\mathbf{p} = (  \pi_1 ( 0 , 1) , \cdots , \pi_{\check{G}(0)} ( 0, 1) ) \in \Pi_{\check{G}(0)}$, and then recall Eq.\ (\ref{eq:ategr}). On the other hand, to prove that $\mathcal{T}$ is sharp, pick any $t \in \mathcal{T}$ and let  $( \mathbf{p}_t ,\mathbf{q}_t ) \in \Pi_{\check{G}(0)}  \times \Pi_{\check{G}(1)} $ be the vectors associated with $t$, i.e.,
\begin{equation*}
t   =   \sum_{k =1}^{\check{G}(1)} 
       \biggl\{  \Prob ( D = 0) q_{t,k}  + \Prob ( D = 1) \pi_{k} (1 , 1 )   \biggr\} \check{\mu}_{k} (1)  -     \sum_{j =1}^{\check{G}(0)}  \biggl\{  \Prob ( D = 0) \pi_{j} (0 , 0 )   +   \Prob ( D = 1)  p_{t,j}       \biggr\}  \check{\mu}_{j} (0) .
\end{equation*}
Then, note that $t$ is indeed the ATE of a model having $\bar G = \check{G}(0) \times \check{G}(1)$ innate types and conditional probabilities given by 
\begin{equation*}
\Prob \{  G \in  \mathcal{G}_{j} (0) \cap \mathcal{G}_{k} (1) \mid  D  =0 \}  =    \pi_{j} (0 , 0 ) q_{t,k}  \  \text{and}  \  \Prob \{  G \in  \mathcal{G}_{j} (0) \cap \mathcal{G}_{k} (1) \mid  D  = 1 \}  = \pi_{k} (1 , 1 ) p_{t,j} ,
\end{equation*}
for $j = 1 , \cdots , \check{G}(0)$ and $k = 1 , \cdots , \check{G}(1)$. Finally, the desired characterization follows immediately from the fact that
$\check{\mu}_1 (1)  = \min \{ \check{\mu}_1 (1) , \cdots,  \check{\mu}_{\check G (1)} (1) \}  $,
$\check{\mu}_{\check G (1)} (1)  =  \max \{ \check{\mu}_1 (1) , \cdots,  \check{\mu}_{\check G (1)} (1) \}$,
$ \check{\mu}_1 (0)  = \min \{ \check{\mu}_1 (0) , \cdots,  \check{\mu}_{\check G (0)} (0) \}$, 
and 
$\check{\mu}_{\check G (0)} (0) = \max \{ \check{\mu}_1 (0) , \cdots,  \check{\mu}_{\check G (0)} (0) \} $.
\qed

\subsubsection*{Proof of Lemma \ref{lem:characT}}
To show that $\tau \in \mathcal{T}_{\bm{\iota}_{\mathcal{P}}}$, define the matrices $\mathbf{p}_0$ and $\mathbf{q}_0$ such that their $(j,k)$-entries are given by
\begin{equation*}
{p}_{0 ,jk} =   \Prob \{ G \in\mathcal{G}_{j}(0) \cap\mathcal{G}_{k} (1)  \mid    D = 0\}  \  \text{and}  \  {q}_{0 ,jk} =   \Prob \{ G \in\mathcal{G}_{j}(0) \cap\mathcal{G}_{k} (1)  \mid    D = 1\}  ,
\end{equation*}
respectively. The desired result is then obtained by noting that $(\mathbf{p}_0 , \mathbf{q}_0) \in  \mathcal{M}_{\bm{\iota}_{\mathcal{P}}} (0) \times \mathcal{M}_{\bm{\iota}_{\mathcal{P}}} (1) $ and due to the fact that
\begin{multline*}
\tau  
= 
\sum_{k =1}^{\check{G}(1)} 
       \biggl\{  \Prob ( D = 0) \sum_{j=1}^{\check{G}(0)} p_{0,jk}  + \Prob ( D = 1) \pi_{k} (1 , 1 )   \biggr\} \check{\mu}_{k} (1) \\ 
-   
\sum_{j =1}^{\check{G}(0)}  \biggl\{  \Prob ( D = 0) \pi_{j} (0 , 0 )   +   \Prob ( D = 1)  \sum_{k=1}^{\check{G}(1)} q_{0,jk}       \biggr\}  \check{\mu}_{j} (0) ,
\end{multline*}
which follows by Eq.\ (\ref{eq:ategr}) and the Law of total probability.

To prove that $\mathcal{T}_{\bm{\iota}_{\mathcal{P}}}$ is sharp, pick any $t \in \mathcal{T}_{\bm{\iota}_{\mathcal{P}}}$ and let $( \mathbf{p}_t ,\mathbf{q}_t ) \in \mathcal{M}_{\bm{\iota}_{\mathcal{P}}} (0) \times \mathcal{M}_{\bm{\iota}_{\mathcal{P}}} (1) $ be the matrices associated with $t$, i.e.,
\begin{multline*}
t 
=  
\sum_{k =1}^{\check{G}(1)} 
       \biggl\{  \Prob ( D = 0) \sum_{j=1}^{\check{G}(0)} p_{t,jk}  + \Prob ( D = 1) \pi_{k} (1 , 1 )   \biggr\} \check{\mu}_{k} (1) \\ 
-     
\sum_{j =1}^{\check{G}(0)}  \biggl\{  \Prob ( D = 0) \pi_{j} (0 , 0 )   +   \Prob ( D = 1)  \sum_{k=1}^{\check{G}(1)} q_{t,jk}       \biggr\}  \check{\mu}_{j} (0) .
\end{multline*}
Then, observe that the identified mixtures of Eqs.\ (\ref{eq:mix0})--(\ref {eq:mix1}) can be rationalized by the following underlying structure that satisfies Assumption \ref{ass:setup}: 
\begin{itemize}

\item $\bar G = \sum_j \sum_k \iota_{\mathcal{P}, jk}$;

\item $(\mathcal{P}(0) , \mathcal{P} (1) ) = \phi ( {\bm{\iota}_{\mathcal{P}}} ) $ with conditional probabilities given by
\begin{equation*}
\Prob \{ G \in\mathcal{G}_{j}(0) \cap\mathcal{G}_{k} (1)  \mid    D = 0\}  =   {p}_{t,jk} \  \text{and}  \    \Prob \{ G \in\mathcal{G}_{j}(0) \cap\mathcal{G}_{k} (1)  \mid    D = 1\} =  {q}_{t,jk} ,
\end{equation*}
where $\phi$ has been defined in the proof of Lemma \ref{lem:partsgen} and $\mathcal{P} (d )  = \{ \mathcal{G}_{1} (d) , \cdots ,  \mathcal{G}_{\check{G}(d)} (d) \}$;

\item For $g \in \{ 1 , \cdots , \bar G  \}$, $\bm\mu_g  = ( \check\mu_j (0)  ,  \check\mu_k (1) )$ with $(j,k)$ being such that $g \in {\mathcal{G}}_{j} ( 0 ) \cap {\mathcal{G}}_{k} (1)$.

\end{itemize}
Observe that $\pi_j ( 0 , 0 ) = \sum_k p_{t,jk}$ and $\pi_k (1 , 1) = \sum_j q_{t,jk}$ follow by construction of $\mathcal{M}_{\bm{\iota}_{\mathcal{P}}} (0) $ and $\mathcal{M}_{\bm{\iota}_{\mathcal{P}}} (1) $, respectively. Thus, the proof is complete after noting that $t$ is the ATE of this underlying structure, which follows from Eq.\ (\ref{eq:ategr}).
\qed

\subsubsection*{Proof of Theorem \ref{thm:main0}}

\paragraph{\emph{(1)}} This part follows by combining Lemmas \ref{lem:partsgen} and \ref{lem:characT}, as the former ensures that every admissible pair of partitions of $\{1, \dots, \bar{G}\}$ is uniquely represented in $\mathscr{I}$.

\paragraph{\emph{(2)}} As a staring point, note that $\min \{ \check{G} (0) ,  \check{G} (1)  \} = 1$ implies that $\bar{G}_L, \bar{G}_C$, and $\bar{G}_U$ are all equal to $\max \{ \check{G} (0) ,  \check{G} (1)  \} $. Consequently, we have that $\bar{G} = \max \{ \check{G} (0) ,  \check{G} (1)  \}$ by Lemma \ref{lem:Gbar} and $\mathcal{T} = \mathcal{T} \bigl( \max \{ \check{G} (0) ,  \check{G} (1)  \} \bigr)$ by part (1), from which the desired results follow immediately. Hence, we will assume $\min \{ \check{G} (0) ,  \check{G} (1)  \} \geq 2$ for the rest of this proof.

\paragraph{\emph{(2)(a)}} It follows from Lemma \ref{alem:shapeT}; specifically, by choosing any matrix $\bm{\iota} \in\mathscr{I} (m) $ that has ones in all entries of its border, i.e., $ \iota_{jk}  =   1$  whenever $j \in  \{ 1  , \check{G} (0) \}$ or $k \in \{ 1 , \check{G} (1) \}$.

\paragraph{\emph{(2)(b)}} The proof of this part is divided into two cases:
\begin{enumerate}

\item[(c1)] $m = \bar{G}_C$.

\item[(c2)]  $\bar{G}_C < m  < 2 \bar{G}_C - 2$.

\end{enumerate}
Starting with (c1), note first that $\ushort{\bm{\iota}}^{(m^{\prime})} \in \mathscr{I} (   \bar{G}_C  )$ and $\bar{\bm{\iota}}^{(m^{\prime\prime})} \in \mathscr{I} (   \bar{G}_C )$ for any $m^{\prime} = 0,1 , \cdots , \check{G} (0)-1$ and $m^{\prime\prime} = 0 , 1,  \cdots , \check{G} (1)-1$, where $\ushort{\bm{\iota}}^{(m^{\prime})}$ and $\bar{\bm{\iota}}^{(m^{\prime\prime})} $ are matrices defined in Lemma \ref{alem:infsup}. As a result,
\begin{equation*}
\left( \bigcup_{m^{\prime}=0}^{\check{G} (0)-1}  \mathcal{T}_{\ushort{\bm{\iota}}^{(m^{\prime})} }   \right)
 \cup 
  \left( \bigcup_{m^{\prime\prime}=0}^{\check{G} (1)-1}  \mathcal{T}_{\bar{\bm{\iota}}^{(m^{\prime\prime})} }   \right)  
   \    \subseteq \ \mathcal{T} \left(  \bar{G}_C  \right)      \    \subseteq \ \mathcal{T}  
\end{equation*}
and therefore the proof for the case (c1) is complete because combining together these inclusions with that of Lemma \ref{alem:infsup}.(4) yields
\begin{equation*} 
    \mathcal{T} \backslash  \mathcal{T} \left(  \bar{G}_C  \right)   \,  \subseteq  \,   \left( \bigcup_{m^{\prime}=0}^{\check{G} (0)-1}  \left\{ \inf \mathcal{T}_{\ushort{\bm{\iota}}^{(m^{\prime})} } ,   \sup  \mathcal{T}_{\ushort{\bm{\iota}}^{(m^{\prime})} } \right\} \right)
 \cup 
  \left( \bigcup_{m^{\prime\prime}=0}^{\check{G} (1)-1}   \left\{ \inf \mathcal{T}_{\bar{\bm{\iota}}^{(m^{\prime\prime})} } ,   \sup  \mathcal{T}_{\bar{\bm{\iota}}^{(m^{\prime\prime})} } \right\}  \right) .
\end{equation*}
The proof of (c2) follows by using the result of the case (c1) and from the fact that the remaining  entries of the matrices $( \text{$\mathbf{p}$}   , \text{$\mathbf{q}$}  ) \in \mathcal{M}_{\bm{\iota}} (0) \times \mathcal{M}_{\bm{\iota}} (1)$, where here $\bm{\iota} \in \mathscr{I} (m)$ with $m  >  \{   \check{G} (0) + \check{G} (1) - 1 \} $, can be arbitrarily close to zero.

\paragraph{\emph{(2)(c)}} For this part, we only prove $\inf \mathcal{T} (  m+1)   <   \inf \mathcal{T} (  m )$ as the other strict inequality, $\sup \mathcal{T} (  m )  <   \sup \mathcal{T} (  m+1 ) $, follows by symmetric arguments. To establish the desired result, letting $m \leq  \check{G}( 0) + \check{G}( 1) - 2$, it suffices to show that for any ${\bm{\iota}}^\prime \in \mathscr{I}(m)$ there is $\bm{\iota}^{\prime\prime} \in \mathscr{I}(m+1)$ such that $\inf \mathcal{T}_{\bm{\iota}^{\prime\prime}}  <  \inf \mathcal{T}_{\bm{\iota}^{\prime}}$ because $\inf \mathcal{T} (  m ) =  \min\{ \inf \mathcal{T}_{\bm{\iota}}  : \bm{\iota} \in \mathscr{I}(m) \}$. So, choose any $\bm{\iota}^\prime \in \mathscr{I}(m)$ and note that there must be a row $j^\prime$ such that $\iota_{j^\prime  1}^\prime= 0$; otherwise, since $m \leq  \check{G}( 0) + \check{G}( 1) - 2$, there would be a column $k^\prime >1$ such that $\iota_{k^\prime   j}^\prime = 0$ for all $j = 1 , \cdots , \check{G} (0)$, violating Condition (2) of Definition \ref{def:matpos0}. Then, $\bm{\iota}^{\prime\prime}$ can be constructed by setting  ${\iota}_{jk}^{\prime\prime} = {\iota}_{jk}^{\prime}$ for all $(j ,k) \neq (  j^\prime , 1 )$, whereas $\iota_{j^\prime , 1}^{\prime\prime} = 1$. Finally, $\inf \mathcal{T}_{\bm{\iota}^{\prime\prime}}  <  \inf \mathcal{T}_{\bm{\iota}^{\prime}}$ follows from Lemma \ref{alem:shapeT}.(1) because $1 =  \ushort{k}_{j^\prime} (\bm{\iota}^{\prime\prime})  <  \ushort{k}_{j^\prime} ( \bm{\iota}^\prime ) $ and $ \ushort{k}_{j} (\bm{\iota}^{\prime\prime})  =  \ushort{k}_{j} ( \bm{\iota}^\prime )$ for $j \neq j^\prime$, while $\bar{j}_1 ( \bm{\iota}^{\prime\prime} )  \geq \bar{j}_1 ( \bm{\iota}^\prime )$ and $\bar{j}_k ( \bm{\iota}^{\prime\prime} )  = \bar{j}_k ( \bm{\iota}^\prime )$ for $k >1$.
\qed

\subsubsection*{Proof of Lemma \ref{lem:sampleselect}}
Choose any $x$ in the support of $X$ and $d \in \{ 0,1 \}$. Recall that the $\eta_0$, $\eta_1$, $\beta( d )$, and $\sigma_{\epsilon\nu} (d)$ are identified due to Assumption \ref{ass:sampleselect}, and let $\mathscr{F}$ denote the family of density functions on $\mathbb{R}$ of the form
\begin{equation*}
f ( \cdot ;  s , m ) : = \frac{1}{  \Phi \left( \eta_0 +  x' \eta_1 \right) } \Phi \left( \frac{ \eta_0 +  x' \eta_1  + \frac{\sigma_{\epsilon \nu}(d) }{s^2}( \cdot  -  m )}{\sqrt{1  - \frac{\sigma_{\epsilon \nu}(d) }{s^2}} } \right)  \frac{1}{s} \phi \left(  \frac{ \cdot  - m}{s} \right)  
\end{equation*}
and indexed by $m \in \mathbb{R}$ and $s >  \max\{ \sigma_{\epsilon \nu}(d)  , 0 \}^{1/2}$. By the theorem of Section 3 and Proposition 6 in \cite{yak68id}, the desired result emerges once we show that the following implication holds for any arbitrary $L \in\mathbb{N}$, $(a_1 , \cdots , a_L) \in \mathbb{R}^L$, and distinct $(s_1 , m_1), \cdots (s_L , m_L) $:
\begin{equation*}
\forall \ y \in \mathbb{R} , \  \  \sum_{l=1}^L a_l \, f ( y ; m_l , s_l ) = 0  \quad \rightarrow \quad  a_1 = 0  , \cdots , a_L  = 0.
\end{equation*}
To prove this implication, consider first the case where $\sigma_{\epsilon \nu}(d) \geq 0 $ and, without loss of generality, assume that $(s_1 , m_1) \prec \cdots \prec (s_L , m_L) $, noting that
\begin{equation*}
\forall \  l < p ,    \  \lim_{y \rightarrow \infty} \frac{f ( y ;  s_l , m_l )} {f ( y ;  s_p , m_p )}   = 0 .
\end{equation*}
Then, observe that $a_L =0$ follows from the equality
\begin{equation*}
\underbrace{ \lim_{y \rightarrow \infty}  \sum_{l<L}  \frac{f ( y ;  s_l , m_l )} {f ( y ;  s_L , m_L )} }_{= 0}  \ + \ a_L = 0
\end{equation*}
and that $a_1 = \cdots = a_{L-1} = 0$ can be obtained by sequential elimination. To complete the proof, we must also consider the case $\sigma_{\epsilon \nu}(d) < 0 $. In this situation, the desired result follows by an analogous argument; specifically, by reversing the lexicographic order with respect to $m$ and by taking the limit as $y \rightarrow -\infty$ instead of $y \rightarrow \infty$.
\qed

\subsubsection*{Proof of Lemma \ref{lem:gen}}

Note that we can write
\begin{align*}
F_{Y| D } (y | d )  
&=  
\sum_{g=1}^{\bar{G}}  \Prob (  G = g  \mid   D = d  )  \Prob \{  Y_g (d)  \leq y \}       \\
&=    
\sum_{g=1}^{\bar{G}}  \Prob (  G = g  \mid   D = d  )  F  \left( y ; \theta_{g} (d) \right)  \\
&=    
\sum_{j=1}^{\check{G}(d)}  \pi_j ( d , d)  F  \left( y ; \check\theta_{j} (d) \right) .
\end{align*}
The first equality follows by combining together the specification in Eq.\ (\ref{eq:ygen}) with the Law of total probability and independence between $Y_g (d) $ and $(D, G)$. The second equality follows by the second condition of Assumption \ref{ass:exten}, while the third by the definitions of $\check\theta_{j} (d)$ and $\pi_j ( d , d) $. Finally, identification of the mixture parameters is an immediate consequence of Assumption \ref{ass:exten}.(2).
\qed

\vspace{.25in}

\clearpage

\section*{Supplement}

\renewcommand{\thepage}{S.\arabic{page}}\setcounter{page}{1}

\renewcommand{\thesection}{S.\arabic{section}}\setcounter{section}{0}

\renewcommand{\thesubsection}{S.\arabic{subsection}}\setcounter{subsection}{0}

\renewcommand{\thelemma}{S.\arabic{lemma}} \setcounter{lemma}{0}

\renewcommand{\thetheorem}{S.\arabic{theorem}} \setcounter{theorem}{0}

\renewcommand{\theequation}{S.\arabic{equation}} \setcounter{equation}{0}

\renewcommand{\thetable}{S.\arabic{table}}  \setcounter{table}{0}

\renewcommand{\thefigure}{S.\arabic{figure}}  \setcounter{figure}{0}

\renewcommand{\theassump}{S.\arabic{assump}} \setcounter{assump}{0}

This supplement is divided into two sections. The first one provides the proofs of the auxiliary lemmas stated in the Appendix, while the  second presents the asymptotic properties of the proposed estimators along with valid inference procedures.

\subsection{Proof of the Auxiliary Lemmas stated in the Appendix}

\subsubsection*{Proof of Lemma \ref*{alem:shapeT}}

Throughout this proof, for a given $\bm\iota \in \mathscr{I}$, consider the affine mapping $ \phi : \mathbb{R}^{\check{G}(0)\times \check{G}(1)} \times \mathbb{R}^{\check{G}(0)\times \check{G}(1)} \rightarrow \mathbb{R}$ defined as follows:
\begin{equation*}
 \phi ( \mathbf{p}   , \mathbf{q}  )  \  : =  \   \lowmathcal{t}  \ + \   \Prob ( D = 0)  \sum_{k =1}^{\check{G}(1)}   \check{\mu}_{k} (1) 
    \sum_{j=1}^{\check{G}(0)}     p_{j, k}         \   -  \  \Prob ( D = 1)   \sum_{j =1}^{\check{G}(0)}    \check{\mu}_{j} (0)   \sum_{k=1}^{\check{G}(1)} q_{j, k}      .
\end{equation*}
Note that $\mathcal{T}_{\bm\iota}$ is equal to the range of $ \phi$ over $\mathcal{M}_{\bm{\iota}} (0) \times \mathcal{M}_{\bm{\iota}} (1)$, i.e., $\mathcal{T}_{\bm\iota} =  \phi ( \mathcal{M}_{\bm{\iota}} (0) \times \mathcal{M}_{\bm{\iota}} (1)  )$.

\paragraph{\emph{(1)}} Observe that $\mathcal{T}_{\bm\iota}$ is an interval because $\mathcal{M}_{\bm{\iota}} (0) \times \mathcal{M}_{\bm{\iota}} (1)$ is convex, and so is the range of $ \phi$ over this set. To obtain the desired expression for $\inf \mathcal{T}_{\bm\iota}$, consider the linear constrained optimization problem
\begin{equation*}
   \inf_{ ( \mathbf{p}   , \mathbf{q}  )  \in \mathcal{M}_{\bm{\iota}} (0) \times \mathcal{M}_{\bm{\iota}} (1)} \  \phi ( \mathbf{p}   , \mathbf{q}  ) ,
\end{equation*}
and note that solving this problem is equivalent to solving
\begin{equation}
\label{eqa:opti}
   \inf_{\mathbf{p}  \in \mathcal{M}_{\bm{\iota}} (0)}   \sum_{j=1}^{\check{G}(0)}   \sum_{k =1}^{\check{G}(1)}    p_{j, k}    \check{\mu}_{k} (1) 
     \quad \text{and}   \quad
\sup_{ \mathbf{q}   \in \mathcal{M}_{\bm{\iota}} (1)} \sum_{k=1}^{\check{G}(1)} \sum_{j =1}^{\check{G}(0)}  q_{j, k}    \check{\mu}_{j} (0)    ,
\end{equation}
separately. The first problem can be solved by assigning to each $j$ the minimum value of $\check{\mu}_{k} (1) $ permitted by Condition (1)(a) in Definition \ref*{def:pmats}, while satisfying Condition (1)(b). Specifically, this value is $\check{\mu}_{\ushort{k}_j (\bm{\iota})} (1)$ and the corresponding condition is $ \sum_{k =1}^{\check{G}(1)}    p_{j, k} = \pi_j ( 0, 0)$. Proceeding in a similar manner with the second problem in (\ref{eqa:opti}), we obtain the desired expression for $\inf \mathcal{T}_{\bm\iota}$. The reasoning to derive the desired expression for $\sup \mathcal{T}_{\bm\iota}$ mirrors the arguments used earlier to obtain the expression for $\inf \mathcal{T}_{\bm\iota}$; hence, it is omitted.

\paragraph{\emph{(2)}}  On the one hand, note that $\sum_{j,k} \iota_{j,k} = \check{G} (0) =  \check{G} (1)$ implies that both $\mathcal{M}_{\bm{\iota}} (0)$ and $\mathcal{M}_{\bm{\iota}} (1)$ are singletons due to Conditions (1)(b) and (2)(b) in Definition \ref*{def:pmats}, respectively. Hence, the range of $ \phi$ over $\mathcal{M}_{\bm{\iota}} (0) \times \mathcal{M}_{\bm{\iota}} (1) )$, which is $\mathcal{T}_{\bm\iota}$, must also be a singleton.

On the other hand, when $\sum_{j,k} \iota_{j,k} > \check{G}(0)$ or $\sum_{j,k} \iota_{j,k} > \check{G}(1)$, or when $\check{G}(0) \neq \check{G}(1)$, the set $\mathcal{M}_{\bm{\iota}}(0) \times \mathcal{M}_{\bm{\iota}}(1)$ is clearly open. Consequently, $\mathcal{T}_{\bm{\iota}} = \phi\big( \mathcal{M}_{\bm{\iota}}(0) \times \mathcal{M}_{\bm{\iota}}(1) \big)$ is also open by the Open Mapping Theorem \cite[][Sections 2.10–2.12]{rud91fun}), for which we emphasize that $\phi$ is onto.

\paragraph{\emph{(3)}} It follows immediately as a consequence of part (1).
\qed

\subsubsection*{Proof of Lemma \ref*{alem:infsup}}

\paragraph{\emph{(1)}} Note first that, by construction of $\ushort{\bm{\iota}}^{(0)}$, all elements in the first column and all elements in the last row of this matrix are equal to 1. Hence, part (3)(a) of Lemma \ref{alem:shapeT} implies that $\inf \mathcal{T}_{\ushort{\bm{\iota}}^{(0)} } = \inf  \mathcal{T}$. Second, observe that
\begin{equation*}
\begin{aligned}
 \ushort{j}_k \left( \ushort{\bm{\iota}}^{(m)} \right) 
 &=
 \left\{  
\begin{array}{ll}
      1             &  \text{if} \  k =1 ,  \\
    \check{G} (0 )  &  \text{if} \  2 \leq k  \leq    \check{G} (1 )-1   , \\
    \check{G} (0 ) - m & \text{if} \  k = \check{G} (1 )  ;
\end{array}   
\right.   \\
\bar{j}_k \left( \ushort{\bm{\iota}}^{(m)} \right)  
&=  
 \left\{  
  \begin{array}{ll}     
        \check{G} (0 )  - m  &  \text{if} \  k = 1   , \\
        \check{G} (0 )  &  \text{if} \ k  >1 ; 
  \end{array}   
\right.   \\
\ushort{k}_j \left( \ushort{\bm{\iota}}^{(m)} \right) 
&=    
  \left\{  
  \begin{array}{ll}
      1               &  \text{if} \  j  \leq \check{G} (0 ) - m   , \\
      \check{G} (1 )  &  \text{if} \   \check{G} (0 ) - m   < j   \leq  \check{G} (0 ) - 1 , \\
      2               & \text{if}\    j = \check{G} (0 )  \ \text{and}  \  m \geq 1 ;
\end{array}   
\right. \\
\bar{k}_j \left( \ushort{\bm{\iota}}^{(m)} \right) 
&=   
\left\{  
  \begin{array}{ll}
      1             &  \text{if} \  j  < \check{G} (0 ) - m   , \\
    \check{G} (1 )  &  \text{if} \  j  \geq \check{G} (0 ) - m .
  \end{array}   
\right. 
\end{aligned} 
\end{equation*}  
As a result, using these subindexes in part (1) of Lemma \ref{alem:shapeT} yields
\begin{multline*}
\sup \mathcal{T}_{\ushort{\bm{\iota}}^{(0)} }    \ =   \  \lowmathcal{t} 
  \ + \   \Prob ( D = 0)
    \left\{ \sum_{j=1}^{\check{G}(0)-1} \pi_j ( 0 , 0)  \check{\mu}_{1} (1)  +\pi_{\check{G}(0)} ( 0 , 0) \check{\mu}_{\check{G}(1)} (1)  \right\}  \\
 -   \  \Prob ( D = 1)     
 \left\{ \pi_1 ( 1 , 1)  \check{\mu}_{1} (0)   +  \sum_{k=2}^{\check{G}(1)}  \pi_{k} ( 1 , 1) \check{\mu}_{\check{G}(0)} (0)  \right\} ,
\end{multline*}
as well as
\begin{multline} \label{aeq:infilow}
\inf \mathcal{T}_{\ushort{\bm{\iota}}^{(m)} }    
=      
\lowmathcal{t}  
+    
\Prob ( D = 0)  
 \left\{
  \sum_{j=1}^{\check{G}(0)-m}  \pi_j ( 0 , 0)  \check{\mu}_{1} (1)   +   \sum_{j=\check{G}(0)-m+1}^{\check{G}(0)-1} \pi_j ( 0 , 0)  \check{\mu}_{\check{G}(1)} (1)  +\pi_{\check{G}(0)} ( 0 , 0) \check{\mu}_{2} (1)  \right\} \\
-    
\Prob ( D = 1) 
\left\{ \pi_1 ( 1 , 1)  \check{\mu}_{\check{G} (0 ) - m } (0)   +  \sum_{k=2}^{\check{G}(1)}  \pi_{k} ( 1 , 1) \check{\mu}_{\check{G}(0)} (0)  \right\}
\end{multline}
and 
\begin{multline} \label{aeq:supilow}
\sup \mathcal{T}_{\ushort{\bm{\iota}}^{(m)} }    
=     
\lowmathcal{t} 
+   
\Prob ( D = 0)  
 \left\{\sum_{j=1}^{\check{G}(0)-m-1}  \pi_j ( 0 , 0)  \check{\mu}_{1} (1)   +   \sum_{j=\check{G}(0)-m}^{\check{G}(0)} \pi_j ( 0 , 0)  \check{\mu}_{\check{G}(1)} (1)    \right\} \\
-    
\Prob ( D = 1) 
\left\{ \pi_1 ( 1 , 1)  \check{\mu}_{1} (0)   +  \sum_{k=2}^{\check{G}(1)-1}  \pi_{k} ( 1 , 1) \check{\mu}_{\check{G}(0)} (0)  +  \pi_{ \check{G}(1) } ( 1 , 1)  \check{\mu}_{\check{G}(0)-m} (0)    \right\}
\end{multline}
for $m \geq 1$. Then, $\inf \mathcal{T}_{\ushort{\bm{\iota}}^{(m)} }  \leq \sup \mathcal{T}_{\ushort{\bm{\iota}}^{(m-1)} } \leq   \sup \mathcal{T}_{\ushort{\bm{\iota}}^{(m)} }$ follows from equations (\ref{aeq:infilow}) and (\ref{aeq:supilow}).

\paragraph{\emph{(2)}} This part follows arguments similar to that used in the proof of part (1). Specifically, note that $\sup \mathcal{T}_{\bar{\bm{\iota}}^{(0)} } = \sup  \mathcal{T}$ follows by part (3)(b) of Lemma \ref{alem:shapeT} and that
\begin{equation*}
\begin{aligned}
\ushort{j}_k \left( \bar{\bm{\iota}}^{(m)} \right) 
&=  
\left\{  
  \begin{array}{ll}
    1            &  \text{if}\  k  \leq \check{G} (1 ) - m ,  \\
    \check{G}(0) & \text{if}\  \check{G}(1) - m < k   \leq  \check{G} (1 ) - 1   , \\
    2           & \text{if} \    k = \check{G} (1 )  \ \text{and}  \ m \geq 1    ;
\end{array}   
\right.   \\
\bar{j}_k \left( \bar{\bm{\iota}}^{(m)} \right)  
&=  
 \left\{ 
  \begin{array}{ll}
      1             &  \text{if} \  k  <  \check{G} (1 ) - m   , \\
    \check{G} (0 )  &  \text{if} \ k  \geq  \check{G} (1 ) - m ; 
  \end{array}   
\right.   \\
\ushort{k}_j \left( \bar{\bm{\iota}}^{(m)} \right) 
&=    
    \left\{  
    \begin{array}{ll}
        1             &  \text{if} \  j  = 1   , \\
      \check{G} (1 )  &  \text{if} \   1 <  j   \leq  \check{G} (0 ) - 1 , \\
  \check{G} (1 )  - m & \text{if} \   j = \check{G} (0 ) ;
\end{array}   
\right. \\
\bar{k}_j \left( \bar{\bm{\iota}}^{(m)} \right) 
&=   
  \left\{  
    \begin{array}{ll}
        \check{G} (1 )  - m   &  \text{if} \  j  =1    , \\
              \check{G} (1 )  &  \text{if} \  j  > 1 .
    \end{array}   
\right. 
\end{aligned}
\end{equation*}
Then, part (1) of Lemma \ref{alem:shapeT} implies that
\begin{multline*}
\inf \mathcal{T}_{\bar{\bm{\iota}}^{(0)} }   
=    
\lowmathcal{t} 
+ 
\Prob ( D = 0)
\left\{ \pi_1 ( 0 , 0)  \check{\mu}_{1} (1)   +   \sum_{j=2}^{\check{G}(0)} \pi_j ( 0 , 0)  \check{\mu}_{\check{G}(1)} (1)  \right\}  \\
 -   \Prob ( D = 1)   \left\{  \sum_{k=1}^{\check{G}(1)-1} \pi_k ( 1 , 1 )  \check{\mu}_{1} (0)  + \pi_{\check{G}(1)} ( 1 , 1)  \check{\mu}_{\check{G}(0)} (0)  \right\}    ,
\end{multline*}
as well as
\begin{multline}\label{aeq:infiup}
\inf \mathcal{T}_{\bar{\bm{\iota}}^{(m)} }   
=     
\lowmathcal{t} 
+    
\Prob ( D = 0)  
 \left\{ \pi_1 ( 0 , 0)  \check{\mu}_{1} (1)   +   \sum_{j=2}^{\check{G}(0)-1} \pi_j ( 0 , 0)  \check{\mu}_{\check{G}(1)} (1)  +\pi_{\check{G}(0)} ( 0 , 0) \check{\mu}_{\check{G}(1)-m} (1)  \right\} \\
-    
\Prob ( D = 1) 
\left\{ \sum_{k=1}^{\check{G}(1)-m-1}  \pi_k ( 1 , 1)  \check{\mu}_{1} (0)   +  \sum_{k=\check{G}(1)-m}^{\check{G}(1)}  \pi_{k} ( 1 , 1) \check{\mu}_{\check{G}(0)} (0)  \right\}
\end{multline}
and 
\begin{multline} \label{aeq:supiup}
\sup \mathcal{T}_{\ushort{\bm{\iota}}^{(m)} }    
=      
\lowmathcal{t} 
 +    
\Prob ( D = 0)  
\left\{ \pi_1 ( 0 , 0)  \check{\mu}_{\check{G}(1)-m} (1)   +   \sum_{j=2}^{\check{G}(0)} \pi_j ( 0 , 0)  \check{\mu}_{\check{G}(1)} (1)    \right\} \\
-   
\Prob ( D = 1) 
\left\{\sum_{k=1}^{\check{G}(1)-m}\pi_k ( 1 , 1)  \check{\mu}_{1} (0)   +  \sum_{k=\check{G}(1)-m+1}^{\check{G}(1)-1}  \pi_{k} ( 1 , 1) \check{\mu}_{\check{G}(0)} (0)  +  \pi_{ \check{G}(1) } ( 1 , 1)  \check{\mu}_{2} (0)    \right\} 
\end{multline}
for $m \geq 1$. Then, the desired inequalities follows from equations (\ref{aeq:infiup}) and (\ref{aeq:supiup}).

\paragraph{\emph{(3)}} This part emerges after setting $m= \check{G} (0 ) -1$ and $m = \check{G} (1 ) -1$ in Eqs.\ (\ref{aeq:supilow}) and (\ref{aeq:infiup}), respectively, which yields
\begin{multline*}
\sup \mathcal{T}_{\ushort{\bm{\iota}}^{(\check{G} (0 ) -1)}} \ = \ 
 \lowmathcal{t} 
 \  + \   \Prob ( D = 0)   \check{\mu}_{\check{G}(1)} (1)    \\
- \   \Prob ( D = 1) 
\left\{ \pi_1 ( 1 , 1)  \check{\mu}_{1} (0)   +  \sum_{k=2}^{\check{G}(1)-1}  \pi_{k} ( 1 , 1) \check{\mu}_{\check{G}(0)} (0)  +  \pi_{ \check{G}(1) } ( 1 , 1)  \check{\mu}_{1} (0)    \right\} .
\end{multline*}
and
\begin{multline*}
\inf \mathcal{T}_{\bar{\bm{\iota}}^{(\check{G} (1 ) -1)}} \ = \   \lowmathcal{t} 
\  +  \   \Prob ( D = 0)  
 \left\{ \pi_1 ( 0 , 0)  \check{\mu}_{1} (1)   +   \sum_{j=2}^{\check{G}(0)-1} \pi_j ( 0 , 0)  \check{\mu}_{\check{G}(1)} (1)  +\pi_{\check{G}(0)} ( 0 , 0) \check{\mu}_{1} (1)  \right\}  \\
 \quad  - \   \Prob ( D = 1)   \check{\mu}_{\check{G}(0)} (0) .
\end{multline*}

\paragraph{\emph{(4)}}  It follows as a consequence of the next inequalities, which can be derived from previous parts:
\begin{equation*}
\begin{aligned}
& \inf \mathcal{T}_{\ushort{\bm{\iota}}^{(0)} } = \inf  \mathcal{T} \leq  \inf \mathcal{T}_{\ushort{\bm{\iota}}^{(1)} }  \leq  \sup \mathcal{T}_{\ushort{\bm{\iota}}^{(0)} } \leq  \sup \mathcal{T}_{\ushort{\bm{\iota}}^{(1)} }  ,    \cdots, 
\\
& \inf \mathcal{T}_{\ushort{\bm{\iota}}^{(\check{G} (0 ) -1)} }  \leq \sup \mathcal{T}_{\ushort{\bm{\iota}}^{(\check{G} (0 ) -2)} } \leq   \sup \mathcal{T}_{\ushort{\bm{\iota}}^{(\check{G} (0 ) -1)} } ,  \ \inf \mathcal{T}_{\bar{\bm{\iota}}^{(\check{G} (1 ) -1)}} \leq \sup \mathcal{T}_{\ushort{\bm{\iota}}^{(\check{G} (0 ) -1)}}, 
\\
& \inf \mathcal{T}_{\bar{\bm{\iota}}^{(\check{G} (1 ) -1)} }  \leq \sup \mathcal{T}_{\bar{\bm{\iota}}^{(\check{G} ( 1 ) -2)} } \leq   \sup \mathcal{T}_{ \bar{\bm{\iota}}^{(\check{G} ( 1) -1)} }   , \,   \cdots  , \ 
 \inf \mathcal{T}_{\bar{\bm{\iota}}^{(0)} }   \leq \sup \mathcal{T}_{\bar{\bm{\iota}}^{(1)} }  \leq \sup  \mathcal{T} =  \sup \mathcal{T}_{\bar{\bm{\iota}}^{(0)} }. 
\end{aligned}
\end{equation*}
\qed

\subsection{Asymptotics and Inference} 	\label{sec:ai}

In this section, we establish the asymptotic properties of the proposed estimators and provide valid inference procedures. Specifically, we present a bootstrap procedure to perform valid inference on $\ell$. All asymptotic results are derived with the sample size $n$ growing to infinity.
Proofs of the lemmas and theorems stated in this section are relegated to Section \ref{sec:proofsupple} below.

% : therefore, inference on conservative measures such as MLB $\ell_{\bar{G}_C}$ or ALB $\tau_{ALB}$ can be performed by using it

% The proofs of the results stated in this section are provided in Section \ref{sec:proofsupple} below.

As a starting point, let
\begin{equation*}
\Upsilon \ : = \ \left( \, \Prob ( D = 0 ), \,    \bm{\pi} (0) ,  \check{\bm{\mu}} (0)  ,  \,  \bm{\pi} (1) ,  \check{\bm{\mu}} (1)  \, \right)  
\end{equation*}
denote the vector of parameters that characterize the identified sets for $\tau$, where
\begin{eqnarray*}
\bm{\pi} ( d ) &  : =  &  \bigl( \pi_1(d,d),\cdots, \pi_{\check{G}(d)-1}(d,d) \bigr)  , \\
\check{\bm{\mu}} (d)    &  : =  & 	 \bigl(  \check{\mu}_1(d),\cdots, \check{\mu}_{\check{G}(d)} \bigr) .
\end{eqnarray*}
Let $\hat\Upsilon$ be an estimator of $\Upsilon$ constructed using some estimators $(\hat{G}(0),\hat{G}(1))$ of the mixture orders $(\check G(0),\check G(1))$.
Then, write
\begin{equation}
\label{eqsupptrue}
\ell = \mathcal{M} \circ \mathcal{L}(\Upsilon) 
\end{equation}
and, for the sake of generality, consider any estimator of the form
\begin{equation}
\label{eqsuppesti}
\hat\ell : = \hat{\mathcal{M}} \circ \hat{\mathcal{L}} (\hat\Upsilon) , 
\end{equation}
where, with a slight abuse of notation, here we modify the definitions of $\mathcal{L}$ and $\hat{\mathcal{L}}$ in Section \ref{sec:esti} so that they take $\Upsilon$ and $\hat\Upsilon$ as their corresponding arguments, rather than $\Lambda$ and $\hat\Lambda$. respectively. Recall also that $\mathcal{M}$ and $\hat{\mathcal{M}}$ has been defined in Section \ref{sec:esti}.

We now impose the following high-level assumption on the estimators $(\hat{G}(0),\hat{G}(1))$ and on the infeasible counterpart $\tilde\Upsilon$ of $\hat\Upsilon$, obtained by replacing $(\hat{G}(0),\hat{G}(1))$ with the true mixture orders $(\check G(0),\check G(1))$. We formulate the assumption in terms of $\tilde\Upsilon$ rather than $\hat\Upsilon$ because the resulting condition is typically easier to verify in practice when the mixture orders are treated as fixed rather than estimated. This formulation is also clearer and aligns with standard textbook treatments, where the mixture order is usually taken to be known.

\begin{assump}		\label{ass:gralesti}
The following conditions hold.
\begin{enumerate}

\item For $d=0,1$, $\hat{G} (d)$ satisfies $\Prob ( \hat{G} (d) = \check{G} (d) ) \rightarrow 1$

\item $\sqrt{n} ( \tilde{\Upsilon}  - \Upsilon ) \convd N ( 0 , \mathrm{V}  )$ for some positive definite matrix $\mathrm{V}$.

\end{enumerate}
\end{assump}

The next lemma shows that Assumption \ref{ass:gralesti} is satisfied by the estimation procedure proposed in Section \ref{sec:esti} under Assumption \ref{ass:setup}. Specifically, this result applies when $\hat{G}(d)$ is estimated by \citet{chen09order}'s method, while $(\bm{\pi}(d)  , \check{\bm{\mu}}(d) )$ is estimated by infeasible MLE over a suitably chosen parameter space $\Gamma(\check{G}(d))$ as specified in the proof of the lemma.\footnote{We remark that, with appropriate modifications to the proof, this result can also be extended to the settings covered by Assumptions \ref{ass:sampleselect} and \ref{ass:exten}.}

\begin{lemma}		\label{lem:gaussok}
Under Assumption \ref{ass:setup}, the estimation procedure proposed in Section \ref{sec:esti} satisfies Assumption \ref{ass:gralesti}.
\end{lemma}

It follows from Assumption \ref{ass:gralesti} and the Delta method that 
\begin{equation}		\label{eq:prenorm}
\sqrt{n} \bigl(  \mathcal{L}( \tilde\Upsilon)   -  \mathcal{L}(\Upsilon) \bigr)  \convd  \mathbb{G}, 
\end{equation} 
where $\mathbb{G} : = ( \mathbb{G}_{\bm\iota}  )_{\bm\iota \in \mathscr{I}_{LC}}\sim N  \bigl( 0 , \mathrm{J}_{\mathcal{L}} ( \Upsilon ) \, \mathrm{V}  \, \mathrm{J}_{\mathcal{L}} ( \Upsilon )^\tr \bigr)$, and $\mathrm{J}_{\mathcal{L}} $ represents the Jacobian matrix of $\mathcal{L}$. We remark that $\mathrm{J}_{\mathcal{L}} ( \Upsilon ) \, \mathrm{V}  \, \mathrm{J}_{\mathcal{L}} ( \Upsilon )^\tr$ is not neccesarily positive definite. However, the marginal distributions of $\mathbb{G}$ have positive variance because $\mathrm{V}$ is positive definite and every row of $\mathrm{J}_{\mathcal{L}}$ contains at least one non-zero entry.

We are now ready to describe the asymptotic distribution of $\hat\ell$.

\begin{lemma}	\label{lem:asymp}
Suppose that one of Assumptions \ref{ass:setup}--\ref{ass:exten} holds and that Assumption \ref{ass:gralesti} is satisfied. Then, $\sqrt{n}\bigl(\hat\ell - \ell \bigr) \convd \mathbb{M} := \mathcal{M}^\prime ( \mathbb{G}  )$, where $\mathcal{M}^\prime$ denotes the Hadamard directional derivative of $\mathcal{M}$ at $ \mathcal{L} ( \Upsilon )$, which is given by
\begin{equation*}
\mathcal{M}^\prime ( v ) =  \left(  \mathcal{M}_{\bar{G}_L}^\prime  (v) , \cdots  ,  \mathcal{M}_{\bar{G}_C}^\prime   (v)  \right)  \in \mathbb{R}^{\bar{G}_C - \bar{G}_L +1}
\end{equation*}
with
\begin{equation*} 
\mathcal{M}^\prime_m ( v ) 
=
\left\{
\begin{aligned} 
&\min \Bigl\{v _{\bm\iota} :\ \bm\iota\in {\mathscr{I}}(m),\ {\mathcal{L}}_{\bm\iota}(\Upsilon) = \ell_m   \Bigr\}  && \text{for $\bar G_L\leq m < \bar G_C$} , \\
&v_{ {\ushort{\bm\iota}}  }  && \text{for $m=\bar G_C$} .
\end{aligned}
\right.
\end{equation*} 
In particular, the MLB estimator is asymptotically normal in that $\sqrt{n}(\hat{\ell}_{\hat{G}_{C}} - {\ell}_{\bar{G}_{C}}) \convd  \mathbb{G}_{\ushort{\bm\iota}}$.
\end{lemma}

% \SJtodo{should we comment somewhere that this is not degenerate? Since the variance matrix of $\mathbb{G}$ is generally only positive semi-definite, one may be concerned that transformation of $\mathbb{G}$ may yield something degenerate. My understanding is that this does not happen because all the marginals of $\mathbb{G}$ is not degenerate. Right?} 

The proof of Lemma \ref{lem:asymp} is based on decomposing $\hat\ell - \ell$ into two components:
\begin{equation}
\label{eq:explalemma}
\sqrt{n}\bigl(\hat\ell - \ell \bigr)  
= 
\underbrace{\sqrt{n}\bigl\{ \hat{\mathcal{M}} \circ \hat{\mathcal{L}}  (\hat\Upsilon) -  \mathcal{M} \circ \mathcal{L}  (\tilde\Upsilon)  \bigr\}}_{\convp\ 0}   
+   
\underbrace{\sqrt{n}\bigl\{  \mathcal{M} \circ \mathcal{L}  (\tilde\Upsilon)   -   \mathcal{M} \circ \mathcal{L}  (\Upsilon)\bigr\}}_{\convd \ \mathcal{M}^\prime ( \mathbb{G}  ) }  .
\end{equation}
The first term on the right-hand side represents the estimation error that arises from using $\bigl( \check G(0), \check G(1) \bigr)$ instead of the true mixture orders. This term converges to zero in probability because the only difference between $\hat\ell = \hat{\mathcal{M}} \circ \hat{\mathcal{L}}  (\hat\Upsilon)$ and $\mathcal{M} \circ \mathcal{L}  (\tilde\Upsilon)$ is whether $\bigl( \hat G(0), \hat G(1) \bigr)$ or $\bigl( \check G(0), \check G(1) \bigr)$ is used, but $\hat G(d)$ converges to $\check G(d)$ at an arbitrarily fast rate by Assumption \ref{ass:gralesti}. Regarding the second term, the convergence result follows from an extension of the Delta method applicable to Hadamard directionally differentiable functions. We refer to the proof of Lemma \ref{lem:asymp} below for further details.

%As a simplified version of the problem, consider estimating $\min\{ \Exp(X_1), \Exp(X_2) \}$ by using a random sample of size $n$ that is drawn from the distribution of $(X_1,X_2)$. The natural estimator will be $\min(\bar X_1, \bar X_2)$ with $\bar X_j$ being the corresponding sample average.  Let $\mathbb{Z} := (\mathbb{Z}_1, \mathbb{Z}_2)$ be the Gaussian distribution representing the limiting distribution of $\sqrt{n}\bigl( \bar X_1 - \Exp(X_1), \bar X_2 - \Exp(X_2) \bigr)^\tr$. Then, it is not difficult to see that $\sqrt{n}\bigl[ \min(\bar X_1,\bar X_2) - \min\{\Exp(X_1), \Exp(X_2) \} \bigr]$ converges in distribution to $\mathbb{Z}_1$ if $\Exp(X_1) <\Exp(X_2)$, $\mathbb{Z}_2$ if $\Exp(X_1) >\Exp(X_2)$, or $\min( \mathbb{Z}_1, \mathbb{Z}_2)$ if $\Exp(X_1) = \Exp(X_2)$. Put differently, $\sqrt{n}\bigl[ \min(\bar X_1,\bar X_2) - \min\{\Exp(X_1), \Exp(X_2) \} \bigr] \convd \min\bigl\{ \mathbb{Z}_j:\ j \in \argmin\{\Exp(X_1), \Exp(X_2)  \} \bigr\}$. The expression in Lemma \ref{lem:asymp} is an adaptation of this reasoning in our context.

Lemma \ref{lem:asymp} also shows that the asymptotic distribution of $\sqrt{n}(\hat{\ell}_{\hat{G}_{C}} - {\ell}_{\bar{G}_{C}})$ is mean-zero normal with positive variance. For the other cases of $\bar G_L \leq m < \bar G_C$, the min functions in $\mathcal{M}$ cannot be ignored in general, making the asymptotic distribution in Lemma \ref{lem:asymp} generally non-Gaussian: it may not be even mean-zero. Therefore, although we can use the asymptotic distribution $\mathbb{M}$ for inference in principle, it is not a convenient approach because the limiting distribution is non-standard. We propose using the bootstrap instead.

Let $s_n > 0$ be a (possibly stochastic) sequence that satisfies $s_n = o_p(1)$ and $ s_n^{-1} = o_p(\sqrt{n})$, which is a slackness parameter proposed in \citet{cht07}. We then introduce the following estimator of $\mathcal{M}^\prime$: $\hat{\mathcal{M}}^\prime (v)  : =  ( \hat{\mathcal{M}}^\prime_{ \hat G_L } (v ), \cdots ,  \hat{\mathcal{M}}^\prime_{ \hat G_C } (v ) )$, where
\begin{equation*} 
\hat{\mathcal{M}}^\prime_m ( v ) 
:=
\left\{
\begin{aligned} 
&\min \Bigl\{v _{\bm\iota} :\ \bm\iota\in \hat{\mathscr{I}}(m),\ \hat{\mathcal{L}}_{\bm\iota}(\hat\Upsilon) \leq \hat\ell_m + s_n  \Bigr\}  && \text{for $\hat G_L\leq m < \hat G_C$} , \\
&v_{ \hat{\ushort{\bm\iota}}  }  && \text{for $m=\hat G_C$} ,
\end{aligned}
\right.
\end{equation*} 
and we write $\hat{\ushort{\bm\iota}}$ and $\hat{\mathscr{I}}(m)$ when $(\hat G(0)  ,\hat G(1))$ is used instead of $(\check G(0)  ,\check G(1))$ in their corresponding definitions.
Now let $\hat{\Upsilon}^\ast$ be the infeasible bootstrap analog of $\hat\Upsilon$, constructed via either the nonparametric or parametric bootstrap, without resampling the estimated mixture orders $\hat G(0)$ and $\hat G(1)$. By equation (\ref{eq:prenorm}) and Lemma \ref{lem:asymp}, it is natural to consider
\begin{equation*}
\mathbb{M}^\ast:= \hat{\mathcal{M}}^\prime \left( \sqrt{n}  \left\{ \hat{\mathcal{L}} (\hat\Upsilon^\ast )  - \hat{\mathcal{L}} ( \hat\Upsilon ) \right\}  \right) 
\end{equation*}
as a bootstrap analog of $\mathbb{M}$.

The next theorem establishes the asymptotic validity of the proposed bootstrap method, for which we impose the following high-level assumption on the infeasible bootstrap analog. Let $\tilde{\Upsilon}^\ast$ denote the bootstrap analog of $\tilde\Upsilon$, i..e., the infeasible version of $\hat{\Upsilon}^\ast$ that is constructed with the same bootstrap algorithm as $\hat{\Upsilon}^\ast$ but using the true mixture orders $\check G(0)$ and $\check G(1)$.

\begin{assump}		\label{ass:gralboot}
The conditional distribution function of $\sqrt{n} ( \tilde{\Upsilon}^\ast -  \tilde\Upsilon)$ given the sample converges in probability to to the distribution function of a $N(0,\mathrm{V})$ random vector.
\end{assump}

\begin{theorem} \label{thm:boot}
Suppose that one of Assumptions \ref{ass:setup}--\ref{ass:exten} holds and that Assumption \ref{ass:gralesti}--\ref{ass:gralboot} are satisfied. Then, the conditional distribution function of $\mathbb{M}^\ast$ given the sample converges in probability to the distribution function of $\mathbb{M}$ at every continuity point of the distribution of $\mathbb{M}$. 
\end{theorem}

The proof of this theorem proceeds by verifying that the conditions of Theorem 3.2 in \cite{fang19inf} are satisfied. In particular, we invoke \citet[Theorem 3.1.(1)]{cht07} to show that $\hat{\mathcal{M}}^\prime$ is a consistent estimator of $\mathcal{M}^\prime$, along with additional supporting results.

Theorem \ref{thm:boot} is useful because it enables us to construct valid confidence intervals and to perform hypothesis tests. For example, a pointwise confidence interval for each $\ell_{(m)}$ can be constructed by the percentile method as follows. For $a \in (0,1)$, let $\hat{\lowmathcal{q}}_{(m)}(a)$ be the $a$--quantile of $\mathbb{M}^{\ast}_{(m)}$, which can be obtained from the bootstrap algorithm. Then, a two-sided $1-\alpha$ confidence interval for $\ell_{(m)}$ can be constructed by 
\begin{equation*}
\mathrm{CI}_{(m)}( 1-\alpha) 
:=  
\left[   
  \hat\ell_{(m)}   -  \frac{\hat{\lowmathcal{q}}_{(m)}(1 - \alpha/2)}{\sqrt{n}},\ 
  \hat\ell_{(m)}   -  \frac{\hat{\lowmathcal{q}}_{(m)}(\alpha/2)}{\sqrt{n}}  
\right]. 
\end{equation*}
We know from Lemma \ref{lem:asymp} and Theorem \ref{thm:boot} that $\Pr \{ \ell_{(m)} \in \mathrm{CI}_{(m)}( 1-\alpha)  \} \rightarrow 1 - \alpha$.  Finally, we remark that inference for various summary statistics such as $\tau_{ALB}$, as well as the construction of uniform confidence bands, can be similarly carried out since the bootstrap procedure is valid for the joint distribution of $\mathbb{M}$.

\subsubsection{Proofs}	\label{sec:proofsupple}

\subsubsection*{Proof of Lemma \ref{lem:gaussok}} \label{sec:proofgaussok}
The first assertion of Assumption \ref{ass:gralesti} is immediately satisfied by Theorem 2 in \cite{chen09order}. To show that the second assertion is also satisfied, letting $\tilde{\Lambda}$ be the infeasible MLE estimator of $\Lambda$ that uses the true mixture orders instead of their estimates,  it suffices to prove that $\sqrt{n} (  \tilde{\Lambda}   -  \Lambda )$ converges in distribution to a non-degenerate normal distribution.

For this purpose, let $\Gamma_0$ be a compact subset of $( 0 ,1)$ such that $\Pr ( D  = 0) \in \mathrm{interior}(\Gamma_0)$, and let $\Gamma $ be a set-valued mapping that assigns to each $k \in \mathbb{N}$ a nonempty and compact set satisfying the next conditions: $\Gamma (1) \subset \mathbb{R} \times \mathbb{R}_{++}$,
\begin{equation*}
\Gamma (k) \subset \left\{  \left( p_1 , \cdots , p_{k-1}  \right)  \in (0,1)^{k-1} : \sum_{j = 1}^{k-1} p_j < 1  \right \}  
   \times  \left\{  \left( m_1 , \cdots , m_{k}  \right)  \in \mathbb{R}^{k} :  m_1 < \cdots < m_{k}  \right\}  \times  \mathbb{R}_{++}
\end{equation*}
for $k \geq 2$, and $\gamma (d) \in \mathrm{interior} \left( \Gamma \left(  \check{G} (d) \right)  \right)$ for $d = 0 , 1$. Observe that $\tilde{\Lambda}$ is indeed the (infeasible) maximum likelihood estimator of the joint distribution of $( D, Y)$; specifically,
\begin{multline*}
\tilde{\Lambda}  
=  
\argmax \ \sum_{i=1}^n  (1 - D_i) \left[ \log(\tilde{p}) +  \log \left\{ \sum_{j=1}^{G(0)} p_j (0) \phi \left( \frac{Y_i -  m_j (0) }{\sqrt{s (0)}} \right) \right\} \right]  \\
+    
\sum_{i=1}^n  D_i \left[ \log(1-\tilde{p}) +  \log \left\{ \sum_{j=1}^{G(1)} p_j (1) \phi \left( \frac{Y_i -  m_j (1) }{\sqrt{s (1)}} \right) \right\} \right]  ,
\end{multline*}
where the maximization is taken over $ \tilde{p} \in \Gamma_0$ and
\begin{equation*}
\left( p_1 (d), \cdots  , p_{\check{G} (d)-1} (d) ,  m_1 (d), \cdots  , m_{\check{G} (d)} (d)   ,  s(d) \right) \in \Gamma \left(  \check{G} (d) \right) \ \ \text{for}  \ d = 0 , 1,
\end{equation*}
setting $p_{\check{G} (d)} = 1 - \sum_{j=1}^{G(d)} p_j (d) $ when $\check{G}(d)\geq2$, while $p_{\check{G} (d)} = 1$ when $\check{G}(d)=1$.
Then, the desired result follows by noting that the conditions of Theorem 3.3 in \cite{NeweyMcFadden94} are satisfied in this context. In particular, we remark that the Fisher information matrix is positive definite because the Fisher information of a $\check{G}(d)$-order normal mixture is itself positive definite; see, e.g., \citet[Section 2]{ch09inf}.
\qed

\subsubsection*{Proof of Lemma \ref{lem:asymp}}		\label{sec:proofthmasymp}

Note that \begin{equation*}
\sqrt{n}\bigl\{  \mathcal{M} (\mathcal{L}  (\tilde\Upsilon) )  -   \mathcal{M} ( \mathcal{L}  (\Upsilon)  ) \bigr\}    \convd \ \mathcal{M}^\prime ( \mathbb{G}  )  
\end{equation*}
follows by equation \eqref{eq:prenorm} and \citet[Theorem 2.1]{fang19inf}; see also \cite{sha91asymp}. Then, the desired asymptotic distribution can be obtained from equation \eqref{eq:explalemma}.
Finally, the expression for the Hadamard derivative $\mathcal{M}^\prime $ can be obtained by applying Lemma S.4.9 in \citet[Supplemental Appendix]{fang19inf}, for which we observe that each vector $v = (v_{\bm\iota}   )_{\bm\iota \in \mathscr{I}_{LC}}$ can be represented by a bounded real-valued function on $\mathscr{I}_{LC}$ defined by the map $\bm\iota \overset{v}{\mapsto} v_{\bm\iota} $. 
Finally, the asymptotic the distribution of $\sqrt{n}(\hat{\ell}_{\hat{G}_{C}} - {\ell}_{\bar{G}_{C}})$ can be derived from Lemma \ref{alem:shapeT}.(3).(a) and the characterization of $\mathcal{M}^\prime_{\bar{G}_C}$.
\qed

\subsubsection*{Proof of Theorem \ref{thm:boot}}	\label{sec:proofasymp}

For a vector space $V$ equipped with a norm $\| \cdot \|$, let $\mathrm{BL} ( V ) $ denote the class of real-valued functions $\psi$ defined on $V$ that satisfies $| \psi (v) | \leq  1$ and $| \psi (v)  - \psi (w) | \leq \| v - w \|$ for all $v,w \in V$. Denote $\mathcal{S}_n  : = \{ (Y_1, D_1), \dots, (Y_n, D_n) \}$, or $\mathcal{S}_n : = \{ (Y_1, X_1 , D_1), \dots, (Y_n, X_n,  D_n) \}$ in the presence of exogenous covariates (Assumption \ref{ass:sampleselect}).
We establish the desired result by proving that
\begin{equation}
\label{eq:need1}
\sup_{\psi \in \mathrm{BL} \left(  \mathbb{R}^{\bar{G}_C - \bar{G}_L +1} \right) }  \,  \left|  \E \left\{  \psi (  \mathbb{M}^\ast ) \mid \mathcal{S}_n \right\} -  \E \left\{ \psi ( \mathbb{M} )  \right\}  \right| \ = \ o_p (1) .
\end{equation}
With this aim, write $\tilde{\mathcal{M}}^\prime (v)  : =  ( \tilde{\mathcal{M}}^\prime_{  \bar{G}_L  (v )}, \cdots ,  \tilde{\mathcal{M}}^\prime_{   \bar{G}_C } (v ) )$ with
\begin{equation*} 
\tilde{\mathcal{M}}^\prime_m ( v ) 
:=
\left\{
\begin{aligned} 
&\min \Bigl\{v _{\bm\iota} :\ \bm\iota\in \mathscr{I}(m),\ \mathcal{L}_{\bm\iota}(\tilde\Upsilon) \leq \tilde{l}_m + s_n  \Bigr\}  && \text{for $ \bar{G}_L\leq m <   \bar{G}_C$} , \\
&v_{ \ushort{\bm\iota}  }  && \text{for $m=  \bar{G}_C$} ,
\end{aligned}
\right.
\end{equation*} 
and
\begin{equation*}
\tilde{\mathbb{M}}^\ast:= \tilde{\mathcal{M}}^\prime \left( \sqrt{n}  \left\{ \mathcal{L} (\tilde\Upsilon^* )  - \mathcal{L} ( \tilde\Upsilon ) \right\}  \right) ,
\end{equation*}
where $\tilde{l}_m =  \mathcal{M}_m \circ \mathcal{L}  (\tilde\Upsilon) $. Observe that, whenever $( \hat{G} (0) ,  \hat{G} (1) ) = ( \check{G} (0) ,  \check{G} (1) ) $, we have $\hat{\mathcal{M}} = \tilde{\mathcal{M}}$ and also both $\hat{\mathbb{M}}^\ast$ and $\tilde{\mathbb{M}}^\ast$ share the same conditional-on-$\mathcal{S}_n$ distribution because $( \hat{G} (0) ,  \hat{G} (1) )$ are not resampled. Thus, for any $\epsilon >0$, the Law of total probability implies
\begin{multline*}
\Pr \left( \sup_{\psi \in \mathrm{BL} (  \bar{G}_C - \bar{G}_L +1 )}  \,  \left|  \E \left\{  \psi (  \tilde{\mathbb{M}}^\ast ) \mid \mathcal{S}_n \right\} - \E \left\{  \psi (  \mathbb{M}^\ast ) \mid \mathcal{S}_n \right\}  \right|   >  \epsilon  \right) \\
\leq \   \  0 \times \Pr \left\{  ( \hat{G} (0) ,  \hat{G} (1) ) = ( \check{G} (0) ,  \check{G} (1) ) \right\} \  +  \ 1 \times \underbrace{\Pr \left\{ \hat{G} (0) ,  \hat{G} (1) ) \neq ( \check{G} (0) ,  \check{G} (1) ) \right\}}_{= \ o(1) \ \text{by condition (\ref{eq:conG}) } }  
\end{multline*}
and therefore establishing (\ref{eq:need1}) reduces to showing that
\begin{equation}
\label{eq:need2}
\sup_{\psi \in \mathrm{BL}   \left(  \mathbb{R}^{\bar{G}_C - \bar{G}_L +1} \right)  }  \,  \left|  \E \left\{  \psi (  \tilde{\mathbb{M}}^\ast ) \mid \mathcal{S}_n \right\} -  \E \left\{ \psi ( \mathbb{M} )  \right\}  \right| \ = \ o_p (1) .
\end{equation}
In words, in the rest of the proof, we can disregard the estimation errors from the mixture orders.

The result in (\ref{eq:need2}) can be obtained by applying Theorem 3.2 in \cite{fang19inf} to the infeasible estimator $\mathcal{L} (\tilde\Upsilon^* )$ and the estimated derivative $\hat{\mathcal{M}}^\prime$. To see this, consider $\mathcal{L}$ as a function from a compact set that contains $\Upsilon$ in its interior to the collection of real-valued bounded functions on $\mathscr{I}_{LC} $, which we denote by $\ell^\infty ( \mathscr{I}_{LC}  )$. Consider also $\mathcal{M}$ as function from $\ell^\infty ( \mathscr{I}_{LC}  )$ to $ \mathbb{R}^{\bar{G}_C - \bar{G}_L +1}$ and $\mathbb{G}$ as a Gaussian process taking values in $\ell^\infty ( \mathscr{I}_{LC}  )$.

We make three remarks to emphasize that Theorem 3.2 in \cite{fang19inf} can be applied to obtain the result in (\ref{eq:need2}).  First, $\mathcal{M}$ is Hadamard directionally differentiable at $ \mathcal{L} ( \Upsilon ) $ tangentially to $\ell^{\infty} ( \mathscr{I}_{LC} )$; see Lemma \ref{lem:asymp}. Second, the conditional distribution function of $\sqrt{n}  \{ \mathcal{L} (\tilde\Upsilon^\ast )  - \mathcal{L} ( \tilde\Upsilon ) \}  $ given the sample converges in probability to the distribution of $\mathbb{G}$, i.e.,
\begin{equation*}
\sup_{\psi \in \mathrm{BL}   \left(  \ell^\infty ( \mathscr{I}_{LC}  ) \right)  }  \,  \left|  \E \left\{  \psi \left(  \sqrt{n}  \{ \mathcal{L} (\tilde\Upsilon^* )  - \mathcal{L} ( \tilde\Upsilon ) \}  \right) \mid \mathcal{S}_n \right\} -  \E \left\{ \psi ( \mathbb{G} )  \right\}  \right| \ = \ o_p (1)    ;
\end{equation*}
this result follows by applying the bootstrap delta method to Assumption \ref{ass:gralboot}.
Third, $\tilde{\mathcal{M}}^\prime$ is a consistent estimator of $\mathcal{M}^\prime$ in that it satisfies the conditions of Remarks 3.4 in \cite{fang19inf}. To see this, pick any $ \bar{G}_L\leq m <   \bar{G}_C$ and note that 
\begin{equation*}
\left\{  \bm\iota\in \mathscr{I}(m) :  \, \mathcal{L}_{\bm\iota}(\tilde\Upsilon) \leq \tilde{l}_m + s_n  \right\}   \ \underset{\text{w.p.a.1}}{=}  \  \argmin_{ \bm\iota \in \mathscr{I}(m) }    \, \mathcal{L}_{ \bm\iota} (\Upsilon)  \  = \  \left\{  \bm\iota\in \mathscr{I}(m) :  \, \mathcal{L}_{\bm\iota}(\Upsilon) \leq \ell_m \right\} ,
\end{equation*}
with the first equality following from Theorem 3.1.(1) in \cite{cht07} and the second by construction of $\mathcal{L}$. So, for any $v \in \ell^{\infty} ( \mathscr{I}_{LC} )$, $\tilde{\mathcal{M}}^\prime (v) = \mathcal{M}^\prime (v)$ holds w.p.a.1.
\qed

\clearpage

\renewcommand{\thepage}{R.\arabic{page}}\setcounter{page}{1}

\bibliography{bibmixdiscrete}

\end{document}